\documentclass[twocolumn,aps,prd,superscriptaddress,longbibliography,nofootinbib]{revtex4-2}

\usepackage{ORCIDinREVTeX}

\usepackage[T1]{fontenc}
\usepackage{xcolor, soul}
	\definecolor{darkgreen}{rgb}{0.0,0.55,0.0}
\usepackage{graphics}
\usepackage{amssymb, bm}
\usepackage{minibox}
\usepackage[colorlinks=true, allcolors=darkgreen]{hyperref}

\usepackage{empheq}
\usepackage{enumerate}
\usepackage{braket}
\usepackage{multirow}
\usepackage{ulem}
\usepackage{float}
\usepackage{cancel}

\usepackage{cleveref}
\usepackage{relsize}
\usepackage{lipsum}

\usepackage[tmargin=1.5cm, bmargin=1.5cm, lmargin=1.37cm, rmargin=1.37cm]{geometry}

\usepackage{lmodern}
\usepackage{enumitem}

\newcommand{\nc}{\newcommand}

\nc{\non}{\nonumber}
\nc{\hsp}{\hspace{0.5cm}}
\nc{\lsp}{\hspace{1cm}}
\nc{\Lsp}{\hspace{2cm}}
\nc{\LLsp}{\lsp\lsp}
\nc{\lra}{\longrightarrow}
\nc{\p}{\prime}
\nc{\dd}{\mathrm{d}}
\nc{\sgn}{\text{sgn}}
\nc{\ph}{\varphi}
\nc{\op}{{\cal O}}
\nc{\cL}{{\cal L}}
\nc{\tr}{{\text{Tr}}}
\nc{\eq}{\text{Eq.~}}
\nc{\cg}{{\cal G}}
\nc{\ch}{{\bm h}}
\nc{\cZ}{\mathbb Z}
\nc{\cw}{\cos\theta_{\textsc w}}
\nc{\sw}{\sin\theta_{\textsc w}}
\nc{\cwsq}{\cos^2\theta_{\textsc w}}
\nc{\swsq}{\sin^2\theta_{\textsc w}}
\def\zBB{{\mathbbm Z}}
\def\z2{\zBB_2}
\nc{\beq}{\begin{equation}}  \nc{\eeq}{\end{equation}}
\nc{\bea}{\begin{eqnarray}}  \nc{\eea}{\end{eqnarray}}
\nc{\baa}{\begin{array}}     \nc{\eaa}{\end{array}}
\nc{\bit}{\begin{itemize}}   \nc{\eit}{\end{itemize}}
\nc{\ben}{\begin{enumerate}} \nc{\een}{\end{enumerate}}
\nc{\bce}{\begin{center}}    \nc{\ece}{\end{center}}
\nc{\bpm}{\begin{pmatrix}}   \nc{\epm}{\end{pmatrix}}
\nc{\bvt}{\begin{verbatim}}  \nc{\evt}{\end{verbatim}}
\def\lsim{\mathrel{\raise.3ex\hbox{$<$\kern-.75em\lower1ex\hbox{$\sim$}}}}
\def\gsim{\mathrel{\raise.3ex\hbox{$>$\kern-.75em\lower1ex\hbox{$\sim$}}}}

\def\udots{\mathinner{\mkern1mu\raise1pt\vbox{\kern7pt\hbox{.}}\mkern2mu\raise4pt\hbox{.}\mkern2mu\raise7pt\hbox{.}\mkern1mu}}
\def\<#1>{\mathinner{\langle#1\rangle}}

\nc{\hc}{\hbox {H.c.}}
\nc{\noi}{\noindent}
\nc{\barx}{\bar{x}}
\nc{\pbarn}{\;\hbox {pb}}
\nc{\fbarn}{\;\hbox {fb}}
\def\mev{\;\hbox{MeV}}
\def\gev{\;\hbox{GeV}}

\def\mpl{M_{\rm Pl}}

\definecolor{agray}{rgb}{0.95, 0.95, 0.99}

\def\sm{\textsc{\rm SM}}
\def\dm{\textsc{\rm DM}}

\def\cxh{{\cal C}_{X}^{\bm h}}
\def\cxp{{\cal C}_{X}^\phi}
\def\mx{m_X}
\def\ghp{g_{h\phi}}
\def\app#1{Appendix~\ref{#1}}
\def\eq#1{Eq.~(\ref{#1})}
\def\fig#1{Fig.~\ref{#1}}
\def\sec#1{Sec.~\ref{#1}}
\def\tab#1{Table~\ref{#1}}
\def\rcite#1{Ref.~\cite{#1}}
\begin{document}

\title{Higgs boson induced reheating and ultraviolet frozen-in dark matter}

\author{Aqeel Ahmed}
\orcid{0000-0002-2907-2433}
\affiliation{Max-Planck-Institut für Kernphysik, Saupfercheckweg 1, 69117 Heidelberg, Germany}

\author{Bohdan Grzadkowski}
\orcid{0000-0001-9980-6335}

\author{Anna Socha}
\orcid{0000-0002-4924-9267}
\affiliation{Faculty of Physics, University of Warsaw, Pasteura 5, 02-093 Warsaw, Poland}

\date{\today}

\begin{abstract}
A reheating phase in the early universe is an essential part of all inflationary models during which not only the Standard Model (SM) quanta are produced but it can also shed light on the production of dark matter.
In this work, we explore a class of reheating models where the reheating is induced by a cubic interaction of the inflaton $\phi$ to the SM Higgs boson ${\bm h}$ of the form $\ghp \mpl \phi |{\bm h}|^2$ adopting the $\alpha$-attractor T-model of inflation. Assuming inflaton as a background field such interaction implies a $\phi$-dependent mass term of the Higgs boson and a non-trivial phase-space suppression of the reheating efficiency. As a consequence, the reheating is prolonged and the maximal temperature of the SM thermal bath is reduced. In particular, due to oscillations of the inflaton field the $\phi$-dependent Higgs boson mass results in periodic transitions between phases of broken and unbroken electroweak gauge symmetry. The consequences of these rapid phase transitions have been studied in detail.
A purely gravitational reheating mechanism in the presence of the inflaton background, i.e., for $\ghp=0$, has also been investigated. It turned out that even though it may account for the total production of SM radiation in the absence of $\ghp$, its contribution to the reheating is subdominant for the range of $\ghp$ considered in this work. Approximate analytical solutions of Boltzmann equations for energy densities of the inflaton and SM radiation have been obtained.  
As a dark matter candidate a massive Abelian vector boson, $X_\mu$, has been considered. Various production mechanisms of $X_\mu$ have been discussed including (i) purely gravitational production from the inflaton background, (ii)  gravitational freeze-in from the SM quanta, (iii) inflaton decay through a dim-5 effective operator, and (iv) Higgs portal freeze-in and Higgs decay through a dim-6 effective operator. 
Parameters that properly describe the observed relic abundance have been determined. 
\end{abstract}
\maketitle
\tableofcontents

\section[intro]{Introduction}

The most successful theory of the early Universe is cosmic inflation which results in a period of exponential expansion~\cite{Guth:1980zm,Linde:1981mu}. 
The theory of inflation can be effectively described by a slowly-rolling single scalar field, called the inflaton $\phi$, with an approximately flat potential. 
Inflation successfully explains puzzles of the early Universe, e.g., the horizon problem, flatness problem, and seeds for Large Scale Structures (for a review see~\cite{Baumann:2009ds}).
Many of the non-trivial features of the inflationary paradigm can be tested in cosmological observations. 
For instance, the slow-roll single field inflation naturally leads to a nearly scale-invariant power spectrum, which matches very well with the recent observations from the cosmic microwave background (CMB) measurements~\cite{Planck:2018jri}. 
During the inflationary phase, the Universe expands exponentially, and therefore, inflation ends with an empty (no radiation/matter) and cold (non-thermal) Universe with total energy density stored in the inflaton field. 
To populate the Universe, one needs a mechanism that converts the inflaton energy density to the Standard Model (SM) and possibly to the dark sector. 
The process of transferring energy density from the inflaton field to the SM through perturbative decays is referred to as {\it reheating}~\cite{Albrecht:1982mp,Dolgov:1982th,Abbott:1982hn,Kofman:1994rk,Shtanov:1994ce,Kofman:1997yn}.

The perturbative reheating can be realized through some interactions between the inflaton and the SM fields. The lowest dimensional SM gauge singlet operator is $|{\bm h}|^2\!=\!{\bm h}^\dag{\bm h}$, where ${\bm h}$ is the SM Higgs doublet. Hence, in generic scalar field inflationary models, one would expect the leading inflaton--SM interaction is through the $\phi |{\bm h}|^2$ term. Whereas interactions of the scalar inflaton with the SM gauge singlet operators involving fermions and gauge bosons are higher dimensional $(\geq 5)$ and, therefore, would be suppressed by the inflationary scale $\Lambda$. 
In order not to spoil the flatness of the inflaton potential, such interactions are expected to be subdominant during the inflationary phase. However, after the end of inflation, the perturbative reheating predominantly follows through the inflaton-Higgs interaction\,\footnote{We note that non-perturbative effects can potentially become relevant for larger inflaton-Higgs couplings, which lead to tachyonic resonant production of Higgs modes, i.e., the so-called preheating regime. However, the tachyonic resonant Higgs production is suppressed in the presence of relatively large quartic SM Higgs coupling, see \sec{s.tachyonic_Higgs}. Therefore, the dominant mechanism of energy transfer from the inflaton field to the Higgs field remains the perturbative decay, see also~\cite{Lebedev:2021tas,Ahmed:2021fvt}.}. 
Regarding the inflaton field as a classical background field, the $\phi |{\bm h}|^2$ term is also a source of $\phi$-dependent Higgs mass that oscillates in time due to coherent oscillations of the inflaton field.

In this work, we aim to study in detail consequences of Higgs dynamics due to inflaton--Higgs interaction during the reheating phase. In particular, our goal is to analyze the implications of the $\phi$-dependent Higgs mass, which oscillates and results in rapid transitions between phases of broken and unbroken electroweak symmetry. We notice that this non-trivial Higgs mass leads to the suppression of perturbative decays of the inflaton field to Higgs boson pairs, which not only leads to elongations of the reheating period but also suppresses the production of the SM radiation energy density. As a result, evolution of the temperature of the SM bath is modified, which in turn can significantly affect the freeze-in production of dark matter (DM) during the reheating phase, see also~\cite{Garcia:2020wiy, Co:2020xaf, Barman:2022tzk, Banerjee:2022fiw}.
The reheating dynamics due to this non-trivial $\phi$-dependent Higgs mass is referred to as the {\it massive reheating scenario}, whereas for a comparison we consider the case where such mass effects are neglected, hence referred to as the {\it massless reheating scenario}. In this work we consider reheating through the SM Higgs boson, however, the results obtained here are straightforward to generalize to any other scalar field which interacts with the inflation field. 

As an example, we employ the $\alpha$-attractor T-model of inflation \cite{Kallosh:2013hoa, Kallosh:2013yoa} whose potential is approximately flat for large inflaton field values suitable for inflation and it has a monomial shape of the form $\propto \phi^{2n}$ during the reheating phase. In this work, we consider generic $n$ which leads to an effective equation of state $w\!=\!(n-1)/(n+1)$ during the reheating phase and determines the evolution of the Universe during this period\,\footnote{This work is a continuation and significant extension of the research described in our earlier paper~\cite{Ahmed:2021fvt} which was limited to the $n\!=\!1$ case only and focused on implications of time-dependent inflaton decay width.}. We provide analytic and numerical results for the dynamics of the inflaton field during the inflationary and reheating phases in the presence of inflaton--Higgs interaction. 

Furthermore, to investigate non-trivial implications of the Higgs-induced reheating on the DM production, we consider a model with the DM candidate being a massive vector field~$X_\mu$ of a dark Abelian gauge symmetry $U(1)_{X}$. The vector DM $X_\mu$ interacts with the SM as well as with the inflaton field through gravity and higher dimensional operators suppressed by the Planck mass. We study production of such DM particles during the reheating phase. Since the DM interactions with the SM and inflaton are assumed to be Planck mass suppressed, therefore vector DM is not in thermal equilibrium with the SM bath. We study DM production through its gravitational interactions with the inflaton background field~\cite{Ema:2015dka,Ema:2016hlw,Ema:2018ucl,Ema:2019yrd,Mambrini:2021zpp,Haque:2021mab,Lebedev:2022ljz, Clery:2021bwz,Haque:2022kez,Clery:2022wib,Barman:2021ugy,Ahmed:2021fvt,Garcia:2022vwm,Aoki:2022dzd} as well as through the annihilation of SM particles, also known as the gravitational freeze-in mechanism~\cite{Aoki:2022dzd,Garny:2015sjg,Tang:2017hvq,Garny:2017kha,Bernal:2018qlk,Redi:2020ffc,Chianese:2020yjo}. Moreover, DM production through direct inflaton decay is also investigated. In the case of vector DM, the leading contribution from the inflaton decay turns out to be triggered by a dim-5 operator suppressed by the Planck mass.  Another source of DM production is through an effective dim-6 operator suppressed by the Planck mass squared involving the Higgs doublet and DM fields~\cite{Kolb:2017jvz,Bernal:2018qlk,Chianese:2020yjo,Chianese:2020khl}. In this case, the production can occur due to the annihilation and decay of the Higgs bosons. 

The paper is organized as follows. In \sec{s.model} we describe details of our model, which include the inflaton, the SM Higgs, and the vector DM.
In \sec{s.inflaton}, we study inflaton dynamics during the inflationary phase as well as during the early stages of reheating, where the SM radiation energy density is negligible compared to that of the inflaton field. 
The reheating dynamics is presented in \sec{s.reheating} where we analyze non-trivial effects of the inflaton-induced Higgs mass on the production of SM radiation quanta, including effects of rapid electroweak phase transitions. In particular, we present the analytic and numerical results for the SM radiation energy density and the SM bath temperature evolution during this phase. 
In \sec{s.DM-effects} we discuss implications of the Higgs-induced reheating on the gravitational production of DM via graviton exchange from the inflaton background and SM radiation. Moreover, we study the DM production due to the inflaton decay and through the direct annihilation and decays of Higgs bosons. 
We summarize our findings in \sec{s.conclusions}. In Appendix~\ref{s.constraints}, we present recent constraints on the inflationary model parameters due to recent CMB measurements by Planck collaboration~\cite{Planck:2018jri}. 
Moreover, we supplement our results with a detailed derivation of the direct production of the SM and DM particles in the presence of oscillating inflaton background as well as the inflaton-induced gravitational production in \app{s.graviational_production}. 

\section[model]{The model \label{s.model}}

In this section, we present our model to describe the inflationary/reheating dynamics and dark matter production. It has been assumed that the mass scale for interactions between the SM, inflaton, and/or DM sector is set by the Planck mass~$\mpl$. 
We consider the following action where the SM, inflaton, and vector DM interact minimally with gravity, 
\begin{equation}
S=\int \dd^4x \sqrt{-g}\bigg[\frac{\mpl^2}{2}R+{\cal L}_{\phi}+{\cal L}_{\rm SM}+{\cal L}_{\rm DM}+{\cal L}_{\rm int}\bigg],	 \label{eq:action}
\end{equation}
with ${\cal L}_{\phi}$, ${\cal L}_{\rm SM}$ and ${\cal L}_{\rm DM}$ being the Lagrangian densities for the inflaton, SM, and DM, respectively, whereas, ${\cal L}_{\rm int}$ describes interactions among the inflaton, SM, vector DM, and graviton.
Above $\mpl\!=\!2.4\times10^{18}\gev$ denotes the reduced Planck mass, while $R$ is the Ricci scalar for background metric $g_{\mu\nu}$ and $g$ denotes its determinant. We consider the background metric as the FLRW metric, i.e.,
\beq
ds^2=dt^2-a(t)^2d{\bm x}^2, 
\eeq
where the ${\bm x}$ vector denotes the three spatial coordinates and $a(t)$ is the scale factor.

The Lagrangian density for the inflaton field reads
\beq
\cL_{\rm \phi}=\frac{1}{2} \partial_{\mu} \phi\, \partial^{\mu} \phi-V(\phi) ,	\label{eq:Linf}
\eeq
where $V(\phi)$ denotes the inflaton potential. Note that in the above Lagrangian, and hereafter, the indices are raised and lowered via the background metric $g_{\mu\nu}$.
In this work, we consider the $\alpha$-attractor T-model of inflation \citep{Kallosh:2013hoa, Kallosh:2013yoa}, such that the inflaton potential is
\beq
\begin{aligned}
V(\phi) &=\Lambda^{4} \tanh ^{2 n}\bigg(\!\frac{|\phi|}{M}\!\bigg) \simeq \begin{dcases}
\Lambda^{4} & |\phi| \gg\! M\\
\Lambda^{4}\Big\vert\frac{\phi}{M}\Big\vert^{2 n}  & |\phi| \ll\! M
\end{dcases},
\end{aligned} \label{eq:inf_pot}
\eeq
where $\Lambda$ determines the scale of inflation, whereas $M$ is related to the reduced Planck mass with the $\alpha$ parameter of the $\alpha$-attractor T-model as $M\equiv\sqrt{6 \alpha}\, \mpl$. The potential $V(\phi)$ has a minimum at $\phi\!=\!0$ for positive values of~$n$. 
The current experimental data \cite{Aghanim:2018eyx, Planck:2018jri} constrain values of the potential parameters. The inflation scale $\Lambda$ can be limited by the amplitude of the scalar perturbations, $A_s$, and the spectral tilt $n_s$ from inflationary observables of the CMB. Planck 2018 data \citep{Aghanim:2018eyx} sets the upper limit \mbox{$\Lambda \lesssim1.4\times 10^{16}\gev$}, see Appendix~\ref{s.constraints}. Furthermore, the current upper bound on the tensor to scalar power spectrum ratio, $r\lesssim 0.032$, limits the value of the $\alpha$ parameter or $M$ from above, such that, $M\lesssim 10 \mpl$.  
Hereinafter, without loss of generality, we fix $\alpha\!=\! 1/6$, such that $M\!=\!\mpl$, and  $\Lambda\!=\!3 \times 10^{-3}\mpl$. 

\begin{figure}[t!]
\begin{center}
\includegraphics[width=0.77\linewidth]{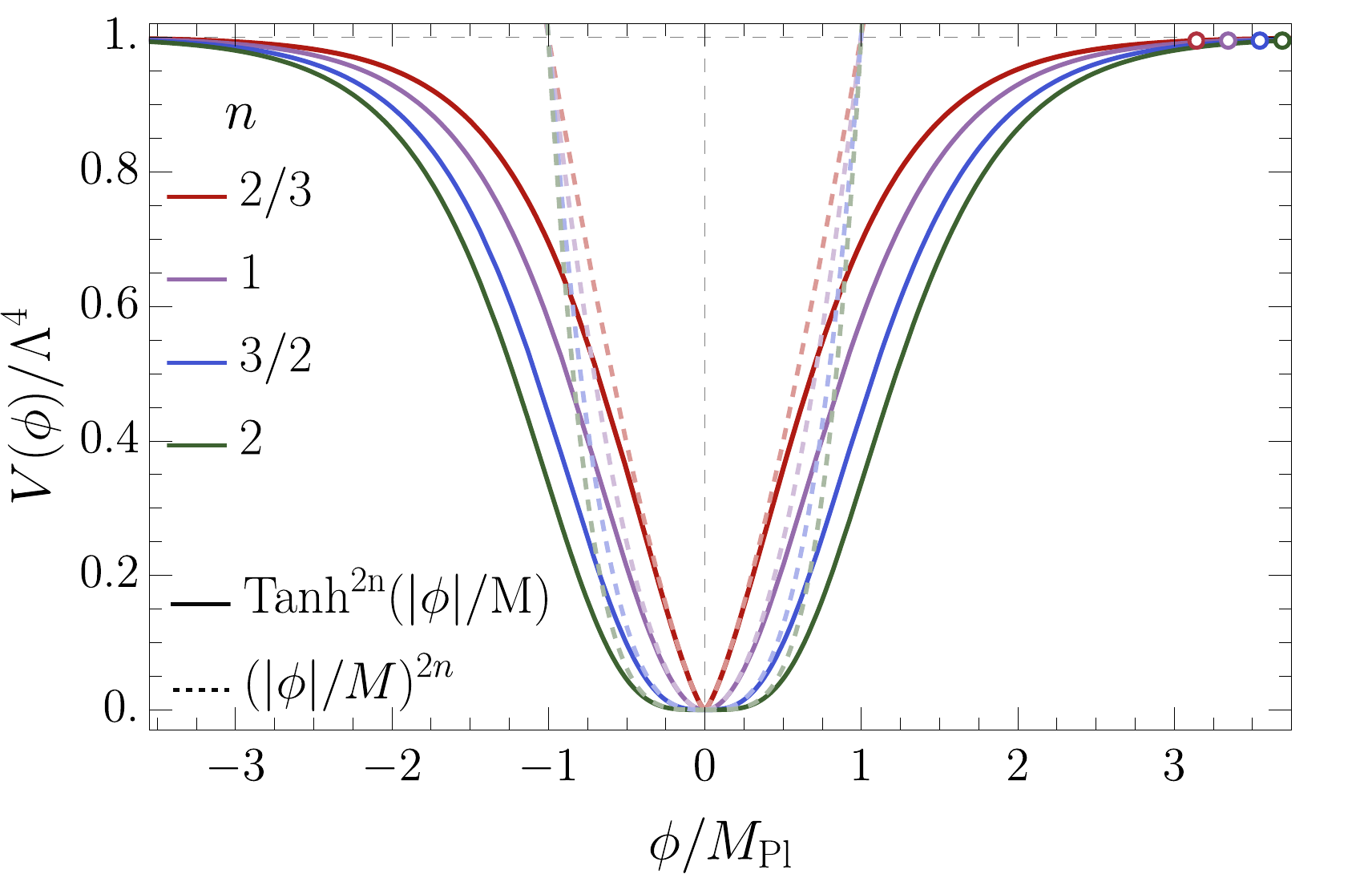}
\caption{The $V(\phi)/\Lambda^4$ term as a function of $\phi/\mpl$ for different values of $n$ and $\alpha \!=\!1/6$. Solid curves present the full form of $V(\phi)$, while dashed curves show the small-field expansion of $V(\phi)$ (the lower line of Eq.~\eqref{eq:inf_pot}). The empty, colored dotes indicate the initial value of the $\phi$ field, for which one gets 50 e-folds of inflation. }
\label{fig:VphiPlot}
\end{center}
\end{figure}

In \fig{fig:VphiPlot} we have plotted $V(\phi)/\Lambda^4$ as a function of the $\phi$ field for $\alpha\!=\!1/6$ with benchmark values of \mbox{$n\!=\! \{ 2/3, 1, 3/2, 2\}$}.  For large field values, i.e., $|\phi| \gg\! M$, the potential has a plateau $V(\phi) \simeq \Lambda^4 $. This region is suitable for cosmic inflation, as the flatness of $V(\phi)$ guarantees that cosmic inflation lasts long enough. In particular, requiring 50 e-folds of inflation leads to an initial condition on $\phi$ (indicated by the colored, empty dots) at the scale factor $a_\star$, which gives $\phi(a_\star) \! \simeq \! \{ 3.14, 3.35, 3.55, 3.69\}\, \mpl$ for $n \!=\! \{ 2/3, 1, 3/2, 2\},$ respectively. When $\phi$ rolls down to smaller values, i.e., $|\phi| \sim\! M$, inflation ends, and $\phi$ starts to oscillate around the bottom of its potential. For smaller field values, $|\phi| \ll\! M$, i.e., during the reheating phase, the potential can be well approximated by the power-law form $\Lambda^{4}\big(|\phi|/M\big)^{2 n}$ (dashed curves). Note that for $n\!=\!1$ and $n\!=\!2$ we reproduce the standard quadratic and quartic inflaton potential, respectively. 

The Lagrangian density for the Abelian vector DM $X_{\mu}$ is
\begin{align}
\cL_{\dm}&=-\frac14X_{\mu\nu} X^{\mu\nu}+\frac12m_X^2 X_\mu X^\mu ,  \label{eq:LDM}
\end{align}
where $X_{\mu \nu} \equiv \partial_{\mu}X_{\nu}-\partial_{\nu}X_{\mu}$ denotes the field strength tensor. 
The mass for the DM vector boson, $m_X$, is generated via an Abelian Higgs mechanism with a large expectation value of a dark Higgs field $\Phi$ so that the radial Higgs mode is heavy and therefore is integrated out.

We consider the following form of interaction Lagrangian,
\begin{align}
\cL_{\text{int}}&= - \bigg\{ \frac{h^{\mu\nu}}{\mpl} \Big[T_{\mu\nu}^{\phi}+T_{\mu\nu}^{\sm}+T_{\mu\nu}^{\dm}\Big]+g_{h\phi}^{} \mpl\, \phi |{\bm h}|^2  \non \\
&\qquad +\frac{\cxp\, m_X^2}{2\mpl} \phi X_{\mu} X^{\mu}+\frac{\cxh\, m_X^2}{2\mpl^2} X_\mu X^\mu |{\bm h}|^2 \bigg\},	 
\label{eq:Lint}
\end{align}
where $T_{\mu\nu}^{\phi}$, $T_{\mu\nu}^{\sm}$, and $T_{\mu\nu}^{\dm}$ are the energy-momentum tensors for the inflaton, SM, and DM sectors, respectively. 
Above $h_{\mu \nu}$ denotes the graviton field and ${\bm h}$ is the SM Higgs doublet, which in the linear parametrization can be written as
\beq
{\bm h}=\frac{1}{\sqrt2}\bpm h_2+i h_3\\ h_0+ih_1\epm, \label{eq:Higgs_para} 
\eeq
where $h_i$ $(i\!=\!0,1,2,3)$ are the four real scalar components. The dimensionless constant $g_{h\phi}^{}$ parametrizes interactions between the inflaton and the SM Higgs, while $\cxp$ and $\cxh$ are the dimensionless Wilson coefficients for the DM--inflaton and DM--Higgs effective interactions, respectively. In what follows, we assume that all couplings i.e., $g_{h \phi}$, $\mathcal{C}_X^{\phi}$ and $\mathcal{C}_X^{{\bm h}}$ are real and positive. 
Note that higher powers of $\phi$ could also appear in the interaction Lagrangian \eqref{eq:Lint}; however, for simplicity, we will consider only the lowest-dimensional operators that allow the inflaton to communicate with the SM and DM sectors. In the case of the vector DM discussed here, the lowest order direct interaction between the SM and DM appears through the dim-6 operator $\cxh m_X^2/(2\mpl^2) X_\mu X^{\mu} |{\bm h}|^2$ \citep{Kolb:2017jvz} and therefore must be suppressed by the UV cut-off scale assumed here to be $\mpl$. Moreover, the interaction Lagrangian contains only linear terms in the graviton field $h_{\mu\nu}$ which induces tree-level interactions between the SM and DM via a single graviton s-channel exchange~\cite{Garny:2015sjg, Garny:2017kha, Tang:2017hvq}.

Note that the last two terms in Eq.\eqref{eq:Lint} can be written in a proper gauge invariant form by coupling the inflaton field and the SM Higgs doublet to the kinetic term of the dark Higgs field $\Phi$, i.e.,
\begin{align}
\cL_{\text{int}}\supset- \frac{1}{2} \frac{\cxp}{\mpl}  \phi\, |D_{\mu}\Phi|^2- \frac{1}{2}\frac{\cxh}{\mpl^2}  |{\bm h}|^2 |D_{\mu}\Phi|^2 , \label{eq:Linthphi}
\end{align}
where $D_{\mu} \Phi = ( \partial_{\mu} - i g_X X_{\mu}) \Phi$ is the covariant derivative of the $\Phi$ field and $g_X$ is the $U(1)_X$ gauge coupling. 
When the dark Higgs field $\Phi$ acquires vacuum-expectation-value (vev) $\langle \Phi \rangle $, the dark $U(1)_X$ symmetry is spontaneously broken, and a mass term for the vector DM field is generated with mass $m_X \equiv g_X \langle\Phi \rangle$.
Assuming the radial mode of the dark Higgs is much heavier than the dark gauge boson, we can integrate out so that effective interaction terms for the vector DM with the inflaton and the SM Higgs doublet are generated as in the interaction Lagrangian~\eqref{eq:Lint}. Furthermore, one can also write a dim-5 effective interaction between the inflaton and vector DM of the form, $(\phi/\mpl)X_{\mu\nu}X^{\mu\nu}$. However, note that the corresponding vertex would involve the momentum of the gauge bosons, which is of the same order as the DM mass since the DM is not in thermal equilibrium. Therefore, effectively the above operator is similar to the one considered in \eq{eq:Lint}.

It is also important to emphasize that the graviton coupling to the SM and DM energy-momentum tensors~\eqref{eq:Lint} leads to an indirect interaction between these two sectors proportional to $1/\mpl^2$, which is of the same order as the interaction via the effective DM--SM Higgs portal operator for $\cxh\!\sim\!\op(1)$.
This was one of our primary motivations to consider such effective operators in the model. 
On the other hand, spin-0 and spin-1/2 DM particles would interact with the SM sector through effective operators of dim-4 and dim-5, respectively. Then, DM would couple to the SM sector much stronger than its coupling induced by a single graviton exchange. In such a scenario, one needs tuning of the DM-SM effective couplings to be suppressed to get a similar strength to the gravitational interaction. Therefore, we have found these options less interesting than the spin-1 case. 

Note also that a possible dim-4 mixing of the SM hypercharge and Abelian dark gauge bosons, of the form $\epsilon B_{\mu\nu}X^{\mu\nu}$, is absent in our model due to dark charge conjugation symmetry, which ensures the stability of DM. 
However, for mixing parameter $\epsilon$ small enough, such mixing would be allowed if the DM lifetime was longer than the age of the Universe. 

It is crucial to notice that dynamics of the $\phi$ field during the inflationary epoch could be modified by its couplings to the SM Higgs and DM specified in ${\cal L}_{\rm int}$~\eqref{eq:Lint}. Therefore, let us first discuss perturbativity limits imposed upon the strength of the $\ghp$ coupling and the Willson coefficients $\cxp$ and $\cxh$. Requiring, for instance, that amplitude for ${\bm h}$ scattering in a background of an inflaton classical field with its double insertion is smaller than the corresponding amplitude with a single insertion implies 
\beq
\ghp\lesssim \left(\frac{\Lambda^2}{ \phi\mpl}\right),
\label{ghp_lower_lim}
\eeq
where $\phi$ denotes the strength of the external classical inflaton field determined by its equation of motion (EOM). 
Similar reasoning implies
\beq
\cxp \lesssim \left(\frac{\mpl}{\phi}\right) \left(\frac{\Lambda}{\mx}\right)^2 ,
\label{eq:cxp_lim}
\eeq
where it has been assumed that $\Lambda \gtrsim \mx$. 
On the other hand, perturbativity of the $X_\mu X^\mu |{\bm h}|^2$ operator together with the requirement that
the maximum temperature $T_{\rm max}$ of the thermal bath during reheating remains below the cut-off scale,
$\Lambda$, for the effective operator 
$(D_{\mu}\Phi)^\dag (D^{\mu}\Phi) |{\bm h}|^2$ implies
\beq
\cxh \lesssim \min\left\{\left(\frac{\mpl}{\mx}\right)^2, \left(\frac{\mpl}{T_{\rm max}}\right)^2 \right\}.
\label{cxp_lim}
\eeq
Before moving forward, let us discuss here the role of the SM Higgs field during the inflation and reheating phases in more detail. 
Note that typical energy scales for inflation and reheating are much larger than the scale of electroweak interactions. Therefore, the Higgs potential during inflation and reheating periods can be approximated as
\begin{align}
V({\bm h}) &=\lambda_h\bigg(|{\bm h}|^2-\frac{v_{\rm EW}^2}{2}\bigg)^2\simeq \lambda_h|{\bm h}|^4,		\label{eq:Hpotential}
\end{align}
where $\lambda_h$ is the Higgs-boson quartic coupling. At the electroweak scale $v_{\rm EW}\simeq246\gev$, the value of Higgs quartic coupling is $\lambda_h\simeq0.13$ corresponding to the Higgs mass \mbox{$m_{h_0}^{\tiny{{\rm EW}}}=125\gev$}. Due to quantum corrections, within the SM, the Higgs quartic coupling $\lambda_h$ runs down to negative values at energy scales $\sim10^{10}\gev$ making the Higgs potential unstable~\cite{Espinosa:2015qea}. 
However, in the presence of the new physics interactions as considered in this work in \eq{eq:Lint}, the Higgs stability can be achieved for larger Higgs field values~\cite{Espinosa:2015qea}. In particular, the inflaton--Higgs interaction term, $g_{h\phi}^{} \mpl \,\phi\, |{\bm h}|^2$, generates a $\phi$-dependent Higgs-boson mass $m_{h_0}^2\!=\!g_{h\phi}^{} \mpl |\phi|$, with $|\phi|\!\sim\! \op({\rm few}) M$ during inflation, which leads to the stability/positivity of the Higgs potential. 
The maximal Higgs field strength up to which the potential is stable could be estimated as $|{\bm h}_{\rm max}|^2 \sim \ghp\mpl \phi/\lambda_h$.

During the inflationary phase, besides the inflaton potential $V(\phi)$, one could consider contributions to inflaton dynamics that originate from the SM Higgs quartic interactions and the inflaton--Higgs coupling. However, in this work, we consider a parameter space where the inflationary dynamics is dominated by the cosmological constant term~$\Lambda^4$. Thus, we require that both $\lambda_h |{\bm h}|^4$ and $g_{h\phi}^{} \mpl \,\phi\, |{\bm h}|^2$ terms are smaller than $\Lambda^4$ in the whole region of stability, i.e., up to the largest allowed Higgs-field strength ${\bm h} = {\bm h}_{\rm max}$. It is easy to see that the required condition reads
\beq
\ghp\lesssim \sqrt{\lambda_h} \left(\frac{\Lambda^2}{\phi \mpl}\right).
\label{ghphi_lower_lim}
\eeq 
Which gives a stronger bound than \eq{ghp_lower_lim} for $\lambda_h<1$, and hence we adopt the \eqref{ghphi_lower_lim} constraint in the following analysis. 

Furthermore, as shown in \rcite{Espinosa:2015qea}, for the Higgs mass \mbox{$m_{h_0}\!>\!3H_{I}/2$} the Higgs field fluctuations during inflation are strongly suppressed ensuring stability, where $H_I$ denotes the Hubble parameter during inflation. This condition could be written in terms of a lower limit for $\ghp$, i.e.,
\beq
\ghp\gtrsim \frac34 \sqrt{6\alpha} \left(\frac{\Lambda^2}{\phi \mpl}\right)^2 \left(\frac{\phi}{M}\right).
\label{ghphi_upper_lim}
\eeq
For the benchmark values of $\Lambda\!=\!3\times 10^{-3}\mpl$ and $\alpha\!=\!1/6$, and assuming that $\phi\!\sim\! M\!=\!\mpl$,
we obtain the following consistency region for $\ghp$:
\beq
6 \cdot 10^{-11} \lesssim \ghp\lesssim 3 \cdot 10^{-6}
\label{ghp_num_lim}
\eeq 
Hereinafter, for numerical calculations, we adopt the EW value of the Higgs quartic coupling $\lambda_h= 0.13$.
Note that the above limits are a subject of $\lambda_h$; increasing $\lambda_h$ implies larger $\ghp$ allowed values.

\section[inflaton]{Inflaton dynamics}
\label{s.inflaton}

In this section, we discuss the dynamics of the inflaton field. Based on the arguments above, we ignore Higgs boson contributions to the inflaton dynamics so that the classical equation of motion for $\phi$ in the FLRW background is given by
\beq
\ddot{\phi}+3 H \dot{\phi}+V_{,\phi}(\phi)=0,	\label{eq:inf_eom}\,
\eeq
where $H \equiv \dot{a}/a$ is the Hubble rate, the $\it{overdot}$ denotes derivative w.r.t. the cosmic time, $t$, whereas
$V_{,\phi}(\phi)$ is derivative of the potential w.r.t. the $\phi$ field. In the above EoM, we have neglected spatial derivatives of the inflaton, regarding $\phi$ as a spatially-homogeneous scalar field. 
Later on, for collision terms in the Boltzmann equations, we will calculate $S$-matrix elements in the presence of classical inflaton fields that are solutions of~\eqref{eq:inf_eom}. In other words, we will treat the SM and DM fields as small perturbations as compared to solutions of~\eqref{eq:inf_eom} and their possible back-reaction on the inflaton field will be neglected\,\footnote{This is similar, e.g., to the Rutherford scattering which is an elastic scattering of a charged particle by a static Coulomb point-particle potential that is a solution of Maxwell equations in empty space undisturbed by the presence of any other particles.}.
As discussed above, the inflaton-Higgs interaction term, i.e., $g_{h \phi} \mpl \phi |\bm{h}|^2$, is suppressed during inflation; therefore, Eq.~\eqref{eq:inf_eom} is valid during this period as well as during the early stages of reheating. However, once the reheating resumes, the Higgs field is produced through the coherent oscillations of the inflaton field. As we will see below, the Higgs dynamics will be taken into account in the Boltzmann equation. 
Assuming that the SM radiation and DM energy densities are negligible in the primordial Universe, one can write the first Friedmann equation as
\beq
H^{2} \simeq \frac{\rho_\phi}{3 M_{\mathrm{Pl}}^{2}}.\,	\label{eq:Friedmann_eqn}
\eeq
The energy density, $\rho_\phi$, and the pressure, $p_\phi$, for the homogenous, scalar field are
\begin{align}
\rho_{\phi}&=\frac{1}{2} \dot{\phi}^{2}+V(\phi),  &p_{\phi}&=\frac{1}{2} \dot{\phi}^{2}-V(\phi).	\label{eq:rho_p}
\end{align}
We can now discuss solutions to Eq.~\eqref{eq:inf_eom} during and after inflation. As we have already pointed out above, during the phase of the accelerated expansion, $V(\phi)$ is approximately flat. In this period $\ddot{\phi}$ is negligible and $V(\phi)\! \gg \!\dot{\phi}^2/2$, which in turn implies $H^2 \simeq V(\phi)/(3 \mpl^2)$. Using these two simplifying assumptions during the inflationary phase one finds constant solutions for the inflaton field $\phi_I (t) \simeq \phi(a_\star)$ and the Hubble rate $H_I \simeq \Lambda^2/(\sqrt{3} \mpl)$.
The slow-roll evolution of the inflaton field can be parameterized by the so-called potential slow-roll parameters $\epsilon_V^{}$ and $\eta_V^{}$, defined as
\begin{align}
\epsilon_V^{}&=\frac{M_{\mathrm{Pl}}^{2}}{2}\left(\frac{V_{, \phi}(\phi)}{V(\phi)}\right)^{2},  		&\eta_V^{}&=M_{\mathrm{Pl}}^{2} \frac{V_{, \phi \phi}(\phi)}{V(\phi)}.
\end{align}
Note that the two cosmological observables, the spectral index $n_s$ and the tensor-to-scalar ratio $r$, can be related to the potential slow-roll parameters as
\begin{align}
r&=16 \epsilon_V^{}, 	&n_{s}&=1-6 \epsilon_V^{}+2 \eta_V^{}.	\label{eq:r_ns}
\end{align}
For the $\alpha$-attractor inflaton potential~\eqref{eq:inf_pot}, the slow-roll potential parameters are
\begin{align}
\epsilon_V^{} &= 8 n^2 \left( \frac{\mpl }{M} \right)^2 \operatorname{csch}^{2}\left(\frac{2|\phi|}{M}\right), 	\label{eq:epsilonV}\\
\eta_V^{}&= 8 n \left( \frac{\mpl }{M} \right)^2 \left[2 n-\cosh \left(\frac{2|\phi|}{M}\right)\right] \operatorname{csch}^{2}\left(\frac{2|\phi|}{M}\right).	\label{eq:etaV}
\end{align}
To achieve sufficiently many e-folds ($\sim50-60$) of expansion, the slow-roll parameters are required to be small, i.e., $\epsilon_V,|\eta_V| \ll 1$, for a long enough period. The accelerated expansion of the Universe ends when $\ddot{a} =0$, which corresponds to $\epsilon_{V}\simeq1$. This condition determines the value of $\phi(a_e)$ at the end of inflation, i.e., at $a\!=\! a_e$.  

\begin{figure}[t!]
\begin{center}
\includegraphics[width=0.7\linewidth]{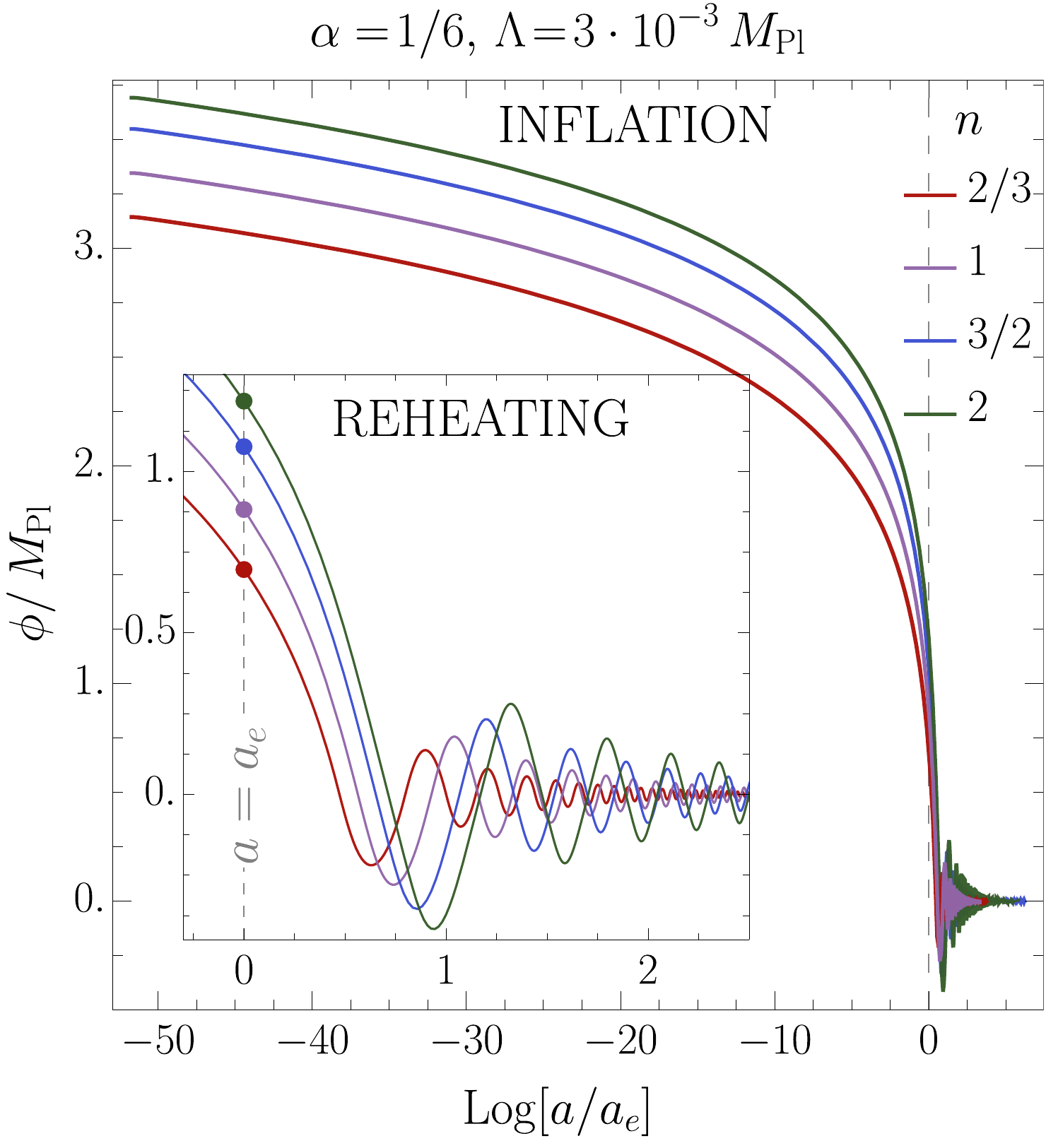}
\caption{Numerical solutions for the inflaton EoM \eqref{eq:inf_eom} and \eqref{eq:Friedmann_eqn} as a function of the scale factor $a$ for various choices on $n$. The vertical dashed line indicates the end of inflation, defined by the condition $\epsilon_V(\phi(a_e))=1$. Colored dots present the value of $\phi(a_e)$ for different values of $n$.}
\label{fig:phiaSolPlot}
\end{center}
\end{figure}

After the end of inflation, the inflaton field $\phi$ starts to oscillate around the minimum of its potential with decreasing amplitude. Moreover, the character of the oscillations strongly depends on the shape of $V(\phi)$ in the vicinity of the minimum. Note that the time scale of the inflaton oscillations is typically much shorter than the variation of its amplitude. Thus, in this phase, a generic solution to Eq.~\eqref{eq:inf_eom} can be written as a product of two functions~\cite{Shtanov:1994ce, Ichikawa:2008iq, Garcia:2020wiy, Clery:2021bwz},
\begin{align}
\phi(t) = \varphi(t) \cdot \mathcal{P}(t), \label{eq:phi_app}
\end{align}
where $\mathcal{P}(t)$ is a quasi-periodic, fast-oscillating function, while $\varphi(t)$ denotes a slowly-varying (w.r.t. the time scale of the oscillations) envelope function defined by the following condition:
\begin{align}
\rho_{\phi} = V(\varphi) = \Lambda^4 \Big(\frac{\varphi}{\mpl}\Big)^{2 n}\,,  \label{eq:envelope}
\end{align}
where the inflaton potential has been expanded in a form applicable during the reheating phase with $\alpha\!=\!1/6$. 
In \fig{fig:phiaSolPlot} we show the numerical solution for $\phi(a)$,  obtained after solving two coupled equations \eqref{eq:inf_eom} and \eqref{eq:Friedmann_eqn}. The rapid oscillations of $\mathcal{P}(a)$ are damped by the decreasing envelope $\varphi(a)$ due to the redshift of the Universe.  The frequency of oscillations as well as their amplitude depends on the slope of the inflaton potential $n$. The explicit form of $\varphi$ and $\mathcal{P}$ will be determined below.

\begin{figure}[t]\begin{center}
\includegraphics[width=0.5\linewidth]{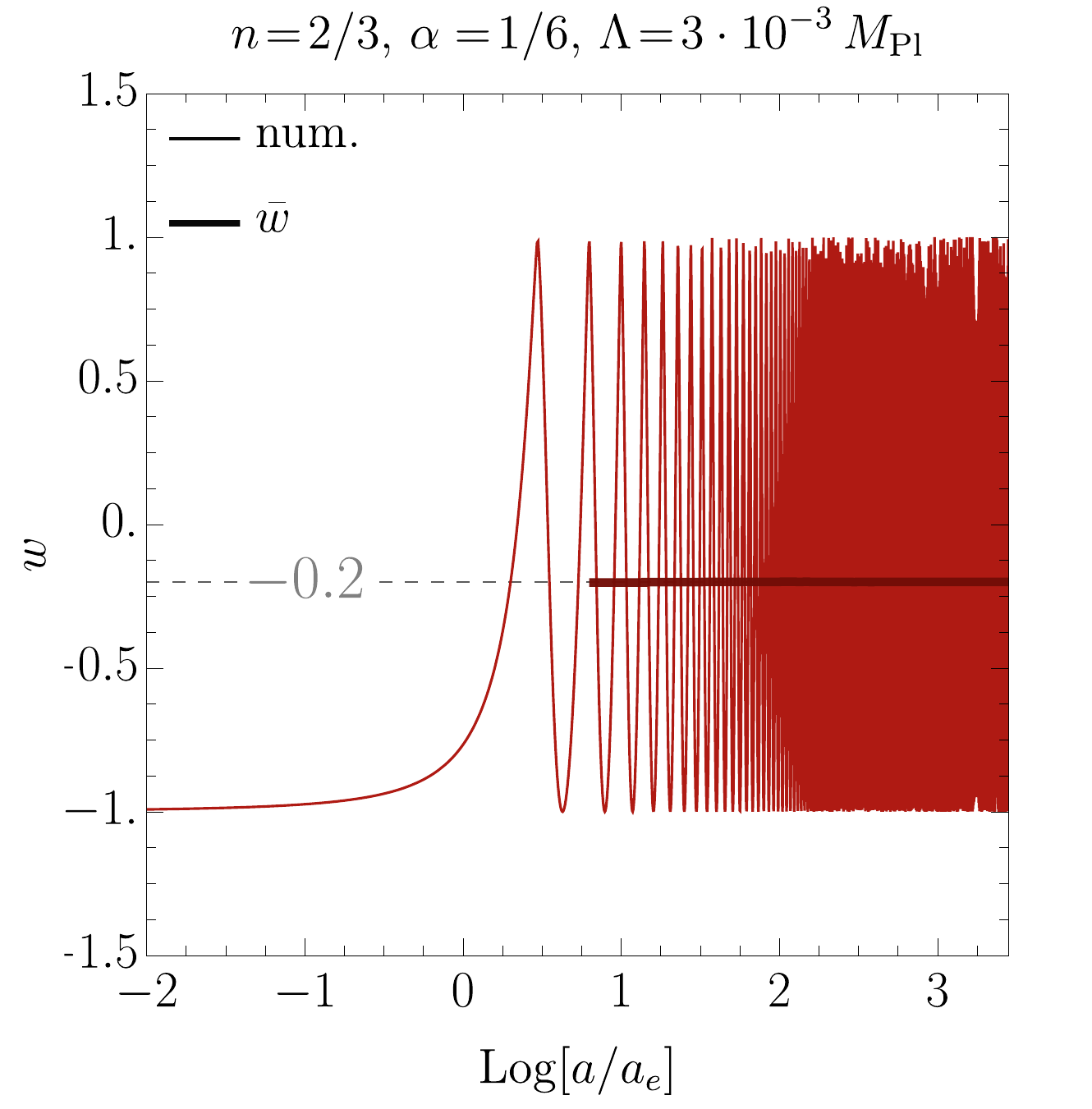}\!\!\!
\includegraphics[width=0.5\linewidth]{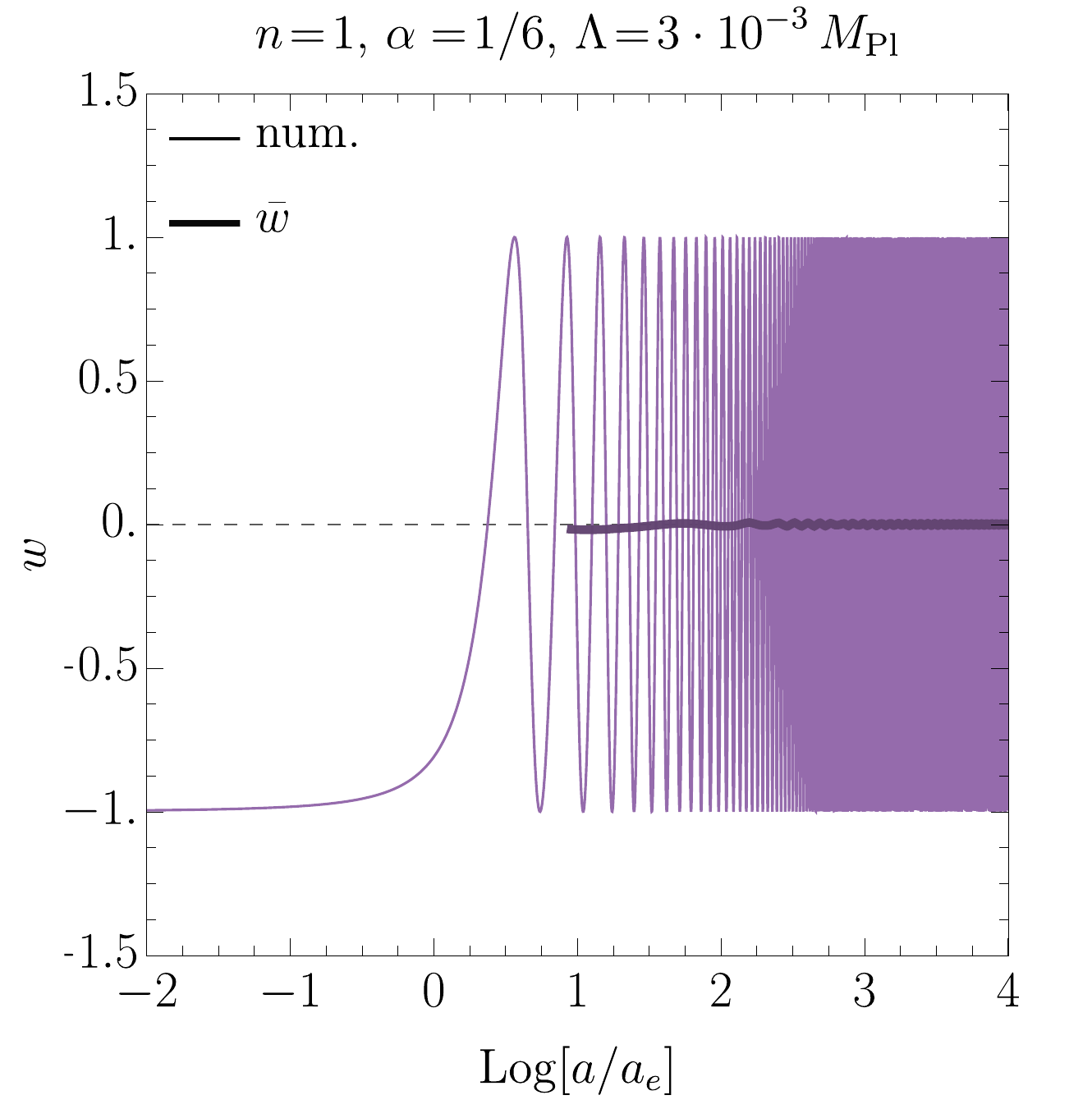}\!\!\!
\includegraphics[width=0.5\linewidth]{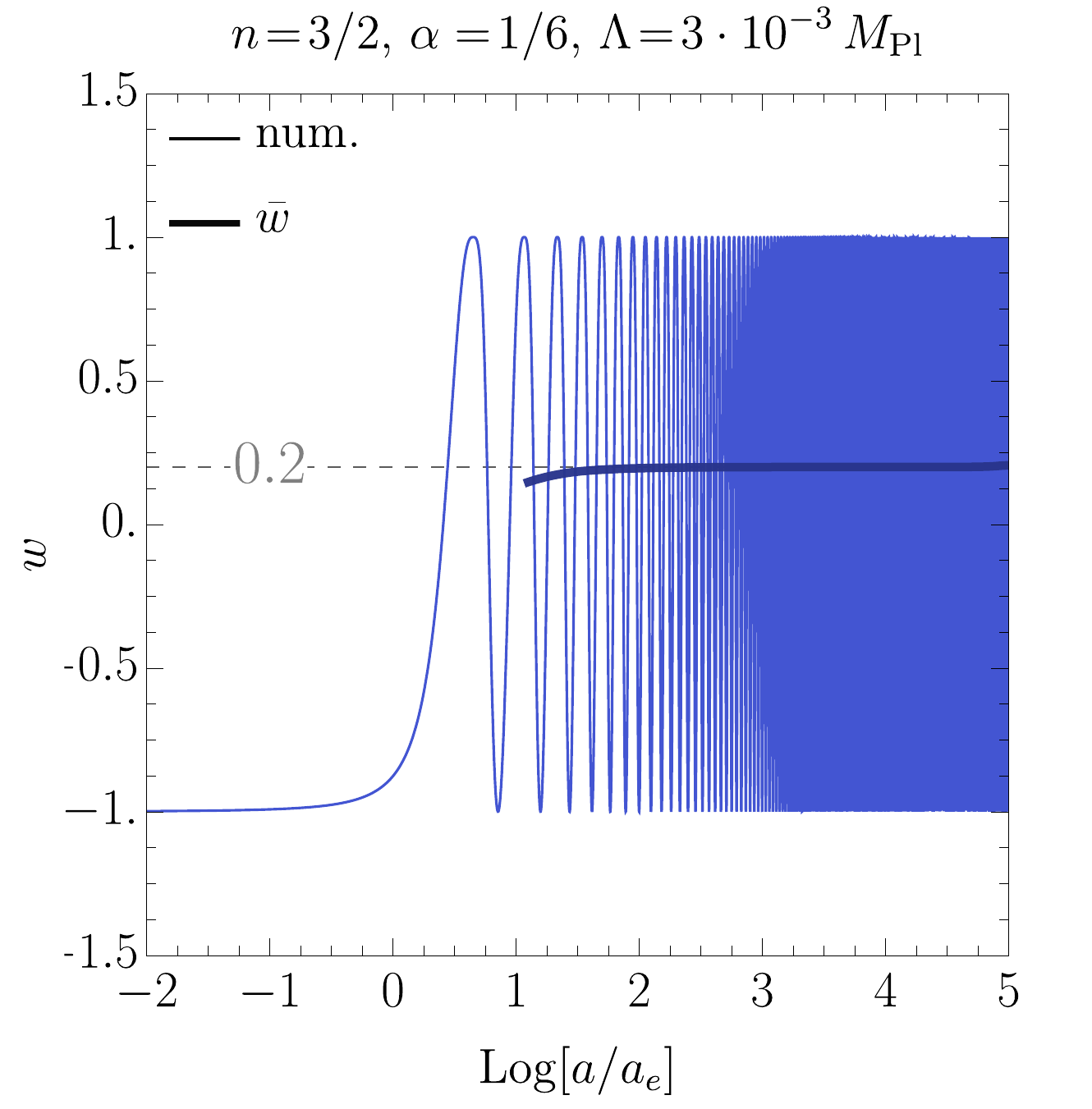}\!\!\!
\includegraphics[width=0.5\linewidth]{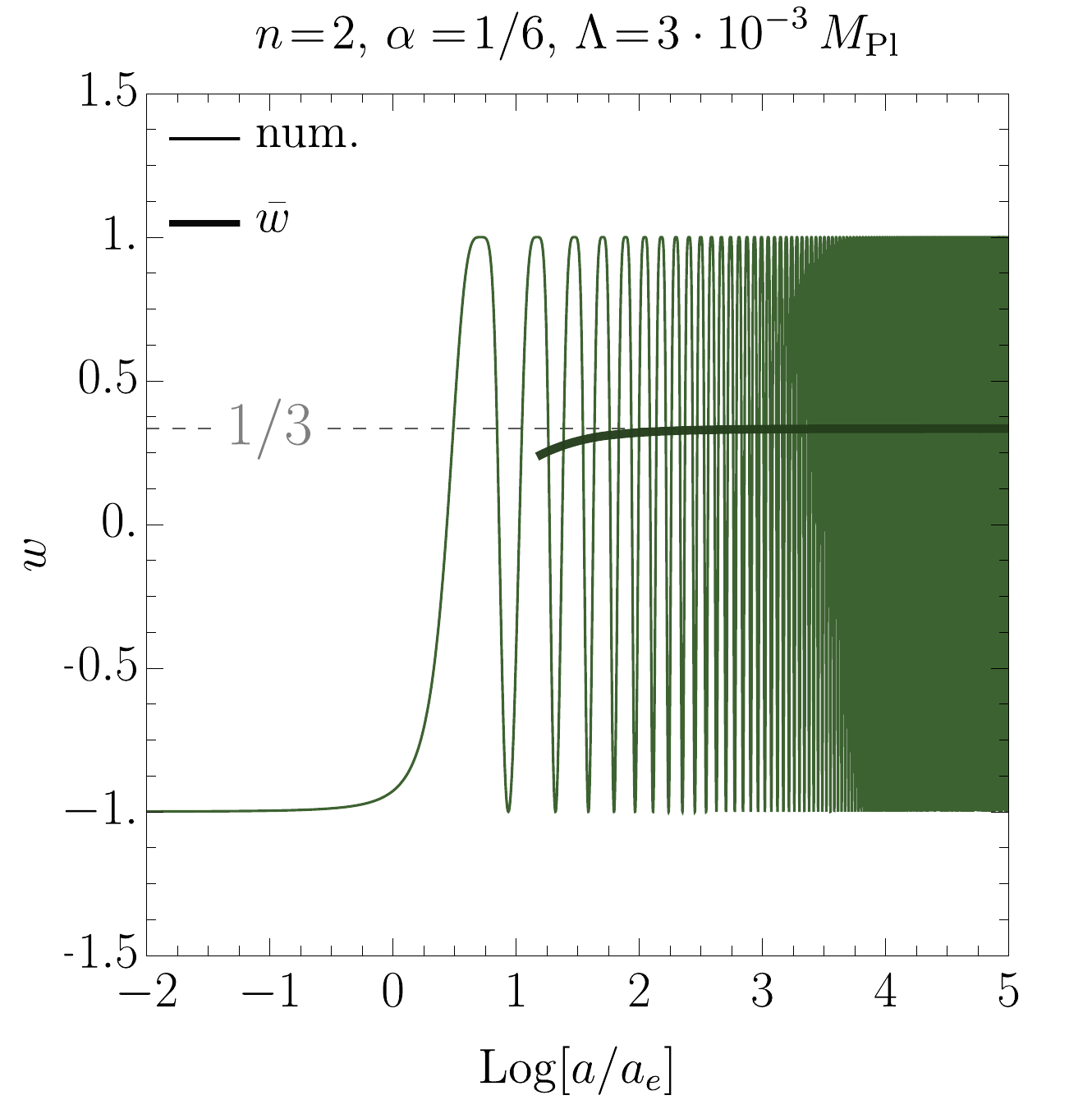}\!\!\!
\caption{The equation-of-state parameter, $w$, as a function of the scale factor $a$. Thin colored curves show the full numerical solutions, i.e.,  $p_\phi/\rho_\phi$, while thick lines present the time-averaged barotropic parameter, i.e., $\bar w\!=\!\langle p_\phi \rangle /\langle \rho_\phi\rangle$ \eqref{eq:w}. The horizontal axes are normalized to the scale factor at the end of inflation $a_e$.}
\label{fig:wAverage}
\end{center}
\end{figure}

Our goal now is to obtain approximate analytical solutions of~\eqref{eq:inf_eom} during the oscillatory phase. Before we do that, let us first find a relation between the inflaton energy density $\rho_\phi$ and pressure $p_\phi$. To that end, we differentiate $\rho_\phi$~\eqref{eq:rho_p} w.r.t. time and using~\eqref{eq:inf_eom} we obtain the continuity equation
\begin{align}
\dot{\rho}_\phi + 3 H (1+w) \rho_\phi&=0,		&w& \equiv p_\phi /\rho_\phi,	\label{eq:continuity-eqn}
\end{align}
where the barotropic parameter~$w$ is, in general, time-dependent. Note that at this stage, the effects of quantum particle production are ignored, so that \eqref{eq:continuity-eqn} holds during inflation and the beginning of reheating as long as the inflaton energy density dominates, i.e., the inflaton decay rate is smaller than the Hubble rate. Ignoring expansion, assuming periodicity, and averaging over one period of oscillations, we can express the inflaton energy density and pressure~\eqref{eq:rho_p} as~\cite{Shtanov:1994ce, Ichikawa:2008iq, Garcia:2020wiy, Clery:2021bwz},
\begin{align}
\langle \rho_\phi \rangle &= V(\varphi) =\rho_\phi \\
\langle p_\phi \rangle &= \frac{n-1}{n+1} V(\varphi) = \frac{n-1}{n+1} \rho_\phi ,
\end{align}
where we have used Eq.~\eqref{eq:envelope} along with the relation $\langle \lvert  \mathcal{P}^{2n} \rvert \rangle\!=\! 1/(n+1)$. Hereinafter, the following definition for the time-average of a quantity $f(t)$ over one period of inflaton oscillation is employed,
\begin{align}
\langle f(t_i) \rangle = \frac{1}{{\cal T}(t_i)} \int_{t_i}^{t_i+{\cal T}(t_i)} \!\!\!d t \;f(t) ,
\end{align}
where $t_i$ is some reference time at a particular instant during the oscillatory phase.
Above, ${\cal T}(t_i)$ denotes the period of the inflaton oscillations, which in general can be time-dependent. 
We can now define the averaged equation-of-state parameter during the reheating phase as
\begin{align}
\bar{w} \equiv \frac{ \langle p_\phi \rangle }{\langle \rho_\phi \rangle } = \frac{n-1}{n+1}. \label{eq:w}
\end{align} 
In particular, we have $\bar w \! \in \! \{ -1/5, 0, 1/5, 1/3\}$ for $n \! \in \! \{ 2/3, 1, 3/2, 2\},$ respectively. The evolution of the equation of state \mbox{$w \equiv \rho_\phi / p_\phi$} and the its time-averaged value $\bar{w}$ is shown in \fig{fig:wAverage} as a function of the scale factor $a$. During inflation, i.e., for $a \!<\! a_e$, the potential term $V(\phi)$ dominates over the kinetic term, $\dot{\phi}^2/2$, and the $\rho_\phi / p_\phi$ ratio is constant, i.e.,  $w\simeq-1$.  After the end of inflation, $w$ starts to oscillate between $-1$ and $1$, while $\bar w$ quickly approaches a constant limit, consistent with the prediction \eqref{eq:w}.

Using the continuity equation \eqref{eq:continuity-eqn} together with the definition \eqref{eq:envelope} and \eqref{eq:w} one can derive the equation of motion for the envelope
\begin{align}
  \dot{\varphi}(t) =- \frac{3}{n+1} H \varphi(t).
\end{align}
Form of the above equation proves that indeed $\varphi$ is a slowly varying function of time and fast oscillations are relegated to the $\mathcal{P}$ function~\footnote{Note that the definition of the envelope function \eqref{eq:envelope} implies that $d\langle \rho_\phi \rangle /dt = \langle \dot{\rho}_\phi \rangle$ assuming periodicity and neglecting cosmological time evolution while averaging over a period.}. The solution to this equation as a function of the scale factor~$a$ reads
\begin{align}
\varphi(a) = \varphi_e \, \Big(\frac{a_e}{a} \Big)^{\frac{3}{n+1}},  \label{eq:envelope_sol}
\end{align}
where $\varphi_e\equiv \varphi (a_e) $ denotes the initial value of the envelope determined by the condition $\epsilon_V (\phi(a_e)) =1$ at the end of inflation, cf. Eq.\eqref{eq:epsilonV}, and for $\alpha=1/6$ it is given by
\begin{align}
\varphi_e \equiv \varphi (a_e) = \frac{\mpl}{2} \operatorname{ArcSinh} \big( 2\sqrt{2}\, n\big). \label{eq:varphie}
\end{align}
Note that for $n\!>\!0$, $\varphi(a)$ decreases with time during the oscillatory phase. 
Using the Friedmann equation \eqref{eq:Friedmann_eqn} together with \eq{eq:envelope}, we obtain the following equation that governs the evolution of the scale factor $a$ with $t$, 
\begin{align}
\frac{ d  a}{a}= \frac{1}{\sqrt{3} \mpl } \Lambda^2  \left(\frac{\varphi}{\mpl} \right)^n  d  t.
\end{align}
 Using Eq.~\eqref{eq:envelope_sol} we get
\begin{align}
a(t) = a_e \bigg[ 1 + \frac{3n}{n+1} \frac{\Lambda^2}{\sqrt{3} \mpl } \Big(\frac{\varphi_e }{\mpl} \Big)^n (t-t_e)\bigg]^{\frac{n+1}{3n}},
\label{eq:at_reh}
\end{align}
where $t_e$ denotes the cosmic time at end of inflation. Note that the above relation implies $\varphi(t)\! \sim \!t^{-1}$ for $n\!=\!1$, whereas, for the quartic potential, i.e., $n\!=\!2$, we get $\varphi(t)\! \sim\! t^{-1/2}$.

In order to find a dynamical equation for the quasi-periodic function $\mathcal{P}(t)$ we differentiate \eqref{eq:phi_app} w.r.t. time, obtaining
\begin{align}
\dot{\phi} = \dot{\varphi} \mathcal{P} + \varphi \dot{\mathcal{P}},
\end{align}
which, in turn, implies
\begin{align}
\dot{\mathcal{P}}(t) = \pm \frac{\sqrt{2(\rho_\phi-V(\phi))}}{\varphi} + \frac{3n}{n+1} H \mathcal{P}(t), \label{eq:P_eqn}
\end{align}
where the $\pm$ sign corresponds to $\dot{\phi}\!>\!0$ and $\dot{\phi}\!<\!0$, respectively.
Note that the second term in the r.h.s. of \eqref{eq:P_eqn} is negligible as long as the time scale of the inflaton oscillations is much shorter than the time scale of expansion. 
Consequently, one can simplify Eq.\eqref{eq:P_eqn} as
\begin{align}
\dot{\mathcal{P}} &\simeq \pm \frac{m_\phi}{\sqrt{n(2n-1)}}\sqrt{1-\lvert \mathcal{P} \rvert^{2n}}, 
\label{eq:p_eqn}
\end{align}
where we have introduced an effective mass of the inflaton field, see also~\cite{Garcia:2020eof, Garcia:2020wiy, Clery:2021bwz},
\begin{align}
m_{\phi}^{2} \!\equiv\!\frac{\partial^2 V(\phi)}{\partial \phi^2}\bigg\vert_{\phi=\varphi} &\!=\!  2n(2n-1) \frac{\Lambda^4}{\mpl^2} \Big( \frac{\rho_\phi }{\Lambda^{4} } \Big)^{\!\!\frac{n-1}{n}}, \label{eq:eff_mass}
\end{align}
which is time-dependent for $n\!\neq\!1$. 
In particular, since $\rho_\phi$ decreases due to the expansion, $m_\phi$ increases for $n \!< \!1$, while for $n\!>\!1$ it decreases with time in the oscillatory phase. 
Using Eqs.~\eqref{eq:envelope}, \eqref{eq:envelope_sol} and  \eqref{eq:eff_mass} we can obtain an explicit formula for $m_\phi$ as a function of the scale factor $a$ 
\begin{align}
m_\phi(a) = \sqrt{2n(2n-1)} \frac{\Lambda^2}{\mpl} \Big(\frac{\varphi_e }{\mpl}\Big)^{n-1} \Big( \frac{a_e}{a}\Big)^{\frac{3(n-1)}{n+1}}\,.
\label{eq:m_phi}
\end{align}
Since $m_\phi$ varies on the time scale much larger than the oscillation time scale, we can solve Eq.~\eqref{eq:p_eqn} assuming that $m_\phi$ is constant during one period of oscillations. Then the generic solution for the quasi-periodic function ${\cal P}(t)$ can be written in terms of the inverse of the regularized incomplete beta function ${\cal I}^{-1}_{z}(i,j)$ as
\begin{align}
\mathcal{P}(a) = \bigg[ {\cal I}^{-1}_{z}\!\Big(\frac{1}{2n}, \frac{1}{2} \Big) \bigg]^{\frac{1}{2n}},	\label{eq.time_average}
\end{align}
with 
\begin{align}
z&\equiv 1 - \frac{4}{\cal T} (t-t_0)		 \notag\\
&\!=\!1-\frac{4}{\mathcal{T}} \frac{n+1}{\sqrt{3}  n} \frac{\mpl}{\Lambda^{2}}\Big(\frac{\mpl}{\varphi_e}\Big)^{\!n}\bigg[\Big(\frac{a}{a_{e}}\Big)^{\frac{3 n}{n+1}}-1\bigg],
\end{align}
where the period of the oscillations, $\mathcal{T}$, is given by
\beq
\mathcal{T}\equiv \frac{2 \pi }{\omega}
=\frac{\sqrt{4 \pi} }{m_\phi} \sqrt{\frac{2n-1}{n}} \frac{\Gamma\left(\frac{1}{2n}\right)}{\Gamma\left(\frac{n+1}{2n}\right)}
\,,
\label{eq:period}
\eeq
with frequency $\omega$.
For the quadratic potential i.e., $n\!=\!1$ we recover the well-known results: $\mathcal{T} \!=\! 2 \pi/ m_\phi$ and $\mathcal{P}(t) \!=\! \cos( m_\phi t)$~\cite{Kofman:1997yn}. Note that the slow variation of $m_\phi$ w.r.t. time is also a source of the variation of the period. Because $m_\phi$ increases (decreases) for $n\!<\!1$ ($n\!>\!1$), $\mathcal{T}$ decreases (increases) with time. 
In the next section, we employ these results obtained for the inflaton profile as well as its time-dependent mass to study the reheating dynamics, including the inflaton interactions with the Higgs boson. 

\section[reheating]{Reheating dynamics}
\label{s.reheating}

The reheating dynamics involve the coherently oscillating inflaton field $\phi$, the SM radiation produced through decays of the inflaton field to the SM Higgs boson, and the vector DM~$X_\mu$. To track the evolution of this system, one has to consider three coupled Boltzmann equations averaged over the inflaton oscillations. The energy or number density of each considered species changes due to the expansion of the Universe, controlled by terms proportional to the Hubble rate $H$, and interactions between those three sectors. In what follows, we assume that $\phi$ can directly ``decay''~\footnote{The inflaton ``decay'' to the SM Higgs boson and vector DM can be understood as the production of the pairs of SM Higgs bosons and DM vector bosons in a quantum process out of the vacuum in the presence of the classical inflaton field~(\ref{eq:phi_app}).} to the SM Higgs pairs, with an averaged (over each oscillation period) decay rate $\langle \Gamma_{\phi \rightarrow h h}\rangle$, and to DM vectors, with an averaged rate $\langle \Gamma_{\phi \rightarrow XX} \rangle$. Due to the fact that the latter process arises from the dim-5 operator, $\phi (D_\mu \Phi)^\dagger(D^\mu \Phi)/\mpl$, the $\phi \! \rightarrow \! XX$ process is subdominant, and the reheating dynamics is dominantly governed by perturbative decays of the inflaton field to the SM Higgs bosons. 

During the reheating period, the Higgs potential can be written as
\begin{align}
V({\bm h}) &=
\mu_h^2 \,|{\bm h}|^2+\lambda_h\, |{\bm h}|^4\,,
\label{eq:H_pot}
\end{align}
where the Higgs mass parameter is
\beq
\mu_h^2(\phi) \equiv g_{h\phi}^{} \mpl\, \phi(t).	\label{eq:muh}
\eeq
In the above expression we have neglected the Higgs mass contributions proportional to the electroweak vev $v_{\rm  EW} \!\simeq\! 246\gev$. 
In the presence of an oscillating inflating field, the Higgs field undergoes rapid phase transitions acquiring a $\phi(t)$-dependent vev during the reheating phase, i.e.,
\begin{align}
v_h &=\begin{dcases}
~0\,, & \phi(t)>\!0\,,		\\
\sqrt{\frac{|\mu_h^2(\phi) |}{\lambda_h}}\,,	& \phi(t)<\! 0.
\end{dcases}	\label{eq:Higgs_vev}
\end{align}
Note that $v_h$ is a function of the inflaton field $\phi$, therefore at the end of the reheating phase, this non-trivial inflaton-induced vacuum-expectation-value (vev) vanishes, and the Higgs field remains in the symmetric phase until the electroweak phase transition at the temperature scale $\op(100)\gev$. 
Furthermore during reheating all SM massive particles receive a non-zero mass due to their coupling to the Higgs field, i.e.,
\begin{align}
m_{\rm{SM}} = \frac{m_{\rm{SM}}^{\rm{EW}}}{v_{\rm{EW}}} v_h,
\end{align}   
where $m_{\rm{SM}}^{\rm{EW}}$ denotes the EW mass of SM particles. 

Before moving ahead, we would like to comment on the gravitational production of the SM radiation from the inflaton background through a graviton exchange~\cite{Clery:2021bwz, Haque:2021mab, Clery:2022wib, Co:2022bgh}. This production mechanism is universal and independent of any direct coupling of the inflaton to the SM sector. However, as we show in Appendix~\ref{s.graviational_production}, in the presence of the inflaton-Higgs interaction~\eqref{eq:Lint} with $\ghp$ satisfying the bound~\eqref{eq:ghphi_con} the gravitational production of SM radiation through the inflaton background field is subdominant. Therefore, we do not discuss the gravitational production of the SM radiation in the following and only consider the production of the SM radiation through inflaton decays to the SM Higgs.

\begin{figure}[t!]
\begin{center}
\includegraphics[width=\linewidth]{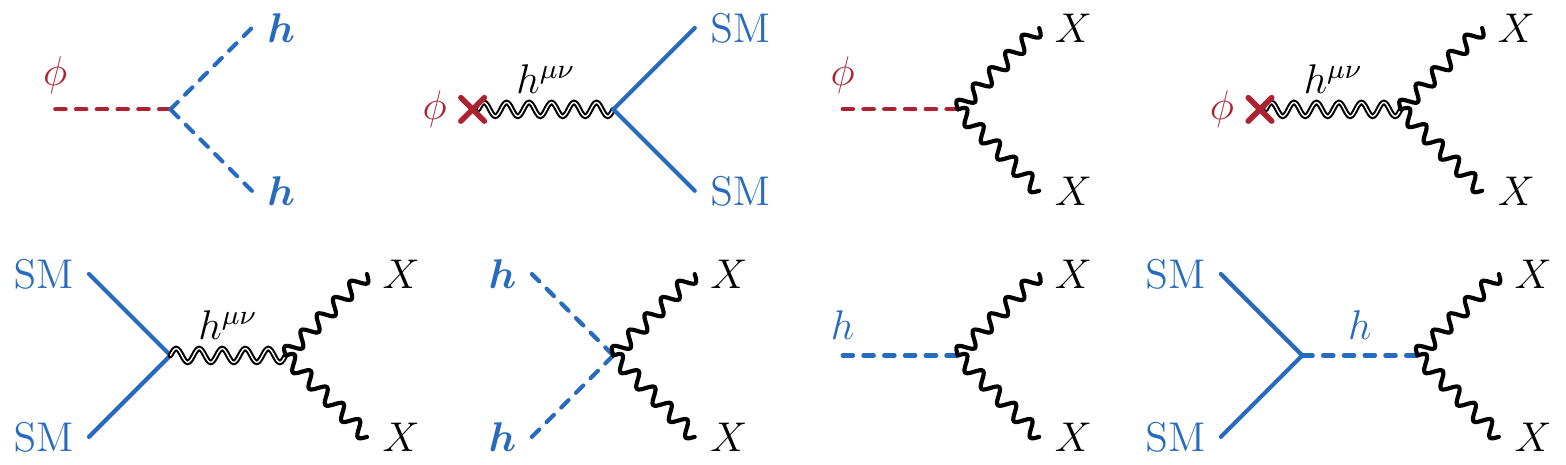}
\caption{Feynman diagrams for all possible interactions between the inflaton, SM, and DM, where ``cross'' indicate the inflaton background interaction with the graviton field.} 
\label{fig:feyn_SM_DM}
\end{center}
\end{figure}

During the reheating epoch the DM particles can be produced through the following types of processes:
\bit[leftmargin=*,itemsep=0pt]
\item {\it Inflaton decays:} In this case, the DM $X_\mu$ particles are produced from the inflaton decays through the effective interaction~\eqref{eq:Lint}. 
\item {\it Gravitational production from the inflaton field}: This is an irreducible DM production mechanism where the inflaton background field transfers its energy density to the DM through the graviton exchange. This production mechanism is a result of graviton universal coupling to energy-momentum tensors of the inflaton and DM fields~\eqref{eq:Lint}. 
\item {\it Gravitational freeze-in:} Annihilation of SM particles to the DM through $s$-channel graviton $h_{\mu \nu}$ exchange can be considered as an irreducible mechanism of DM production, which does not require any additional coupling between these two sectors. An amplitude for this process is suppressed by $1/\mpl^2$ as a result of graviton universal coupling to energy-momentum tensors of the SM and DM sectors~\eqref{eq:Lint}.  
\item {\it Higgs portal freeze-in and Higgs decays:}
In this case, the dark sector is produced through annihilations of Higgs bosons due to the dim-6 operator~\eqref{eq:Lint}. In the \textit{massive reheating scenario}, Higgs boson acquires the inflaton-induced vev \eqref{eq:Higgs_vev} and therefore the contact operator $\cxh\, m_X^2/(2\mpl^2)X_\mu X^\mu\lvert {\bm h} \rvert^2$, after expanding around the vev $v_h$, generates an additional Higgs-DM interaction responsible for $h_0\to XX$ decays, i.e.,
\begin{align}
\frac{\cxh\, m_X^2}{2\mpl^2} v_h h_0  X_{\mu} X^{\mu} \nonumber  + \cdots\,.
\end{align}
Furthermore, the above term opens a new DM production channel, in which $X$ particles are created from the annihilation of SM species with $s$-channel Higgs exchange, i.e., ${\rm SM \, SM} \rightarrow h_0 \rightarrow XX$. Here we should point out that both processes $h_0 \! \rightarrow \! XX$ and ${\rm SM \, SM} \rightarrow h_0 \rightarrow XX$ are possible only in the broken phase, i.e., half-period of the inflaton oscillations when $\phi \! < \! 0$ and the Higgs boson vev is non-zero. Note that in the \textit{massless reheating scenario} DM particles are produced only through the Higgs annihilations via the contact diagram $hh \! \rightarrow XX$.
\eit
In \fig{fig:feyn_SM_DM} we collect all possible interactions among the inflaton, SM and DM fields. Where ``cross'' denotes the inflaton background interaction with the graviton field. 

We write the time-averaged Boltzmann equations (BEq) for the inflaton energy density $\rho_\phi$, the SM radiation energy density $ \rho_{\rm SM}^{}$, and the DM number density $n_X$ during the reheating phase as
\begin{align}
\dot{\rho}_\phi \!+\! \frac{6n}{n\!+\!1} H \rho_\phi &=- \langle \Gamma_\phi \rangle \rho_\phi,  \label{eq:beqn_phi} \\
\dot{\rho}_{\rm SM}^{} + 4 H \rho_{\rm SM}^{} &\!=\! \langle \Gamma_{\phi\to {\rm SM}\,{\rm SM}} \rangle \rho_\phi \!-\! 2 \langle E_X \rangle \mathcal{S}_{\rm{SM}}  \!-\! \langle E_{h_0} \rangle \mathcal{D}_{h_0},
 \label{eq:beqn_r} \\
\dot{n}_X + 3 H n_X &=  \mathcal{D}_\phi + \mathcal{S}_\phi + \mathcal{S}_{\rm{SM}} + \mathcal{D}_{h_0}
 \label{eq:beqn_x},
\end{align}
where the Hubble rate is 
\begin{align}
H^2 = \frac{1}{3 \mpl ^2} \left( \rho_\phi + \rho_{\rm SM} + \rho_X\right). \label{eq:hubble}
\end{align}
Above $\langle \cdots \rangle$ stands for thermal or (and) time-averaged quantity.
The total inflaton decay rate, $\langle \Gamma_\phi \rangle$, contains two terms induced by the contact inflaton interactions with the Higgs field and DM vectors, as well as terms generated by indirect interactions through gravity. For details see \app{s.graviational_production}. 
The first two terms in the third Boltzmann equation, i.e.,  $\mathcal{D}_\phi$ and $\mathcal{S}_{\phi}$, account for the DM pair production from the vacuum in the presence of the inflaton background via the effective dim-5 operator~\eqref{eq:Lint} and through gravity, respectively. The explicit form of the these two terms has been derived in \app{s.graviational_production} where we have used the notation  ${\cal D}_\phi\equiv {\cal D}^{(1)}_{\phi\to XX}$, and ${\cal S}_\phi\equiv {\cal D}^{(2)}_{\phi\to XX}$. Next two terms in the r.h.s. of Eq.\eqref{eq:beqn_x} are defined as 
\begin{align}
\mathcal{S}_{\rm{SM}} & \!\equiv\! \bar{n}_X^2\,\langle \sigma_{XX \rightarrow {\rm  SM \, SM}}^{} \lvert v \rvert \rangle,   &\mathcal{D}_{h_0} &\!\equiv\!  \bar{n}_{h_0}\langle \Gamma_{h_0\to XX} \rangle , \label{eq:collision_decay}
\end{align}
with $\sigma_{XX \rightarrow {\rm  SM \, SM}}^{}$ being the DM annihilation cross-section to the SM, whereas $\Gamma_{h_0\to XX}$ denotes the Higgs decay rate to DM particles. Above $\langle E_X \rangle$ and $\langle E_{h_0} \rangle$ denote averaged energies of the $X$ and $h_0$ particles, respectively.
Note that in the above relations~\eqref{eq:collision_decay} we have assumed that the number density of the Higgs field follows the thermal equilibrium value, i.e., $n_{h_0} \! \simeq \! \bar{n}_{h_0}$, while the dark sector is assumed to be out-of-thermal equilibrium, i.e., $n_X \! \ll \!\bar{n}_X$.  
Ignoring quantum statistics and adopting the Maxwell-Boltzmann distribution function, the equilibrium number density of species $i$ with spin $J_i$ is given by
\begin{align}
\bar{n}_i &=\frac{(2 J_i+1)}{2 \pi^2} m_i^2\, T\, K_2 \!\left(\frac{m_i}{T} \right), 
\end{align} 
where $K_2(x)$ denotes the modified Bessel function of the second kind. 
Finally, let us emphasize that in the above Boltzmann equations we have assumed that during one oscillation period variations of the Hubble rate $H$ and the effective mass of the inflaton field $m_\phi$ could be neglected. 

In this section, we assume that the presence of DM interactions does not modify substantially the reheating dynamics, which is mainly driven by the first two Boltzmann equations~\eqref{eq:beqn_phi}-\eqref{eq:beqn_r}. 
In particular, we assume that inflaton predominantly transfers its energy into the radiation sector through the Higgs portal, since the inflaton's higher dimensional interactions through gravity to the SM and DM as well as direct coupling to the vector DM are negligible as compared to a relevant (dim-3) Higgs portal operator. Such that $\langle \Gamma_{\phi} \rangle\simeq  \langle  \Gamma_{\phi \rightarrow {\rm{SM}}\, {\rm{SM}} } \rangle  \simeq \langle  \Gamma_{\phi \rightarrow  hh} \rangle $. Furthermore, in the evolution of inflation and SM radiation energy densities, we ignore the DM scatting and Higgs decays to DM particles. This is a reasonable approximation for the freeze-in DM scenario with a negligible DM abundance at the onset of the reheating period.

There are several other comments here in order. Firstly, notice that we are assuming that the whole SM is created from Higgs boson pairs emerging in the process of inflaton ``decays'', i.e., the Higgs bosons decay and scatter to produce the rest of the SM.
Secondly, we assume the SM particles produced are thermalized instantaneously. This is a reasonable assumption given that once the Higgs bosons are produced, they would immediately decay/scatter to the rest of the SM particles, leading to instantaneous thermalization~\cite{Harigaya:2013vwa}.
Thirdly, note that the r.h.s. of \eqref{eq:beqn_phi} was not present when we have discussed the inflaton dynamics in \sec{s.inflaton}, cf. \eqref{eq:continuity-eqn}. Here it describes a quantum process of the particle pair production (in other words, a transition from the quantum vacuum to the two-particles final state) in the presence of the oscillating classical inflaton field.
Fourthly, let us define the temperature of the thermal bath, $T$, through the following well-known relation,
\begin{align}
\rho_{\rm SM}^{}= \frac{\pi^2}{30} g_\star(T)\, T^4, \label{eq:rhoRTrel}
\end{align}
where $g_\star(T)$ counts an effective number of relativistic degrees of freedom at the temperature~$T$. 
Here we approximate $g_\star(T)$ by the constant value of $106.75$.

Finally, we would like to comment on possible thermal effects on the Higgs potential~\eqref{eq:H_pot} and in particular thermal corrections to the Higgs mass parameter $\mu_h^2$ during the reheating phase.  In the large temperature limit, i.e., $T\!\gg\! m_{\rm{SM}}$, the thermal corrections to the Higgs mass $\delta\mu_h^2(T)$ are~\cite{Quiros:1999jp},
\begin{align}
\delta\mu_h^2(T)&\simeq \bigg[\frac{3 g^2}{16}+\frac{g^{\prime 2}}{16}+\frac{y_t^2}{4}+\frac{\lambda_h}{2}\bigg] \,T^2 \simeq 0.4 \,T^2, 	\label{eq:thermal}
\end{align}
where $g/g^\prime$ and $y_t$ are the SM electroweak gauge and top Yukawa couplings, respectively. The temperature-dependent contribution to the Higgs mass arises from the Higgs interactions with the high-temperature bath of relativistic particles. 
In the following analysis, we neglect such thermal corrections which become relevant only when $T\!\gtrsim\! |\mu_h^{}(\varphi)|$.

\subsection{Evolution of the inflaton and SM energy densities}
\label{s.inflaton_energy_density}

In the following, our goal is to find approximate analytical solutions for the inflaton and the SM radiation energy densities by solving the first two Boltzmann Eqs.~\eqref{eq:beqn_phi}-\eqref{eq:beqn_r}. After employing the simplifications discussed above, the Boltzmann Eqs.~\eqref{eq:beqn_phi}-\eqref{eq:beqn_r} take the simple form, 
\begin{align}
\dot{\rho}_\phi \!+\! \frac{6n}{n\!+\!1} H \rho_\phi &=0 ,  \label{eq:beqn_phi_simp} \\
\dot{\rho}_{\rm SM}^{} + 4 H \rho_{\rm SM}^{} &= \langle \Gamma_{\phi\to hh} \rangle \rho_\phi .
 \label{eq:beqn_r_simp} 
\end{align}
Above in the first equation, we have neglected the inflaton decay rate $\langle \Gamma_\phi \rangle$ compared to the Hubble rate $H$ during the reheating phase. We use the above set of simplified Boltzmann equations to analytically determine the reheating dynamics, however, note that for the numerical analysis we consider solutions of the exact Boltzmann Eqs.~\eqref{eq:beqn_phi}-\eqref{eq:beqn_r}.
The initial conditions for the inflaton energy density $\rho_\phi$ and the SM energy density $\rho_{\rm SM}$ at the onset of the reheating phase are
\begin{align}
\rho_{\phi_e}  &\equiv \rho_\phi(a_e) = \Lambda^4\Big( \frac{\varphi_e}{\mpl} \Big)^{2 n},		& \rho_{\rm SM}(a_e) &= 0,
\end{align}
where we have employed Eq.~\eqref{eq:envelope}.
Hereinafter, we use the scale factor $a$ as an independent time variable rather than physical time $t$, where the two are related during the reheating phase through~\eq{eq:at_reh}. 
Therefore, the inflaton energy density during reheating evolves as
\begin{align}
\rho_\phi(a)& = \rho_{\phi_e}  \Big(\frac{a_e}{a} \Big)^{\frac{6n}{n+1}},& &\text{for} &a \in [a_e, a_{\tiny \text{rh}}], \label{eq:rhophia}
\end{align}
where $a_{\tiny \text{rh}}$ denotes the scale factor at the end of reheating which we define by the equality of the inflaton and radiation energy densities, i.e., $\rho_\phi(a_{\tiny \text{rh}}) = \rho_{\rm SM}(a_{\tiny \text{rh}}) $. As we see in the following, the end of reheating roughly coincides with the condition $\langle \Gamma_\phi \rangle\! \sim\!H$.

Since the inflaton energy density dominates the total energy density during reheating, we can solve the Friedmann equation \eqref{eq:hubble}, obtaining
\begin{align}
H(a) &= H_e \left( \frac{a_e}{a} \right)^{\frac{3n}{n+1}},& &\text{for} &a\in [a_e, a_{\tiny \text{rh}}], \label{eq:HubbleRate_app}
\end{align}
where $H_e\equiv H(a_e)= \sqrt{\rho_{\phi_e}  }/(\sqrt{3}\mpl)$.
It is instructive to rewrite the Boltzmann equation for the SM energy density~\eqref{eq:beqn_r_simp} as
\begin{align}
\frac{d(a^4\rho_{\rm SM}^{})}{da} & \simeq \frac{\langle \Gamma_{\phi\to  hh}  \rangle }{H}  a^3\rho_\phi. \label{eq:BeqR} 
\end{align}
After using the solutions for the inflaton energy density \eqref{eq:rhophia} and the Hubble rate \eqref{eq:HubbleRate_app} during the reheating phase, we can rewrite Eq.~\eqref{eq:BeqR} as
\begin{align}
\frac{d(a^4\rho_{\rm SM}^{})}{da} \simeq 3\mpl^2 H_e\langle \Gamma_{\phi\rightarrow  hh} \rangle a_e^{3}  \, \Big( \frac{a}{a_e} \Big)^{\frac{3}{n+1} },\label{eq:eqR1}
\end{align}
which is straightforward to solve when the inflaton decay rate to SM Higgs boson pairs is known.
As we will show in the following subsections, in general, the width is a function of time~\cite{Ahmed:2021fvt}.
In the next subsection, we calculate the inflaton ``decay'' to a pair of Higgs bosons, taking into account the non-trivial $\phi$-depenent Higgs mass. 

\subsection{Higgs boson production from oscillating inflaton}
\label{s.inflaton_decay}

In this subsection, we calculate the energy gain $E$ of the SM radiation due to the Higgs pair production at the expense of the inflaton field. The process that is considered is a quantum production from a vacuum state, $\ket{0}$, to the two-Higgs-boson final state, $\ket{ \bm{h} \bm{h}}$, in the presence of the classical inflaton field, see also ~\cite{Ichikawa:2008ne, Garcia:2020wiy, Clery:2021bwz, Garcia:2022vwm, Kaneta:2022gug}. 
Due to the energy conservation, the energy density gained by the Higgs pair production from the vacuum must be equal to the energy loss of the inflaton field. To put it another way, the energy density, initially accumulated in the coherent oscillations of the inflaton field, is transferred to the SM radiation sector during reheating due to the cubic inflaton-Higgs coupling~$\phi |{\bm h}|^2$. Note that in this term, $\phi= \varphi \,\mathcal{P}$ should be interpreted as a given time-dependent coefficient, so effectively it makes the Higgs boson mass vary with time. For interactions linear in the $\phi$ field the energy gain per unit volume $V$ and per unit time due to the pair production of final state particles $f$ with mass $m_{f}$ can be calculated as
\begin{align}
\frac{1}{V} \frac{d E}{dt} &\!=\! \frac{\varphi^2(t)}{8 \pi} \sum_{k=1}^\infty  k \omega \lvert \mathcal{P}_k \rvert^2 \Big\lvert \mathcal{M}_{\phi \rightarrow ff}(k) \Big\rvert^2{\rm Re}\!\bigg[\!\sqrt{1- \frac{4 m_f^2}{k^2\omega^2}}\bigg], \label{eq:E_g}
\end{align}
where $V$ is the volume factor. The matrix element squared $\lvert \mathcal{M}_{\phi \rightarrow ff} \rvert^2$, summed over spin/polarization of final states, describes the quantum process of production of pair of particles out of the vacuum in the presence of the classical inflaton background, while $\mathcal{P}_k$ denotes Fourier coefficients in the expansion of $\mathcal{P}(t)$:
\begin{align}
\mathcal{P}(t) = \sum_{k= - \infty}^{\infty} \mathcal{P}_k e^{ - i k  \omega t}\,, 
\end{align}
 with the oscillation frequency/energy $\omega = 2 \pi/{\cal T}$~\eqref{eq:period} and Fourier mode number $k$,
 \begin{align}
 \mathcal{P}_k = \frac{1}{\mathcal{T}(t_0)} \int_{t_0}^{t_0 + \mathcal{T}(t_0)} \, dt \,  \mathcal{P}(t) e^{i k \omega t}.
 \end{align}
 Note that for $n\!=\! 1$ the only non-zero Fourier coefficient is $\mathcal{P}_1 = 1/2$, while for $n\!=\! 2/3$, $n\!=\!3/2$ and $n\!=\!2$  all even coefficients are zero. Moreover, the value of  $\mathcal{P}_k$ quickly decreases with $k$ such that the sum $\sum_k |\mathcal{P}_k|^2$ converges around $k$ of the order ten. The numerical values of $\sum_k |\mathcal{P}_k|^2$ are collected in \tab{tab:PkSums}.
\begin{table}[t!]
$
\begin{array}{|c||c|c|}
\hline
n & \sum_k |\mathcal{P}_k|^2 & \sum_k |\mathcal{P}_k^{2n}|^2 \\
\hline
 2/3 & 0.2605 & 0.1883\\
 1& 1/4 & 1/16  \\
 3/2 & 0.2377 & 0.1455 \\
 2 & 0.2286  & 0.0635   \\
 \hline
\end{array}
$\vspace{7pt}
\caption{Numerical values of the summation factors appearing in Eq.~\eqref{eq:E_g}  (first column) and Eq.~\eqref{eq:sphiX} (second column) for different values of $n$.}
\vspace{7pt}
\label{tab:PkSums}
\end{table}

It is understood that the above process is possible only if it is kinematically allowed, i.e., when \mbox{$kw\!>\!2m_f$}.
Then, the collision term in Eq.~\eqref{eq:beqn_phi} could be written by defining the inflaton ``decay width'' as 
\beq
\frac{1}{V} \frac{d E}{dt} \equiv \rho_\phi\, \Gamma_\phi.
\label{eq:width_def}
\eeq
Adopting the r.h.s. of the above expression, one can mimic the standard form of the collision term for decaying particles with the ``width'' calculated using the classical solution of \eqref{eq:inf_eom}. It would be an acceptable iterative procedure that starts with the classical solution. Nevertheless, we will adopt an alternative approach. It is important to realize that the ``width'' defined by \eqref{eq:width_def} non-trivially depends on $\rho_\phi$, i.e., on the function we are seeking by solving the respective Boltzmann equation. Therefore we find it more appropriate to express the whole r.h.s. of \eqref{eq:beqn_phi} in terms of the inflaton density $\rho_\phi$ adopting \eqref{eq:envelope}. Solutions of the Boltzmann equations in both cases approximately coincide during reheating when the ``width'' is smaller than the Hubble rate. Formally the difference is of higher order in powers of coupling constant that enters the ``width''.

Our next step is to calculate the time-averaged inflaton decay width, $\langle\Gamma_{\phi\to hh}\rangle$, using the strategy described above. For the cubic inflaton-Higgs interaction, we obtain the following expression by averaging over the fast oscillations of the inflaton field, 
\begin{align} 
 \langle \Gamma_{\phi \rightarrow  hh} \rangle  
 & = \frac{g_{h\phi}^2}{32 \pi} \frac{\mpl^2}{m_\phi} \gamma_h^{} \,, \label{eq:Gam_phi}
\end{align}
with
\begin{align} 
\gamma_h^{} &\!\equiv\!
\frac{8 n \sqrt{\!\pi n (2n\!-\!1)}\Gamma\!\left(\!\frac{n+1}{2n}\!\right)}{\Gamma\!\left(\frac{1}{2n}\right)}\!\!
\sum_{i=0}^{3} 
 \sum_{k=1}^\infty k \lvert \mathcal{P}_k \rvert^2 \bigg\langle \!\!{\rm Re}\!\bigg[\!\sqrt{1- \frac{4 m_{h_i}^2}{k^2\omega^2}}\bigg]\!\bigg\rangle ,		\label{eq:gammah}
\end{align}
where we have employed \eq{eq:E_g} and \eqref{eq:width_def}. Above the inflaton decay is summed over four real components, $h_i$, of the Higgs doublet. Note that $\langle \Gamma_{\phi \rightarrow  hh} \rangle$ has two sources of time-dependence: 
\bit
\item[(i)] the effective mass of the inflaton field $m_\phi$~\eqref{eq:eff_mass},
\item[(ii)] the inflaton-induced mass of the SM Higgs $m_{h_i}$~\eqref{eq:muh}. 
\eit
The time-averaged inflaton decay rate is time-independent only in the \textit{massless reheating scenario}, i.e., $m_{h_i} \!=\! 0$ with the quadratic inflaton potential, i.e., $n\!=\!1$. In this special case $\gamma_h^{}\!=\!4$ due to the four real massless scalar components of the Higgs field. 

Since the energy scale of the electroweak symmetry breaking is much smaller than the energy scale of reheating, the EW mass of the Higgs boson can be neglected during this period. However, all Higgs doublet components, $h_i$, acquire non-zero masses via their coupling to the inflaton field,
 \begin{align}
 m_{h_i}^2 = g_{h\phi} \mpl \,  \varphi 
\begin{cases} \lvert \mathcal{P} \rvert, &  \mathcal{P}>0, \, i= 0,1,2,3,\\
 \!\!2 \lvert \mathcal{P} \rvert, & \mathcal{P}<0,  \, i= 0, \\
 \infty, & \mathcal{P}<0,   \, i= 1, 2, 3. \label{eq:Higgs_mass}
\end{cases}
 \end{align}
We see that $m_{h_i}^2$ is subjected to the short-scale oscillations of the inflaton field through $\mathcal{P}$ and that is why we have to average the square root in \eqref{eq:Gam_phi}. Though, the summation and averaging performed in \eqref{eq:Gam_phi} deserves an explanation.
During one half of the inflaton oscillation period, when $\mathcal{P} \!>\!0$, the electroweak symmetry remains unbroken, and each Higgs doublet component receives a mass $m_{h_i}\!=\! \sqrt{g_{{\bm h}} \mpl  \varphi|\mathcal{P}|}$. In the second half of the period, when $\mathcal{P}$ becomes negative, the physical Higgs mass is $m_{h_0}\!=\! \sqrt{ 2 g_{{\bm h}} \mpl  \varphi |\mathcal{P}|}$, while the three Goldstone bosons become the longitudinal components of the SM gauge bosons and decouple from the inflaton. In the unitary gauge that corresponds to an infinite mass of $h_i$ for $i=1,2,3$, these are non-dynamical modes. 
The averaging performed in \eqref{eq:Gam_phi} takes the above properties of $m_{h_i}^2$ into account.
It is instructive to write the averaged masses of the Higgs boson components as
\beq
\langle m_{h_i}\!\rangle\!=\! \sqrt{\!g_{h\phi} \mpl \,  \varphi} \,
\frac{\Gamma \!\left(\!\frac{3}{4 n}\!\right) \Gamma\! \left(\!\frac{n+1}{2 n}\!\right)}{\Gamma\! \left(\!\frac{2 n+3}{4 n}\!\right) \Gamma\! \left(\!\frac{1}{2 n}\!\right)}\!\!
\begin{cases} \!\frac{1+\sqrt{2}}{2}, &  i\!=\! 0,\\
\! \frac12, &  i\!=\! 1, 2, 3, 		
\end{cases}	\label{eq:Higgs_mass_ave}
\eeq
which depends on time only through the slow varying envelope function $\varphi (a)$ given in \eq{eq:envelope_sol}. 

Note that the non-trivial Higgs mass effects appear only through the phase-space factor $\langle \sqrt{1- (2 m_{h_i}/k\omega)^2} \rangle$ in $\gamma_h^{}$ \eqref{eq:gammah}. 
In particular, the phase-space kinematic effects are only relevant when the energy $k\omega$ of the inflaton mode $k$ is larger than twice Higgs boson mass $m_{h_i}$. We also note that for a general value of $n$ the dominant contribution originates from the first Fourier mode, i.e., $k\!=\!1$, as the amplitude ${\cal P}_k$ drops quickly for larger mode numbers. Therefore it is instructive to write the ratio 
\begin{align}
\frac{\langle m_{h}\rangle}{\omega}&\!\simeq\! \frac{\sqrt{\!g_{h\phi} \mpl \,  \varphi}}{m_\phi}\sqrt{\frac{2n-1}{\pi\,n}}\frac{\Gamma \!\left(\!\frac{3}{4 n}\!\right) }{\Gamma\! \left(\!\frac{2 n+3}{4 n}\!\right) },	\notag \\
&\!= \!\frac{\sqrt{\!g_{h\phi}}}{\sqrt{2\pi}\,n} \frac{\mpl^2}{\Lambda^2}\frac{\Gamma \!\left(\!\frac{3}{4 n}\!\right) }{\Gamma\! \left(\!\frac{2 n+3}{4 n}\!\right) }\Big(\frac{\varphi_e}{\mpl}\Big)^{\!\frac{3-2n}{2}} \Big(\frac{a}{a_e}\Big)^{\!\frac{3 (2 n-3)}{2 (n+1)}}, \label{eq:mhomega}
\end{align}
where for brevity we have neglected the Higgs field $h_i$ component-dependent contribution $(1\!+\!\sqrt{2})/2$ for $i=0$ and $1/2$ for $i=1,2,3$. 
In the second line of \eq{eq:mhomega} we have employed the solution for $\varphi(a)$ \eqref{eq:envelope_sol}, which is valid during reheating, i.e., $\langle \Gamma_{\phi \rightarrow  hh} \rangle \! \ll \! H$, and the expression for the inflaton mass $m_\phi$ \eqref{eq:m_phi}. 
Note that the time-dependence of $\langle m_{h}\rangle/\omega$ through the scale factor $a$ disappears for $n\!=\!3/2$, while for $n\!<\!3/2$ ($n\!>\!3/2$) it decreases (increases) with time during reheating. This implies that the kinematic suppression amplifies over time for $n\!>\!3/2$ and stays constant for $n\!=\!3/2$. However, for $n\!<\!3/2$, the role of the kinematic suppression becomes less and less important as the Universe expands.

\begin{figure}[t!]
\begin{center}
\includegraphics[width=0.5\linewidth]{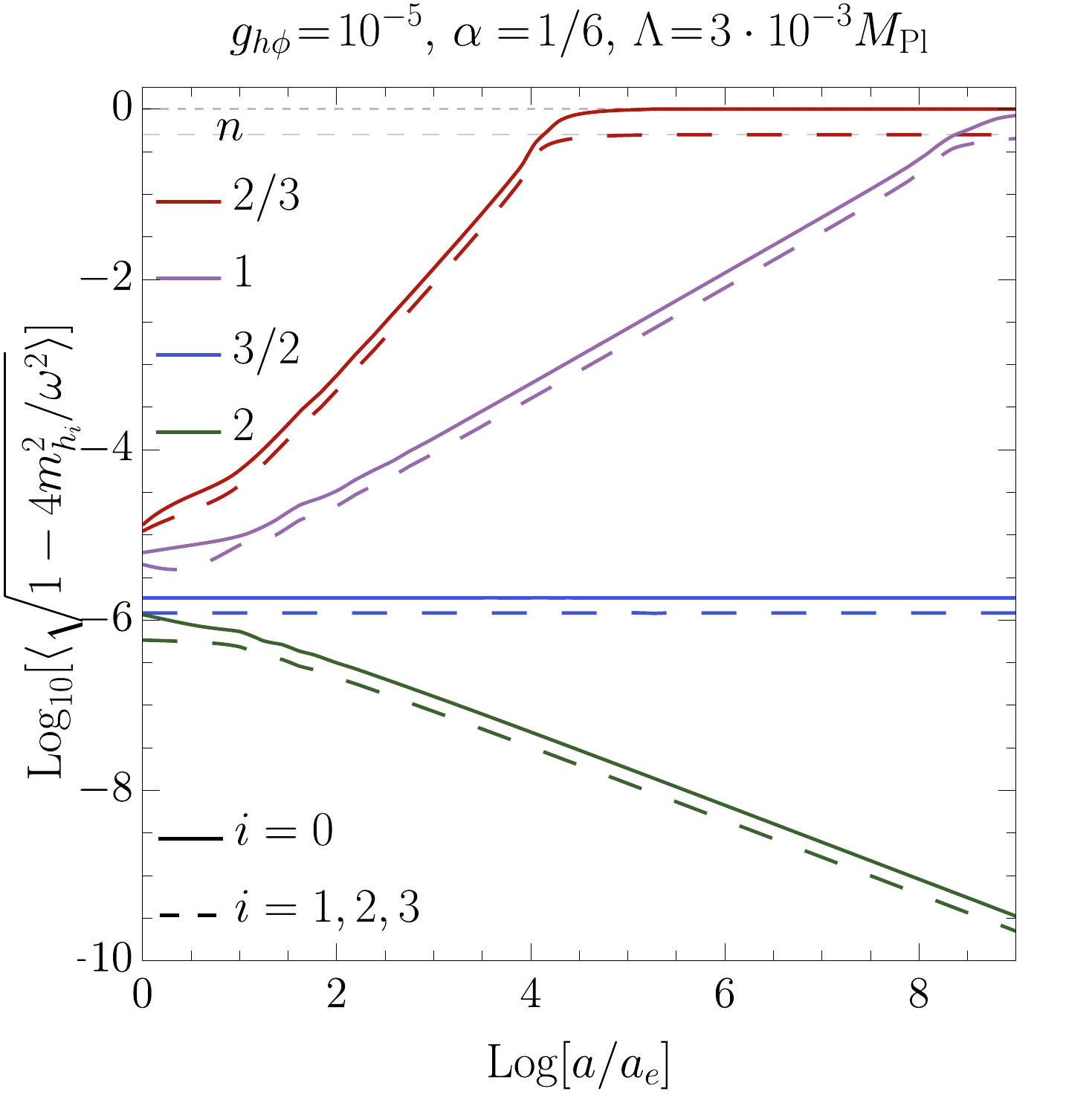}\!\!\!
\includegraphics[width=0.5\linewidth]{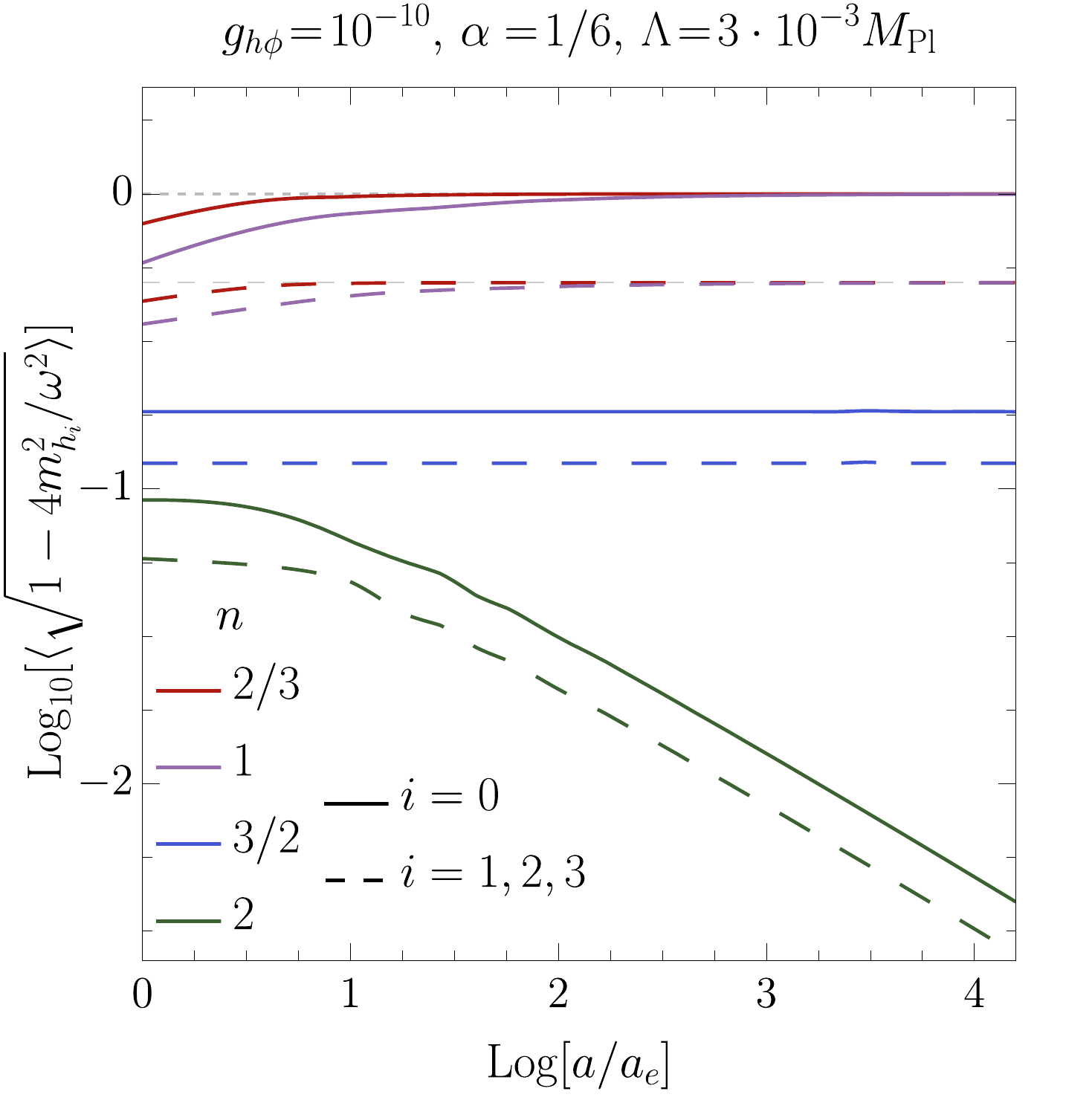}
\caption{Evolution of the phase-space factor $\big\langle\! \sqrt{1- (2 m_{h_i}/k \omega)^2} \big\rangle$ for the inflaton decay to a pair Higgs bosons as a function of the scale factor $a$ for $k\!=\!1$ and $g_{h \phi} \!=\! 10^{-5}$ and $10^{-10}$ in the left- and right-panels, respectively. Solid (dashed) lines are for the $i\!=\!0$ ($1,2,3$) component of the SM Higgs doublet. }
\label{fig:sqAveragePlotgh510}
\end{center}
\end{figure}

The crucial factor in the calculation of the inflaton width \eqref{eq:Gam_phi} is $\langle \!\sqrt{1-(2 m_{h_i}/k\omega)^2} \rangle$, which captures the kinematic phase-space suppression due to non-trivial Higgs mass. We find the phase-space factor after time-averaging over fast inflaton oscillations as
\begin{align}
\bigg\langle \!\!\sqrt{1\!-\! \frac{4 m_{h_i}^2}{k^2\omega^2}}\!\bigg\rangle &\approx \bigg\lceil\!\frac{9\,\kappa_i \,n \Gamma\! \left(\frac{n+1}{2 n}\right)}{16  \sqrt{\pi }\, \Gamma\! \left(\frac{1}{2 n}\right)}\frac{k^2\omega^2}{g_{h\phi}\mpl\varphi}\!\bigg\rceil,	\notag\\
&=\bigg\lceil\frac{9 \sqrt{\pi } \kappa_i n^2  \Gamma \!\left(\frac{n+1}{2 n}\right)^3}{16 (2 n-1) \Gamma\! \left(\frac{1}{2 n}\right)^3}\frac{k^2 m_\phi^2}{g_{h\phi}\mpl\varphi}\bigg\rceil, \label{eq:mhi_sqrt}
\end{align}
where for $i\!=\!0$ component of the Higgs field $\kappa_0\!=\!1$, whereas for $i\!=\!1,2,3$ components $\kappa_{1,2,3}\!=\!1/2$ due to the fact that these modes become Goldstone modes in the broken phase. 
Above `ceiling' function $\lceil\cdots\rceil$ is defined as follows:
\beq
\lceil x \rceil =\begin{cases}
1, & x\geq1,	\\ 
x,	 & x<1,
\end{cases}  \label{eq:ceiling}
\eeq
where $\lceil\cdots\rceil = 1$ implies massless final states, i.e., no kinematic suppression. 
Note that in the above expression, for a generic $n$, the time dependence enters through the inflaton envelope function $\varphi(a)$~\eqref{eq:envelope_sol} and the inflaton mass $m_\phi$~\eqref{eq:m_phi}. We can also rewrite \eq{eq:mhi_sqrt} as a function of the scale factor:
\begin{align}
\bigg\langle \!\!\sqrt{1\!-\! \frac{4 m_{h_i}^2}{k^2\omega^2}}\!\bigg\rangle \!\approx\! \bigg\lceil\frac{g_{k}^{}}{g_{h\phi}}\Big(\!\frac{a_e}{a}\!\Big)^{\!\frac{3 (2 n-3)}{(n+1)}}\bigg\rceil,	\label{eq:sq_factor}
\end{align}
where we have used $M\!=\!\mpl$ ($\alpha\!=\!1/6$), $\varphi_e\! \sim \! \mpl$ and defined 
\begin{align}
g_{k}^{}\!\equiv\! \frac{9\sqrt{\pi}  \kappa_i  k^2 n^3 \Lambda^4\, \Gamma\! \left(\!\frac{n+1}{2 n}\!\right)^3}{8\, \mpl^4\,\Gamma\! \left(\!\frac{1}{2 n}\!\right)^3}.	\label{eq:gkin}
\end{align}
From \eq{eq:sq_factor} we see that the phase-space factor $\langle \!\sqrt{1-(2 m_{h_i}/k\omega)^2}\rangle$ increases (decreases) with time for $n\!<\!3/2$ ($n\!>\!3/2$), whereas it stays constant for $n\!=\!3/2$ as shown in \fig{fig:sqAveragePlotgh510}.

Moreover, one can estimate the scale factor $a_{k}$ at which the kinematic suppression becomes irrelevant (for $n\!<\!3/2$) or relevant (for $n\!>\!3/2$) by the condition $\lceil\cdots\rceil = 1$ in the above expression as 
\beq
a_{k}\equiv a_e \bigg(\frac{g_{k}^{}}{g_{h\phi}}\bigg)^{\frac{(n+1)}{3 (2 n-3)}},\qquad n\neq3/2.	\label{eq:akin}
\eeq
Whereas for the special case $n\!=\!3/2$ the phase-space~\eqref{eq:sq_factor} suppression is a time-independent constant factor when $g_{h\phi}> g_{k}^{}$ for $k=1$.
Note that depending on the strength of $g_{h \phi}$, the kinematic suppression may end during the reheating phase for $n\!<\!3/2$. 
In this case, for scale factor \mbox{$a_{k}\!<\!a\!\leq\! a_{\rm rh}$} the reheating dynamics would be the same as that of massless Higgs final states. 
On the other hand, the kinematic suppression becomes more and more significant with time for $n>3/2$. This is because even if the averaged Higgs mass was much smaller than the inflaton energy at the onset of reheating, it would soon become of the order of $k \omega$, and hence the kinematic effects would be important until the end of the reheating phase.
With the above definition of $a_{k}$, we can recast the phase-phase factor \eqref{eq:sq_factor} as
\begin{align}
\bigg\langle \!\!\sqrt{1\!-\! \frac{4 m_{h_i}^2}{k^2\omega^2}}\!\bigg\rangle\! \approx\!\begin{dcases}
\!\!\bigg\lceil\Big(\!\frac{a_{k}}{a}\!\Big)^{\!\frac{3 (2 n-3)}{(n+1)}}\bigg\rceil, 	&	n\!\neq\!3/2,	\\
\!\!\bigg\lceil\frac{g_{k}^{}}{g_{h\phi}}\bigg\rceil,	&	n\!=\!3/2.
\end{dcases}\label{eq:sq_factor_kin}
\end{align}   
In \fig{fig:sqAveragePlotgh510} we have plotted exact numerical results for the phase-space factor $\langle \!\sqrt{1-(2 m_{h_i}/k\omega)^2} \rangle$ with $k\!=\!1$ as a function of the scale factor $a$ for two values of the inflaton-Higgs coupling $g_{h \phi}=10^{-5}$ (left panel) and $10^{-10}$ (rightpanel). 
First of all, let us note that if the mass of the SM Higgs is small as compared to the inflaton mode energy $k \omega$, the phase-space factor $\langle \!\sqrt{1-(2 m_{h_i}/k\omega)^2} \rangle\!=\!1$. 
For $n\!<\! 3/2$ and sufficient strong inflaton-Higgs coupling, e.g., $g_{h \phi}\!=\! 10^{-5}$, the $\langle \!\sqrt{1-(2 m_{h_i}/k\omega)^2} \rangle$ factor initially grows with the scale factor $a$, until the inflaton-induced mass of the Higgs boson becomes negligible and the kinematical suppression is no longer relevant. 
From the left-panel of Fig.~\ref{fig:sqAveragePlotgh510} for $g_{h \phi }\!=\! 10^{-5}$ the number of e-folds for kinematic suppression is $N_{k}\!\sim\!4.5$ and $9$ for $n\!=\! 2/3$ and $n\!=\!1$, respectively. 
As noted above for $n\!=\!3/2$, the phase-space factor is time-independent and is given by $\langle\! \sqrt{1-(2 m_{h_i}/k\omega)^2} \rangle \!\sim\!\lceil \kappa_i \Lambda^4/(4g_{h\phi}\mpl^4)\rceil$. This means that the kinematical suppression is present during the whole period of reheating unless the $g_{h \phi}$ coupling is very weak and the ceiling function approaches 1. For $n\!>\!3/2$, the role of the non-zero mass of the Higgs boson increases with time so that $\langle \sqrt{1- (2 m_{h_i}/k\omega)^2} \rangle$ decreases as the scale factor increases. This means that even if the effects of the non-zero Higgs mass are negligible at the onset of reheating period with increasing time, the kinematical suppression becomes more and more relevant. The numerical results for the phase-space factor agree with our approximate analytic results~\eqref{eq:sq_factor} to very good precision.

\begin{figure}[t!]
\begin{center}
\includegraphics[width=0.5\linewidth]{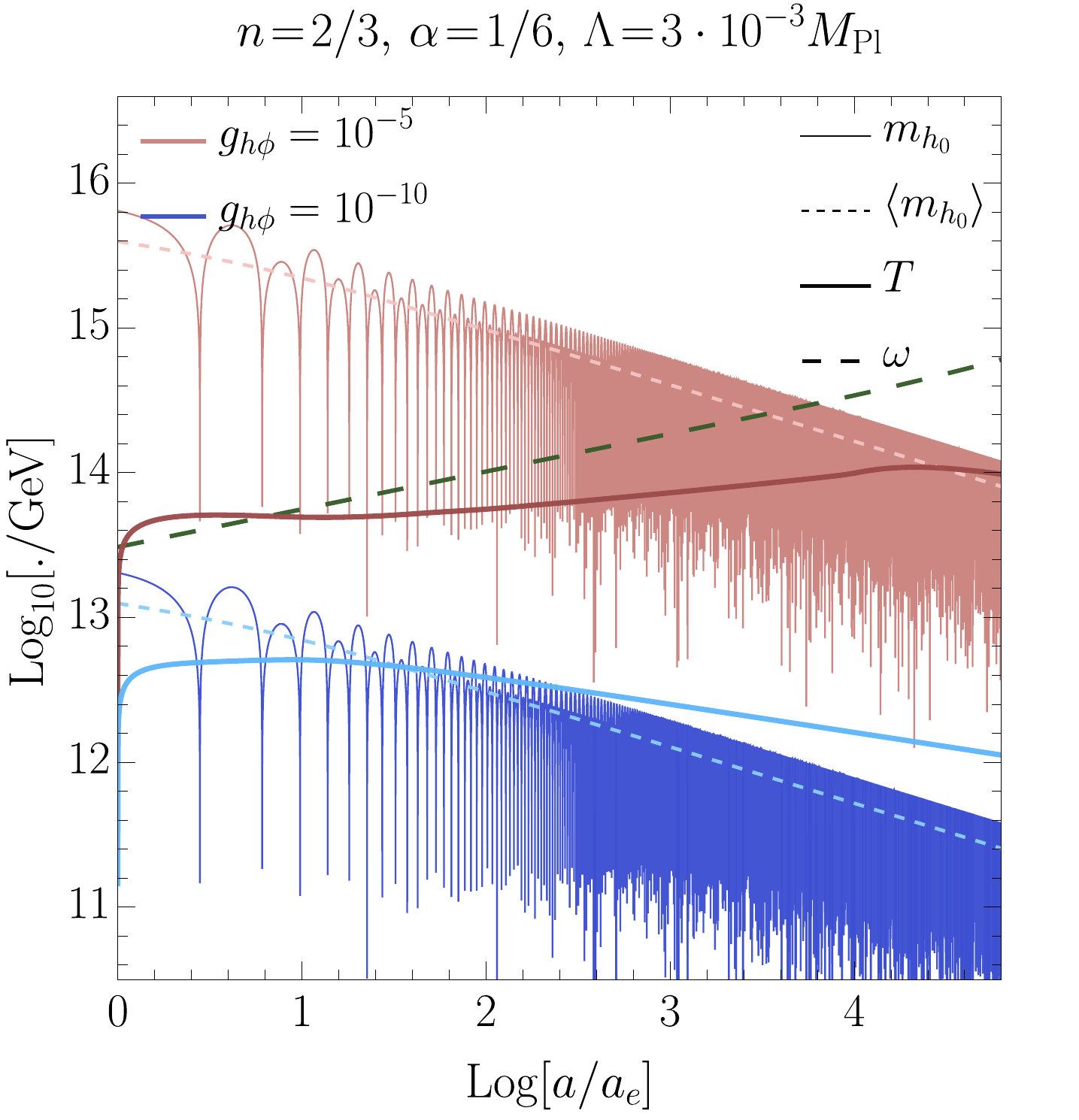}\!\!\!
\includegraphics[width=0.5\linewidth]{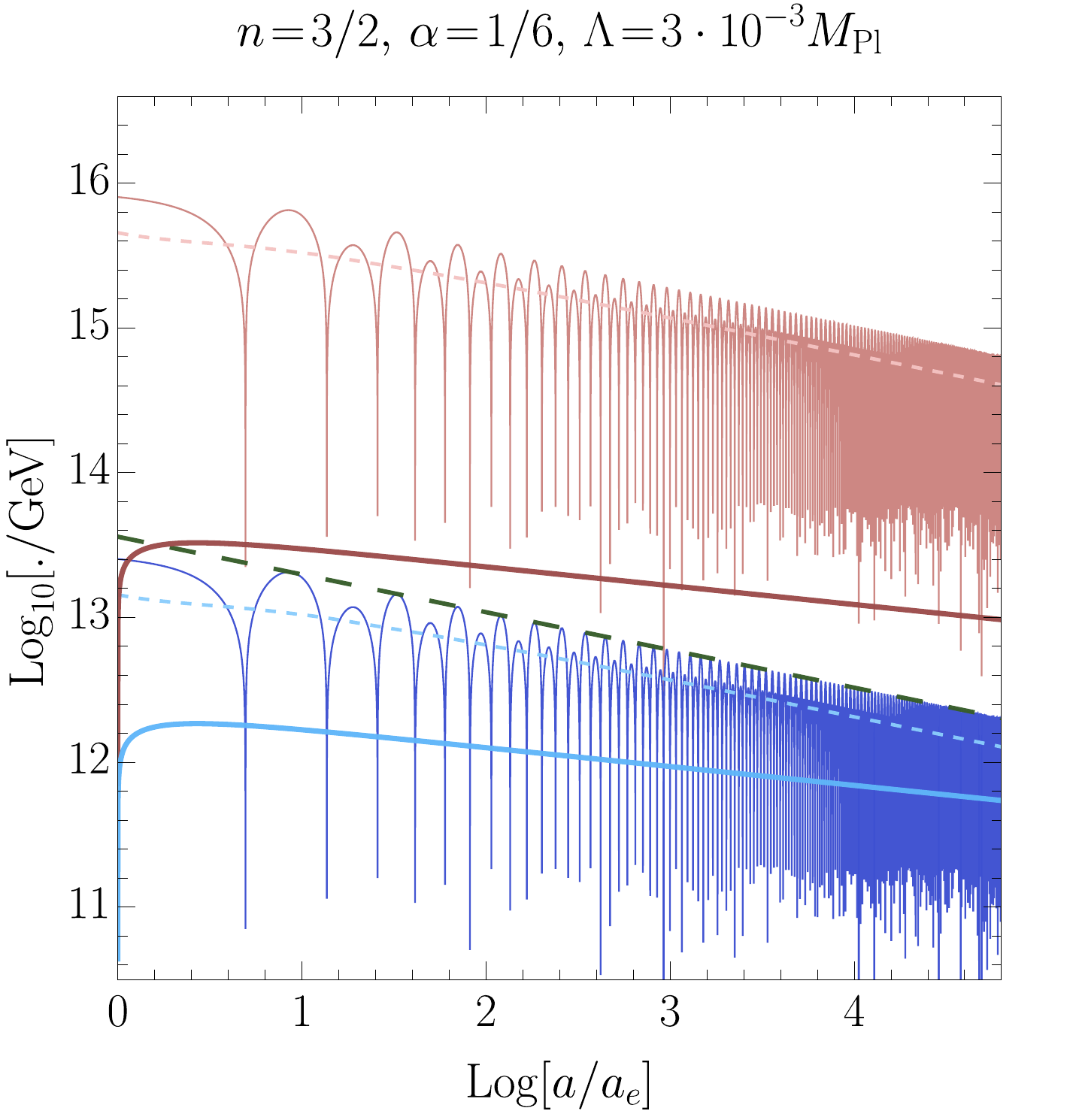}
\caption{The evolution of the Higgs mass $m_{h_0}(a)$ for two values of the inflaton-Higgs coupling: $g_{h \phi} = 10^{-5}$ (blue) and $g_{h \phi} = 10^{-7}$ (red) for $n\!=\!2/3$ and $3/2$ in the left and right-panel, respectively. The colored (blue, red) dashed curves show the evolution of the averaged Higgs mass, i.e., $\langle m_{h_0}(a) \rangle$. The thick solid purple and light blue lines denote $T(a)$ for $g_{h \phi} = 10^{-5}$ and $10^{-7}$, respectively, whereas the loosely dashed green line illustrates the evolution of the inflaton mass $m_\phi$. }
\label{fig:plotmHn23}
\end{center}
\end{figure}

In \fig{fig:plotmHn23} we show the evolution of the mass of the 0-component of the Higgs doublet, $m_{h_0}$~\eqref{eq:Higgs_mass}, its average $\langle m_{h_0} \rangle$~\eqref{eq:Higgs_mass_ave}, and the effective inflaton energy $\omega$~\eqref{eq:period} as a function of the scale factor during the first few e-folds of reheating. In these plots we have fix $n\!=\!2/3$ (left-panel) and $3/2$ (right-panel) for two values of $g_{h\phi}\!=\!10^{-5},10^{-7}$. 
Note that kinematic phase-space suppression discussed above is relevant for regions where $\langle m_{h_0}\rangle\!\gtrsim\!m_\phi$. 
Furthermore, in \fig{fig:plotmHn23} we also have compared $m_{h_0}(a)$ with the evolution of the thermal bath temperature $T$ (see below for details).  When the Higgs mass is larger than the SM bath temperature it behaves as a non-relativistic matter. We note that the $\phi$-dependent mass of the SM Higgs drops below the temperature, roughly corresponding to the $a \simeq a_k$. 

\subsection{Reheating through massless Higgs bosons}
\label{s.massless_reheating}
In this subsection, we consider the reheating dynamics by neglecting the $\phi$-dependent Higgs mass, which we refer to as the {\it massless reheating scenario}. In other words, we neglect the contribution to the Higgs boson mass originating from the cubic interaction term $\phi |{\bm h}|^2$ in the Lagrangian~\eqref{eq:Lint}. This is, of course, merely a reference point introduced to illustrate kinematical mass effects in the realistic case, i.e., with $m_{h_i}\neq 0$. As shown in the following subsection, there are significant differences between the {\it massless scenario} and the case where we consider the inflaton-induced Higgs mass.
Our main goal is to find an approximate analytical solution to the Boltzmann equation for the SM radiation~\eqref{eq:eqR1}, which then allows us to determine the temperature evolution during the reheating phase. In the \textit{massless reheating scenario}, there is no kinematical suppression of the Higgs boson production. Thus, the only source of time-dependence of the inflaton decay rate~\eqref{eq:Gam_phi} is through the effective inflaton mass $m_\phi$~\eqref{eq:m_phi}. In this case, the inflaton decay can be written as
\begin{align}
 \langle\Gamma_{\phi \rightarrow  hh} ^{(0)}(a)\rangle =\widetilde\Gamma_{\phi \rightarrow  hh}^{(0)} \Big( \frac{a}{a_e} \Big)^{\frac{3(n-1)}{n+1}},  \label{eq:Gamma_zero}
\end{align}
with the constant width
\begin{align}
\widetilde\Gamma_{\phi \rightarrow  hh}^{(0)} &\equiv  \frac{g_{h \phi}^2}{\sqrt{2 \pi}} \frac{ n \Gamma \left(\frac{n+1}{2 n}\right)}{\Gamma \left(\frac{1}{2 n}\right)}\frac{\mpl^3}{\Lambda^2} \Big(\frac{\mpl}{\varphi_e}\Big)^{\!n-1}  \sum_{k=1}^\infty k |{\cal P}_k|^2,
\end{align}
where superscript ``$(0)$'' denotes the massless Higgs boson case. 
Note that for a generic value of $n \!\neq\! 1$ the inflaton decay width is a function of time which can lead to non-trivial consequences for reheating dynamics~\cite{Ahmed:2021fvt}. In particular, for $n\!<\!1$, $\Gamma_{\phi \rightarrow  hh}^{(0)}$ decreases with the growth of the scale factor, while for $n\!>\!1$ it increases. Therefore, one can expect that for fixed values of the model parameters ($g_{h \phi},\Lambda$), the reheating is the most efficient for $n\!>\!1$, whereas for $n \!<\!1$, the duration of reheating is prolonged. Moreover, for the quadratic inflaton potential, i.e., $n\!=\!1$ we reproduce the well-known expression for the time-independent decay width, i.e., \mbox{$\Gamma_{\phi \rightarrow  hh}^{(0)} \!=\!(g_{h \phi} \mpl)^2/(8\pi m_\phi)$}, with $m_\phi \!=\! \sqrt{2} \Lambda^2/M$ and ${\cal P}_k\!=\!\delta_{1k}/2$.

Inserting formula \eqref{eq:Gamma_zero} into Eq.~\eqref{eq:eqR1} one obtains the following solution for the SM radiation energy density,
\begin{align}
\rho_{\rm SM}^{(0)} (a) \!=\! 3\mpl^2 H_e\, \widetilde\Gamma_{\phi\rightarrow  hh}^{(0)} \, \frac{n+1}{4n+1}\! \bigg[\!\Big(\frac{a_e}{a} \Big)^{\!\frac{3}{n+1}}-\Big(\frac{a_e}{a} \Big)^{\!4}\bigg], \label{eq:rhoR_sol}
\end{align}
for $a\!\in\! [a_e, a_{\tiny \text{rh}}]$.
Note that for positive $n$, the first term in the square brackets above dominates over the second one during the reheating phase, i.e., $a_e\!<\!a\leq a_{\rm rh}$. 
Consequently, the temperature of the thermal bath~\eqref{eq:rhoRTrel} during reheating evolves as
\begin{align}
 T^{(0)}(a)  &\simeq \bigg[ \frac{90(n+1)\mpl^2 H_e\, \widetilde\Gamma_{\phi\rightarrow  hh}^{(0)}}{ (4n+1)\pi^2\, g_\star(T)}\bigg]^{\!1/4} \Big( \frac{a_e}{a}\Big)^{\frac{3}{4(n+1)}}. 	\label{eq:Tsol0}
\end{align}
For $n\!=\!1$ we obtain standard scaling of temperature w.r.t. the scale factor, i.e., $T \! \propto \! a^{-3/8}$, however for $n\!<\! 1$ ($n\!>\! 1$) the dependence of the temperature on the scale factor is more (less) steeper. The explicit dependence on the scale factor of the SM energy density $\rho_{\rm SM}(a)$ and temperature $T(a)$  for different $n$ is listed in Table~\ref{tab:scaling}.

Another quantity of physical importance is the maximum temperature of the SM bath $T_{{\rm max}}^{(0)}$. In the \textit{massless reheating} scenario, $T_{{\rm max}}^{(0)}$ is typically reached just after the end of inflation, at 
\begin{align}
a_{{\rm max}}^{(0)} = a_e \bigg(\! \frac{4(n+1)}{3}\!\bigg)^{\!\frac{n+1}{4n+1}},	 \label{eq:amax0}
\end{align}
Thus, the maximum temperature in the massless reheating scenario is
\begin{align}
T_{{\rm max}}^{(0)} &\!=\! \bigg[\!\frac{90(n+1)\mpl^2 H_e\, \widetilde\Gamma_{\phi\rightarrow  hh}^{(0)}}{ (4n+1)\pi^2\, g_\star(T)}\!\bigg]^{\!1/4}	\!\!\bigg(\! \frac{3}{4(n+1)}\!\bigg)^{\!\frac{3}{4(4n+1)}}.	\label{eq:Tmax0}
\end{align}

\subsection{Reheating through massive Higgs bosons}
\label{s.massive_reheating}
In this subsection, we discuss the reheating through inflaton decays to Higgs boson pairs, taking into account the effects of the inflaton-induced Higgs boson mass on the reheating dynamics. The important difference of this scenario as compared to the reheating with the massless Higgs, discussed above, is the non-trivial kinematical phase-space suppression of the inflaton decay width, $\langle \Gamma_{\phi \rightarrow hh} \rangle$~\eqref{eq:Gam_phi} through $\gamma_h$ factor~\eqref{eq:gammah}. 
In particular, the main difference is the time-averaged phase-space factor $\langle \!\sqrt{1-(2 m_{h_i}/k\omega)^2}\rangle$ that generates a kinematical suppression due to the non-zero Higgs mass, as discussed above in~\sec{s.inflaton_decay}. The explicit form of the time-dependence of the phase-space factor has been calculated in~\eq{eq:sq_factor} for the four components of the SM Higgs doublet. 
Moreover, as it has been pointed out above, the $\gamma_h$ factor~\eqref{eq:gammah} is time-independent during the reheating phase for $n\!=\!3/2$, whereas for $n\!<\!3/2 \, (n\!>\!3/2)$, $\gamma_h$  increases (decreases) during reheating. For $n\!<\!3/2$ the kinematic suppression gradually disappears as the averaged Higgs mass becomes smaller compared to the inflaton energy, i.e., $\langle m_{h}\rangle\lesssim\omega$. The scale factor at which the kinematic suppression vanishes is denoted by $a_{k}$. Note that, in principle, the kinematic suppression can be present throughout the reheating phase, or it may vanish during this phase, i.e., $a_{k} \!< \! a_{\rm rh}$. 
In the latter case, the inflaton-induced mass of the Higgs field suppresses the radiation production for $a_e\!<\!a\!\leq\! a_{k}$, while for $a_{k}<a\leq a_{\rm rh}$, the kinematic suppression is irrelevant, and the reheating dynamics follows the same evolution as that of the \textit{massless reheating scenario} discussed above. On the other hand, for $n\!>\! 3/2$ it can happen that at the onset of reheating, the kinematical suppression has a negligible impact on the reheating dynamics, but at some point, $a_{k}$, when the averaged Higgs mass approaches the inflaton energy, i.e., $\langle m_{h}\rangle\sim \omega$, it starts to affect the reheating dynamics until the end of reheating phase. 

The averaged inflaton decay width~\eq{eq:Gam_phi} with inflaton-induced Higgs mass is given by,
\begin{align}
 \langle\Gamma_{\phi \rightarrow  hh}(a)\rangle &\simeq\widetilde\Gamma_{\phi \rightarrow  hh} \Big( \frac{a}{a_e} \Big)^{\frac{3(n-1)}{n+1}}  \notag\\
 &\times \sum_{k=1}^\infty k |{\cal P}_k|^2 \begin{dcases}
\!\!\bigg\lceil\!\Big(\!\frac{a_{k}}{a}\!\Big)^{\!\frac{3 (2 n-3)}{(n+1)}}\bigg\rceil, 	&	n\!\neq\!3/2,	\\
\!\!\bigg\lceil\frac{ g_{k}^{}}{g_{h\phi}}\bigg\rceil,	&	n\!=\!3/2,	
\end{dcases}	 \label{eq:Gamma_phi}
\end{align}
where constant width $\widetilde\Gamma_{\phi \rightarrow  hh}$ is defined as
\begin{align}
\widetilde\Gamma_{\phi \rightarrow  hh} &\equiv \frac{g_{h \phi}^2}{\sqrt{ 2\pi}}\frac{ n \Gamma \left(\frac{n+1}{2 n}\right)}{\Gamma \left(\frac{1}{2 n}\right)}\frac{\mpl^3}{\Lambda^2}\Big(\frac{\mpl}{\varphi_e}\Big)^{\!n-1} .
\end{align}
Consequently, in the {\it massive reheating scenario}, the Boltzmann equation for the SM radiation energy density~\eqref{eq:eqR1} can be analytically solved with the inflaton decay rate~\eqref{eq:Gamma_phi}. In the following, we calculate the SM radiation energy density and temperature for three different cases, $n\!\lessgtr\! 3/2$, and $n=3/2$.

In this case when $n\!\lessgtr\!3/2$ for $a\!\gtrless\! a_{k}$ the inflaton decay rate to the SM Higgs boson~\eqref{eq:Gamma_phi} has no phase-space suppression, and it is equal to that of the massless Higgs boson case~\eq{eq:Gamma_zero}. 
Therefore for $n\!\lessgtr\! 3/2$ and $a\!\gtrless\! a_{k}$ the SM radiation energy density is the same as the massless case \eq{eq:rhoR_sol}. 
However, for $a\!\lessgtr\! a_{k}$ with $n\!\lessgtr\!3/2$ we have a phase-space kinematic suppression in the inflaton decay rate~\eqref{eq:Gamma_phi} and, therefore, the SM energy density is modified in comparison to the massless case \eq{eq:rhoR_sol}. Similarly, for the case when $n=3/2$, there is no time-dependent kinematical suppression for the inflaton decay rate; however, there is a constant suppression proportional to $\lceil g_k/g_{h\phi}\rceil$. Therefore, for $n=3/2$, the time dependence of the SM energy density and the temperature would be the same as that of the massless case up to constant kinematical suppression. Hence we can write the SM energy density for different values of $n$ as
\begin{widetext}
\vspace{-10pt}
\begin{align}
\rho_{\rm SM}^{}(a) &\simeq 3\mpl^2 H_e \widetilde\Gamma_{\phi \rightarrow  hh}\begin{dcases} 
 \!\frac{n+1}{2(5-n)} \! \bigg[\!\Big(\frac{a_e}{a} \Big)^{\!\frac{6(n-1)}{n+1}}-\Big(\frac{a_e}{a} \Big)^{\!4}\bigg]\! \sum_{k=1}^\infty k |{\cal P}_k|^2 \Big(\frac{a_{k}}{a_e}\Big)^{\!\frac{3 (2 n-3)}{(n+1)}}, \quad& a\!\lessgtr\! a_k,\, n\!\lessgtr\!3/2,\\
\frac{n+1}{4n+1} \bigg[\!\Big(\frac{a_e}{a} \Big)^{\!\frac{3}{n+1}}-\Big(\frac{a_e}{a} \Big)^{\!4}\bigg] \! \sum_{k=1}^\infty k |{\cal P}_k|^2, & a\!\gtrless\! a_{k},\, n\!\lessgtr\!3/2,\\
\frac{n+1}{4n+1} \bigg[\!\Big(\frac{a_e}{a} \Big)^{\!\frac{3}{n+1}}-\Big(\frac{a_e}{a} \Big)^{\!4}\bigg] \! \sum_{k=1}^\infty k |{\cal P}_k|^2\Big\lceil\frac{g_k}{g_{h\phi}}\Big\rceil, & \qquad n\!=\!3/2,
\end{dcases}	\label{eq:rhoR_sol_mass}
\end{align}
\vspace{-10pt}
\end{widetext}
where the time-dependence through the scale factor is dominated by the first terms in the square brackets above. 
Thus, during the kinematical suppression phase, i.e., $a\!\lessgtr\!a_k$ for $n\!\lessgtr\!3/2$, the SM radiation energy density and the temperature scale as
\begin{align}
\rho_{\rm SM} &\propto a^\frac{6(1-n)}{n+1}\,, 	&T& \propto a^\frac{3(1-n)}{2(n+1)}	,	&a&\!\lessgtr\! a_k,\, n\!\lessgtr\!3/2.
\label{scal_R_T}
\end{align}
Whereas, during the region of parameters where the time-dependent kinematical suppression is absent, we get the scaling as that of the massless Higgs case, i.e.,
\begin{align}
\rho_{\rm SM}\!\sim\! \rho_{\rm SM}^{(0)} &\propto a^\frac{-3}{n+1}\,, 	&T&\!\sim\! T^{(0)} \propto a^\frac{-3}{4(n+1)}, 	& a&\!\gtrless\! a_{k},\, n\!\lesseqgtr\!3/2.
\label{scal_R_T_zero}
\end{align}

It is important to note that the above scaling behavior is very unusual for the SM radiation energy density. Thus, the temperature of the thermal bath during reheating when the kinematical suppression is present. It turns out that in the \textit{massive scenario} for $n\!=\! 2/3$, both $\rho_{\rm SM}$ and $T$ increases with time, whereas for $n\!=\! 1$ they are nearly constant. As we have discussed earlier on, for $n\!=\!3/2$ the inflaton decay rate depends on time only though $m_\phi$ and thus, in both \textit{massless} and \textit{massive} cases, we obtain the same scaling. Finally, for $n\!>\!3/2$, the $\rho_{\rm SM}$ dependence on the scale factor is steeper in the massive reheating scenario in comparison to the massless Higgs case.
The approximate scalings of averaged inflaton decay width, $\langle \Gamma_{\phi \rightarrow  hh} \rangle$, the SM radiation energy density, $\rho_{\rm SM}$, and the temperature, $T$, with the scale factor, $a$, are collected in \tab{tab:scaling} for the \textit{massless} (when the kinematic suppression is irrelevant) and \textit{massive reheating scenarios} (when the kinematic suppression is relevant).  
\begin{table}[t!]
$\begin{array}{|c||c|c|c||c|c|c|}
\hline
n & \langle \Gamma_{\phi \rightarrow  hh}^{(0)} \rangle& \rho_{\rm SM}^{(0)} & T^{(0)}& \langle \Gamma_{\phi \rightarrow  hh} \rangle& \rho_{\rm SM}& T\\
\hline
 2/3 & a^{-3/5} & a^{-9/5}  & a^{-9/20}  & a^{12/5} & a^{6/5} & a^{3/10} \\
 1& a^{0}  & a^{-3/2} & a^{-3/8} &a^{3/2} & a^0 & a^0 \\
 3/2 & a^{3/5} & a^{-6/5}  & a^{-3/10}  & a^{3/5} & a^{-6/5}  & a^{-3/10}  \\
 2 & a^{1}  & a^{-1}   & a^{-1/4} & a^{0} & a^{-2}  &a^{-1/2} \\
 \hline
\end{array}$\vspace{3pt}
\caption{Approximate scaling of the averaged inflaton decay width, $\langle \Gamma_{\phi \rightarrow  hh} \rangle$, the SM radiation energy density, $\rho_{\rm SM}$, and the temperature, $T$, with the scale factor, $a$, in the \textit{massless} (with superscript $(0)$) and \textit{massive reheating scenario} when the kinematic suppression is relevant. }
\label{tab:scaling}
\end{table}

\subsection{Tachyonic resonant production of Higgs bosons}
\label{s.tachyonic_Higgs}
Before we go onto the numerical analysis of the reheating dynamics we would like to make some comments here regarding the possible tachyonic resonant production of Higgs boson during the early phases of reheating. In our model Higgs boson Lagrangian has the form, 
\beq
{\cal L}\supset \sqrt{-g}\Big[g^{\mu \nu}\left(\partial_\nu \bm{h}\right)^{\dagger}\left(\partial_\mu \bm{h}\right)-\mu_h^2|\bm{h}|^2-\lambda_h|\bm{h}|^4\Big]
\eeq
where the inflaton-induced Higgs mass $\mu_h^2\equiv g_{h \phi} M_{\mathrm{Pl}} \phi$ \eqref{eq:H_pot} could be larger than the inflaton effective mass $m_\phi^2$ \eqref{eq:eff_mass} at the onset of reheating phase, i.e. $\phi\!=\!\phi_e\!\sim\!\mpl$ for values of $g_{h\phi}$ satisfying \eqref{ghphi_lower_lim} and \eqref{ghphi_upper_lim}, i.e.
\beq
\sqrt{\lambda_h}\left(\frac{\Lambda^2}{\phi \mpl}\right)\gtrsim \ghp\gtrsim \frac34 \left(\frac{\Lambda^2}{\phi \mpl}\right)^2 \left(\frac{\phi}{\mpl}\right).
\label{ghphi_lim}
\eeq 
for $\alpha=1/6$.
Therefore tachyonic resonant production of Higgs boson can constitute an efficient source of (p)reheating~\cite{Dufaux:2006ee,Abolhasani:2009nb}. This can be easily seen from the equation of motion of the Higgs field in the unitary gauge as, 
\beq
\left(\frac{d^2}{d t^2}-\frac{\nabla^2}{a^2}+3 H \frac{d}{d t}+g_{h \phi} M_{\mathrm{Pl}} \phi\right) h_0=-\lambda_h h_0^3
\eeq
where the non-linear term is present due to Higgs self-interactions. General solution of the above non-linear equation with the expanding background is difficult to handle and require lattice simulations. However, in the following we employ the Hartree approximation to replace the non-linear term by a linear term as $\lambda_h h_0^3\to3\lambda_h \langle h_0^2\rangle h_0$, where $\langle h_0^2\rangle$ is variance of the Higgs field which can be calculated using the linear solution (or iteratively in orders of non-linearity parameter~$\lambda_h$) for the Higgs mode function $h_{0, k}$ with mode momentum $k$ as, 
\beq
\left\langle h_0^2\right\rangle(t)=\frac{1}{2 \pi^2} \int_0^{\infty} dk\, k^2\left|h_{0, k}(t)\right|^2.		\label{eq:Higgs_variance}
\eeq
The Higgs mode function $h_{0, k}$ satisfies the mode equation
\beq
\ddot{h}_{0, k}+3 H \dot{h}_{0, k}+\left(\frac{k^2}{a^2}+g_{h \phi} M_{\mathrm{Pl}} \phi+3 \lambda_h\left\langle h_0^2\right\rangle\right) h_{0, k}=0, 	\label{eq:H_mode}
\eeq
where we have adopted the Hartree approximation.
Applying the field redefinition, $\tilde h_{0, k}\equiv a^{3 / 2} h_{0, k}$, we can recast the above mode equation as
\beq
\ddot{\tilde h}_{0, k}+\omega_k^2 \tilde h_{0, k}=0,
\eeq
where the dispersion relation is
\beq
\omega_k^2 \equiv \frac{k^2}{a^2}+\frac{9}{4} H^2 w+g_{h \phi} \mpl \phi+3 \lambda_h\left\langle h_0^2\right\rangle,
\eeq
with $w=p/\rho$ which can be in the range $w\in[-1,1]$ for our choice of inflaton potential~\eqref{eq:inf_pot}. In the above dispersion relation $\omega_k^2$ second and third terms can be negative and hence source the non-perturbative production of Higgs modes, i.e. preheating dynamics. However, note that second term $H^2 w$ is much smaller compared to the inflaton-induced Higgs mass when $\phi$ is oscillating with an amplitude $\sim\!\mpl$ and $g_{h\phi}$ satisfying \eqref{ghphi_lim}. Therefore tachyonic resonant Higgs production is mainly driven by large negative Higgs mass term $\mu_h^2\sim-g_{h\phi}\mpl |\phi|$. However, the positive large self-interaction contribution term $\lambda_h\left\langle h_0^2\right\rangle$ to mode frequency shuts off the tachyonic production for any $k$ mode if
\beq
3 \lambda_h\left\langle h_0^2\right\rangle\gtrsim |\mu_h^2|.	\label{eq:lamh_bound}
\eeq
This gives a lower-bound on the Higgs quartic coupling for which the tachyonic resonant production is irrelevant for $\mu_h^2<0$. 
We calculate the variance of the Higgs field $\langle h_0^2\rangle$~\eqref{eq:Higgs_variance} for $\mu_h^2<0$ by using the leading linear solution for the mode function $h_{0, k}(t)$, i.e. neglecting $\lambda_h$ term in the mode equation~\eqref{eq:H_mode}, see also~\cite{Dufaux:2006ee}, as
\beq
\langle h_0^2\rangle\sim \frac{|\mu_h|\Lambda^2}{16\pi^2\mpl}\exp\!\bigg(\frac{4|\mu_h|\mpl}{\Lambda^2}\bigg).
\eeq
Hence the lower-bound on the Higgs quartic coupling \eqref{eq:lamh_bound} reads,
\begin{align}
\lambda_h&\gtrsim \frac{16 \pi^2}{3} \frac{\sqrt{g_{h\phi} |\phi|}\mpl^{3/2}}{\Lambda^2}\exp\!\bigg(\!\!-\!\frac{4 \sqrt{g_{h\phi} |\phi|}\mpl^{3/2}}{\Lambda^2}\bigg).
\end{align}
For the values of Higgs quartic coupling employed throughout this work, $\lambda_h\sim0.1$, the lower limit for $g_{h\phi}$ to block the tachyonic resonant production of Higgs boson reads,
\beq
\ghp\gtrsim  \frac{4\Lambda^4}{\mpl^3 |\phi| }.		\label{ghphi_lower_lim_t}
\eeq
Note that the above lower limit for $g_{h\phi}$ is slightly stronger than the one in \eqref{ghphi_lim}. On the other hand we find that for 
\beq
\frac{3}{4} \left(\frac{\Lambda^4}{\mpl^3|\phi|}\right)\lesssim \ghp\lesssim 4\left(\frac{\Lambda^4}{\mpl^3 |\phi| }\right).
\eeq
and $\lambda_h\sim 0.1$ tachyonic resonant production of Higgs modes is possible. A detailed analysis of this possibility is an interesting possibility, however it is beyond the scope of present work.

To summarize, our analysis above (with an order of magnitude approximations) concludes that in the parameter space considered in this work (with $g_{h\phi}$ respecting lower bound \eqref{ghphi_lower_lim_t}) the tachyonic resonant Higgs production is inefficient due to Higgs self-interactions. Therefore in the following numerical analysis we only focus on the perturbative production of Higgs bosons as discussed in the subsections above. 

\subsection{Numerical analysis}
\label{s.numerical_analysis}

In this subsection, we present a numerical analysis of the reheating dynamics due to the inflaton decays to the SM Higgs boson through cubic interaction of the form $g_{h\phi}\mpl\,\phi |{\bm h}|^2$~\eqref{eq:Lint}. In particular, we provide solutions of the first two Boltzmann equations \eqref{eq:beqn_phi} and \eqref{eq:beqn_r} for the inflaton $\rho_\phi(a)$ and SM radiation $\rho_{\rm SM}^{}(a)$ energy densities. The reheating dynamics depend only on the form of the inflaton potential and inflaton-Higgs interaction term\,\footnote{We assume that the inflaton interactions with other SM fields and DM are much smaller than the leading inflaton-Higgs coupling through the dim-3 operator.}. For the numerical analysis, we consider the $\alpha$-attractor T-model of inflaton with potential \eqref{eq:inf_pot}, where we have fixed $\alpha=1/6$, such that $M=\mpl$, and the scale of inflation $\Lambda=3\times10^{-3}\mpl$. Whereas, the benchmark values for $n$ have been chosen to be $2/3,1,3/2,2$ which correspond to the equation of state during the reheating phase $\bar w=-1/5,0,1/5,1/3$, respectively. Furthermore, we consider two values for the inflaton-Higgs interaction $g_{h\phi}\!=\!10^{-5}$ and $10^{-10}$, which are close to its upper and lower bounds, see \eq{ghp_num_lim}.

\begin{figure*}[t!]
\begin{center}
\includegraphics[width=0.25\linewidth]{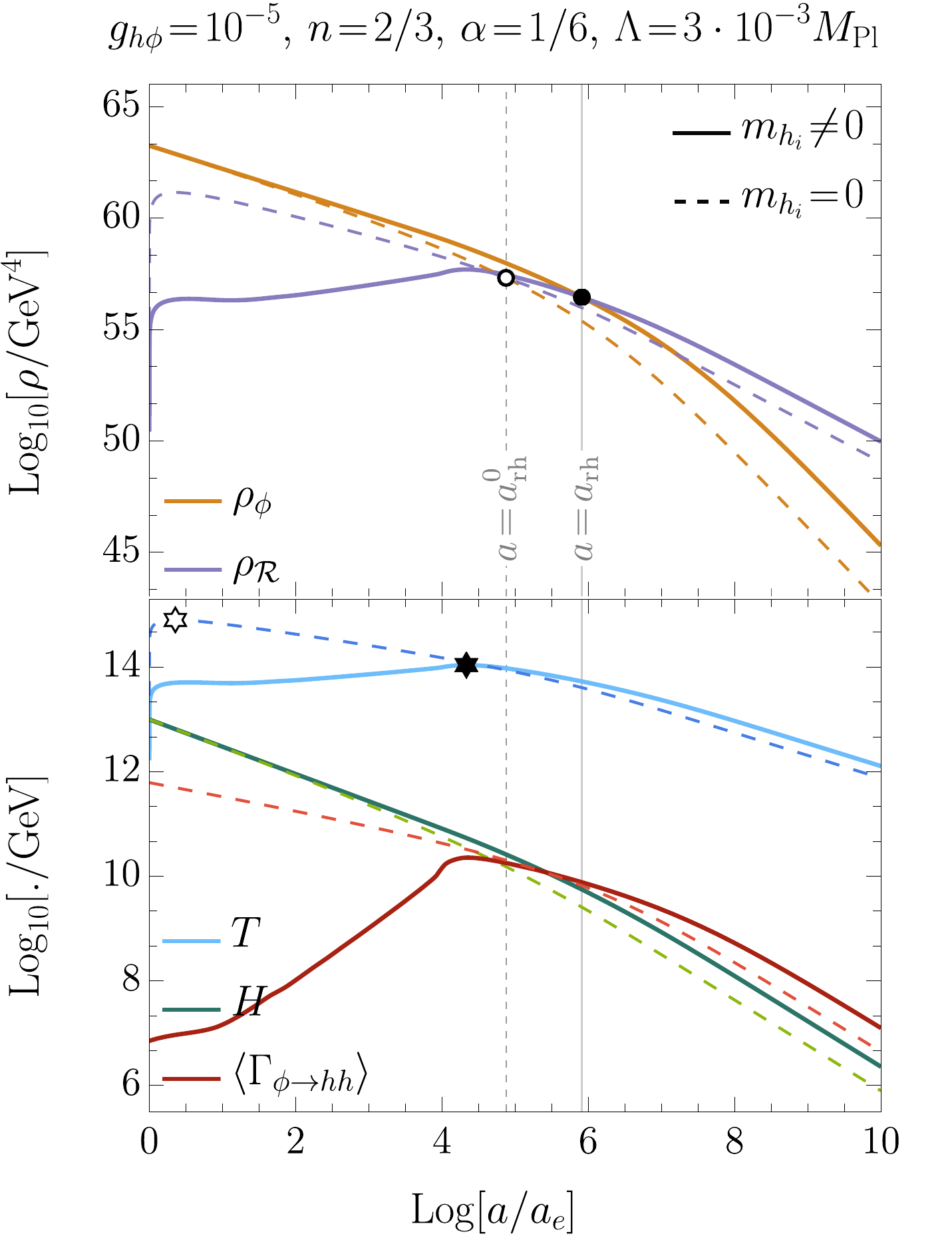}\!\!\!
\includegraphics[width=0.25\linewidth]{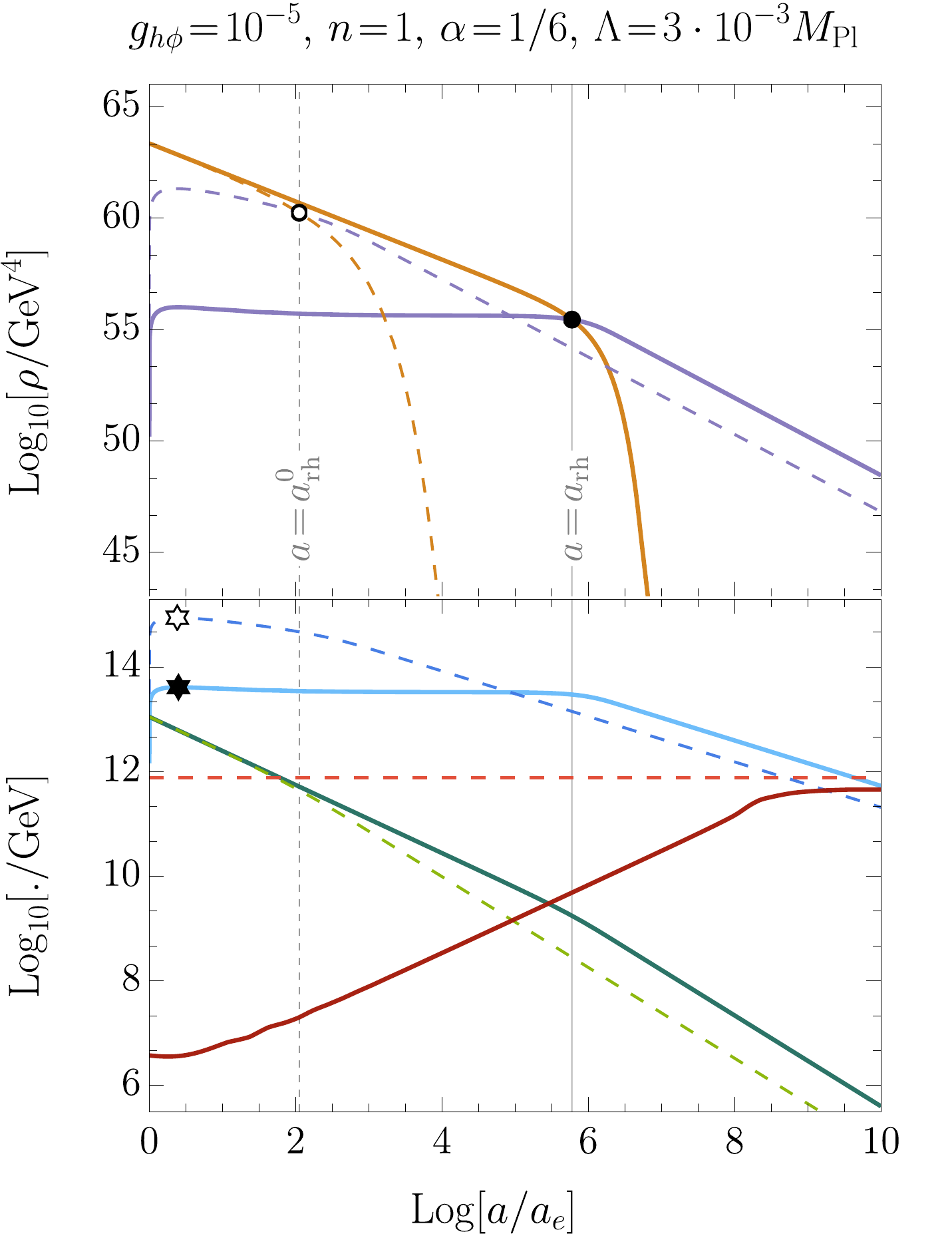}\!\!\!
\includegraphics[width=0.25\linewidth]{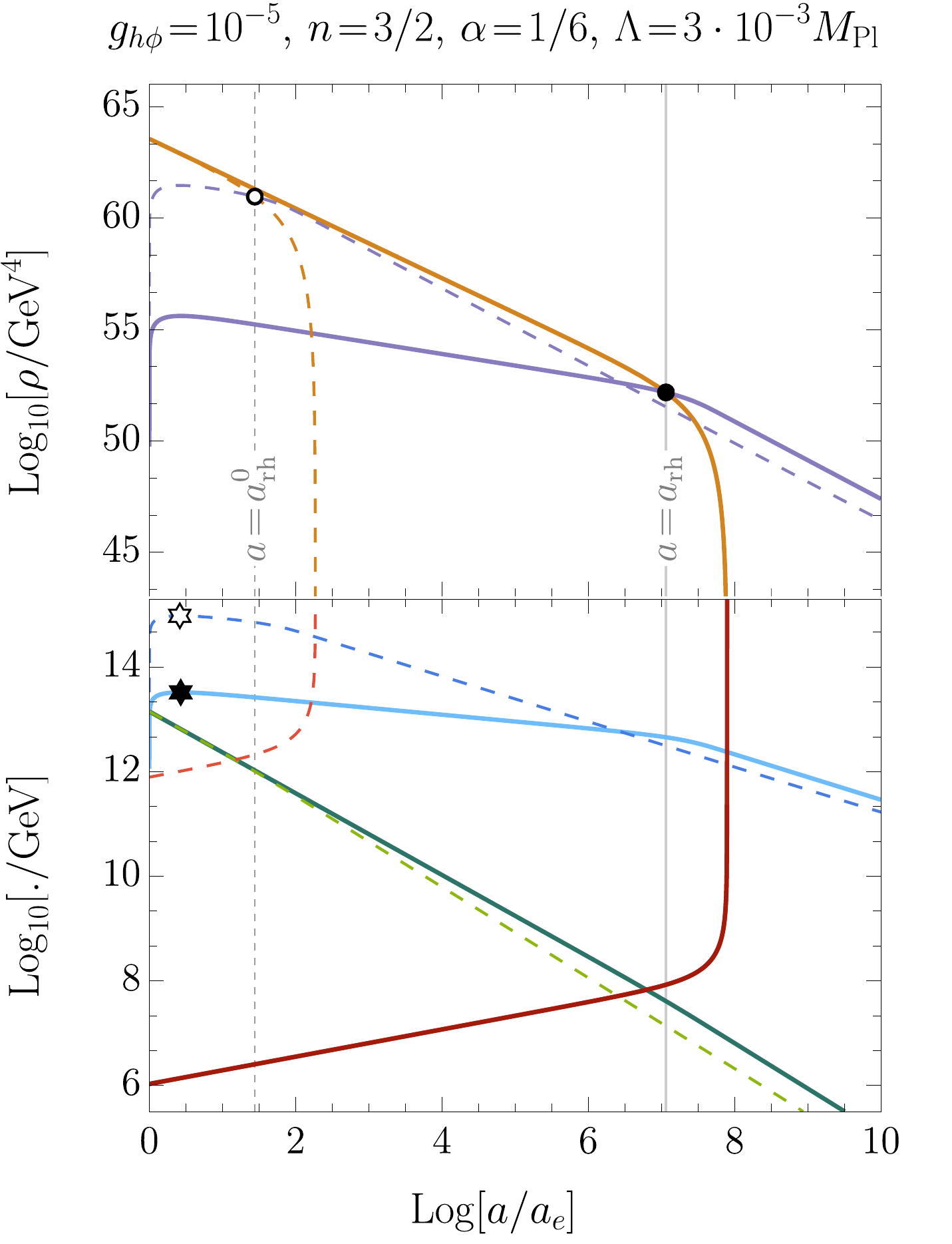}\!\!\!
\includegraphics[width=0.25\linewidth]{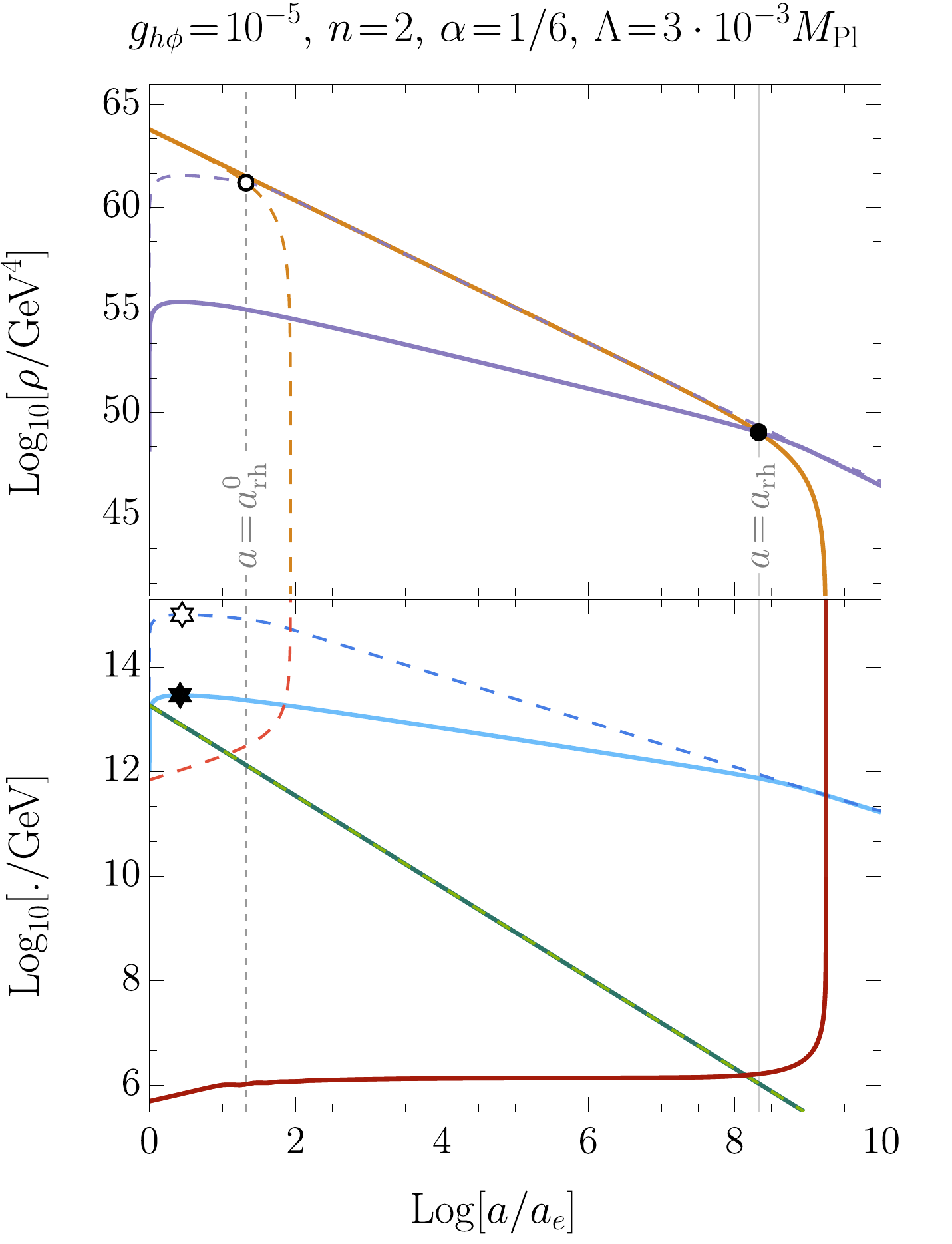}
\caption{Solutions of the first two Boltzmann equations \eqref{eq:beqn_phi} and \eqref{eq:beqn_r} for the inflaton $\rho_\phi$ and radiation $\rho_{\rm SM}$ energy densities are plotted in the upper panel as a function of the scale factor $a$. The empty and filled dots indicate the end of reheating if produced Higgs boson pairs are massless and massive, respectively. The lower panel shows thermal bath temperature $T$ (blue), the Hubble parameter $H$ (green) and the averaged inflaton width $\langle \Gamma_{\phi \rightarrow  hh} \rangle$ (red) for the massless and massive cases.The empty and filled stars show where the maximal temperature during reheating has been obtained for the massless and massive cases, respectively. Adopted parameters are specified above.}
\label{fig:rhoghphi5}
\end{center}
\end{figure*}
\begin{figure*}[t!]
\begin{center}
\includegraphics[width=0.25\linewidth]{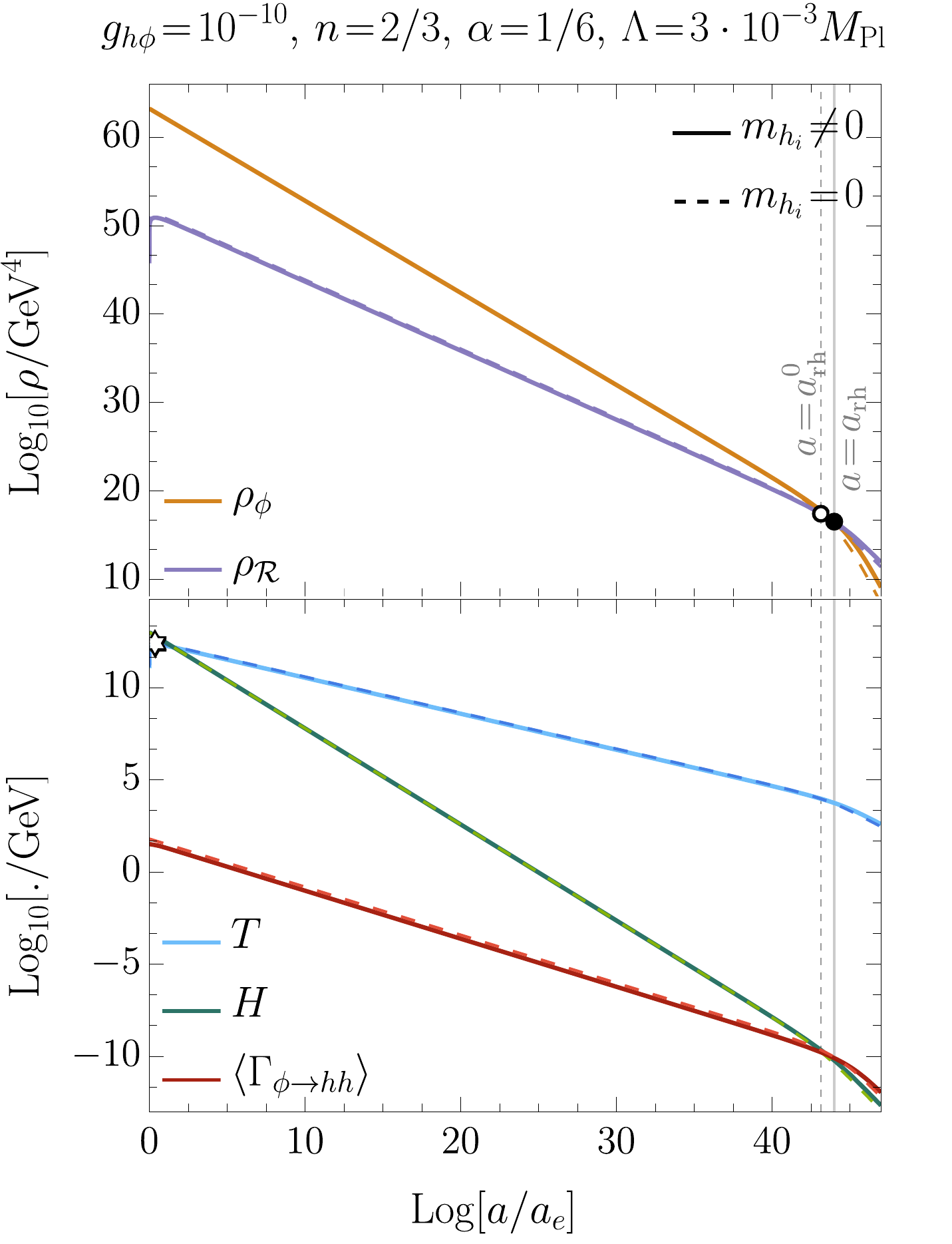}\!\!\!
\includegraphics[width=0.25\linewidth]{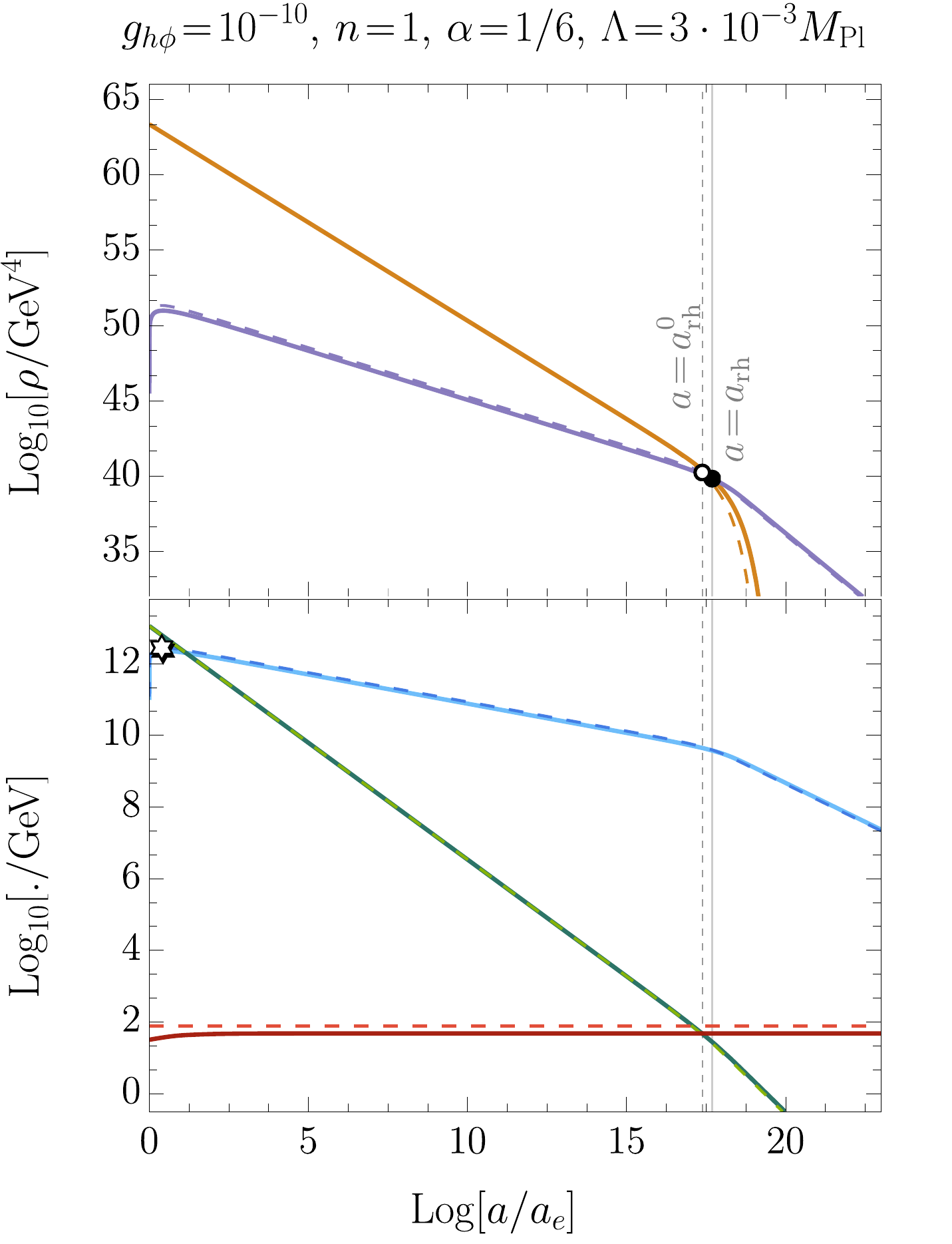}\!\!\!
\includegraphics[width=0.25\linewidth]{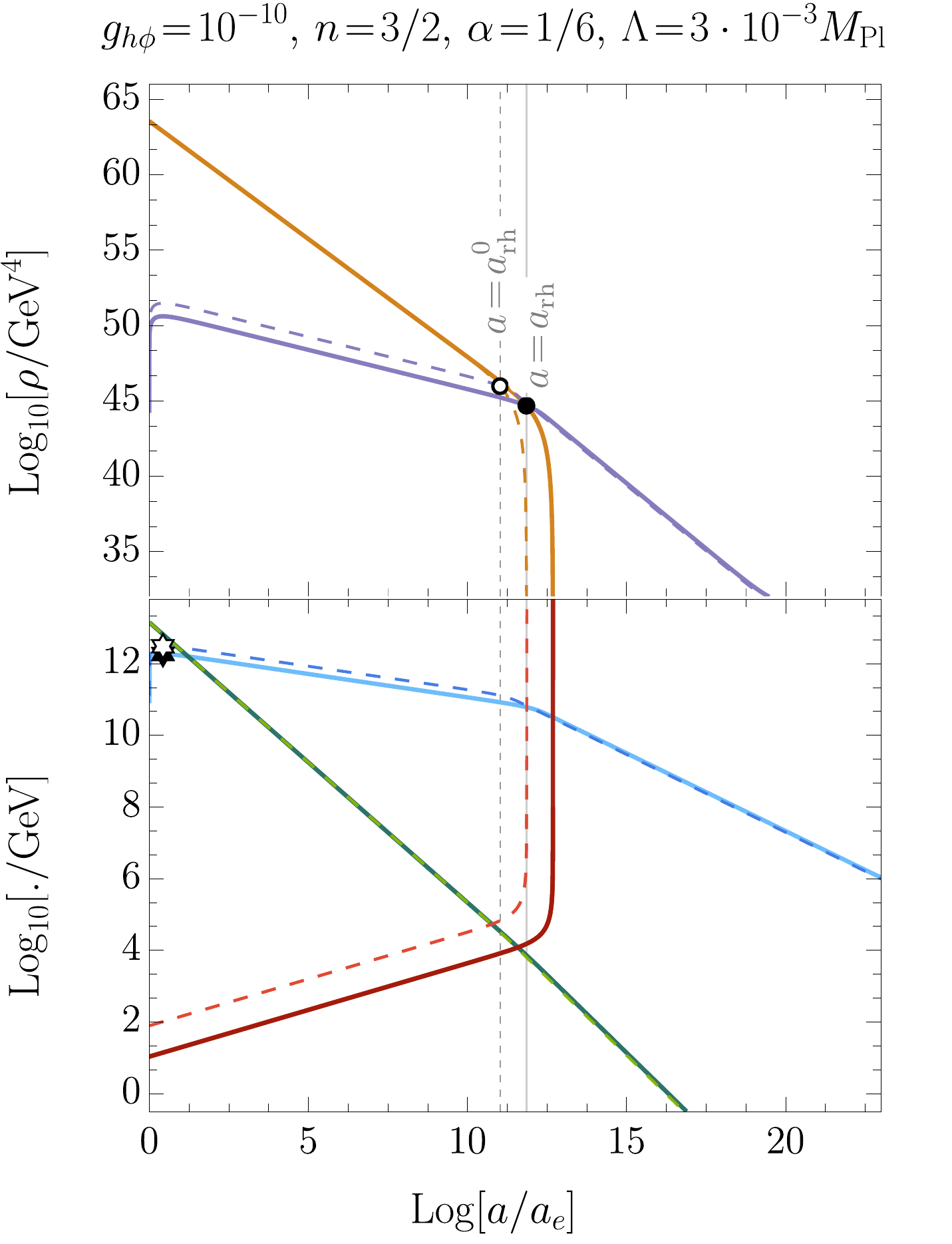}\!\!\!
\includegraphics[width=0.25\linewidth]{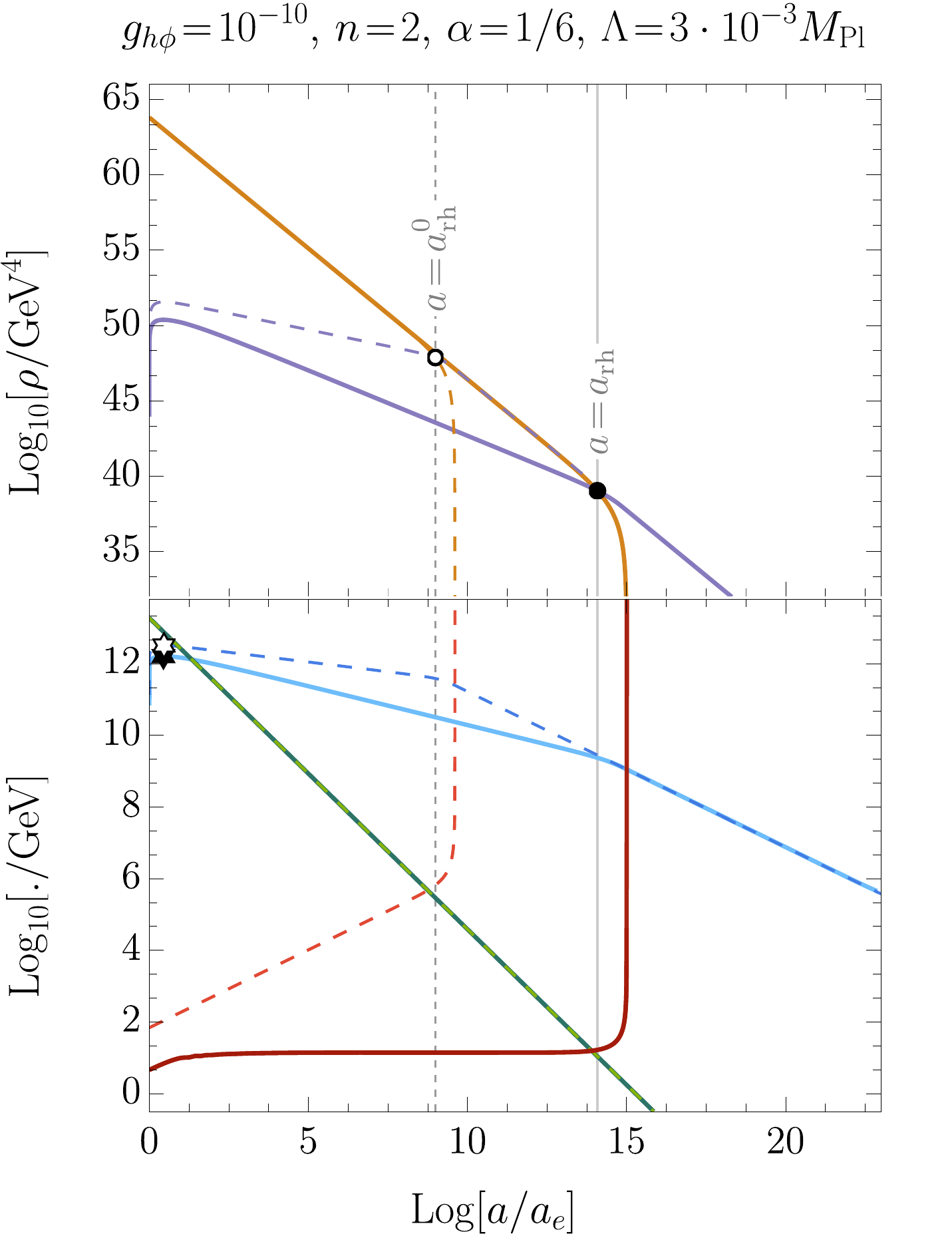}
\caption{Same as in \fig{fig:rhoghphi5} except for $\ghp=10^{-10}$. }
\label{fig:rhoghphi10}
\end{center}
\end{figure*}
In the upper panel of \fig{fig:rhoghphi5} and \fig{fig:rhoghphi10} we have plotted the numerical solutions for the inflaton (orange) and radiation (purple) energy densities for fixed $g_{h\phi}=10^{-5}$ and $10^{-10}$ in \fig{fig:rhoghphi5} and \fig{fig:rhoghphi10}, respectively. The lower panel of each figure presents the evolution of the thermal bath temperature (blue), the Hubble rate (green), and the averaged inflaton decay width (red) as functions of the scale factor. The solid and dashed curves correspond to the {\it massive} and {\it massless reheating scenario}, respectively. 
The empty (filled) dots indicate the end of reheating phase, i.e., the moment at which the inflaton energy density becomes equal to the SM radiation energy density in the \textit{massless} (\textit{massive}) scenario.  Note that the $\rho_\phi(a_\text{rh}) \!=\!\rho_{\rm SM}^{}(a_\text{rh})$ equality roughly coincides with $\langle \Gamma_{\phi \rightarrow  hh} \rangle\!\sim\! H$ in both considered scenarios.

First of all, let us recall that in the both considered reheating scenarios the inflaton energy density scales as $a^{-6n/(n+1)}$~\eqref{eq:rhophia} for $a\!<\!a_\text{rh}$, while after the end of reheating $\rho_\phi$ sharply drops as it is seen in upper panels of \fig{fig:rhoghphi5} and \fig{fig:rhoghphi10}. However, the evolution of the SM radiation energy density during the reheating period is drastically different in the \textit{massive reheating scenario} as compared to the case where the inflaton-induced Higgs mass has been neglected. This non-trivial behavior not only changes the SM radiation energy density but also affects the duration of reheating in the two cases. Note that when inflaton-induced Higgs mass effects are taken into account, the SM radiation energy density is suppressed compared to the \textit{massless case}, which results in elongation of the reheating period for the \textit{massive reheating case}. Such effects are more significant for relatively large inflaton--Higgs coupling $g_{h\phi}$, which is manifestly shown in \fig{fig:rhoghphi5} and \fig{fig:rhoghphi10}.
These numerical results for the SM radiation energy density agree well with our analytic results \eq{eq:rhoR_sol} for \textit{massless} and \eq{eq:rhoR_sol_mass} for \textit{massive reheating cases}. Similarly, the SM bath temperature $T$, shown in the lower panel of these plots, also agrees with our analytic results obtained above. 
Finally, we should also emphasize that after the end of reheating, i.e., $a>a_{\rm rh}$, the energy density of the Universe is mainly dominated by the SM radiation energy density which scales as $a^{-4}$ in both considered scenarios.

As it is shown in the lower panel of these plots, the Hubble rate $H(a)$ scales as $a^{-3n/(n+1)}$ during the reheating phase and as $a^{-2}$ afterward, independent of the Higgs mass effects. However, as discussed above, the most significant effect of the non-trivial Higgs mass during reheating is on the inflaton decay rate to the Higgs boson due to phase space suppression. In the scenario with the massless Higgs boson, the inflaton decay rate scales as $a^{-3(n-1)/(n+1)}$ \eqref{eq:Gamma_zero}, which is constant for $n=1$. Whereas, in the presence of the kinematical suppression the evolution of the inflaton decay rate presented in \fig{fig:rhoghphi5} and \fig{fig:rhoghphi10} agrees well with analytic result \eqref{eq:Gamma_phi}. Note, however, that the above analytical results are only valid during the reheating period ($a_e\leq a\leq a_{\rm rh}$) when the inflaton energy density~\eqref{eq:rhophia} is analytically calculated. Since after the end of reheating $a>a_{\rm rh}$ the inflaton energy sharply vanishes, therefore the inflaton decay rate scale accordingly. 
In particular, as it has been discussed below Eq.~\eqref{eq:width_def}, the inflaton width $\langle \Gamma_{\phi \rightarrow  hh} \rangle $ adopted in the RHS of \eq{eq:beqn_phi} depends on $\rho_\phi$ through this inflaton mass \eqref{eq:eff_mass}. Thus, for $n\!>\!1$, according to \eq{eq:Gam_phi} the vanishing inflaton energy density $\rho_\phi$ after the end of reheating implies divergent averaged ``width'', which is indeed seen in \fig{fig:rhoghphi5} and \fig{fig:rhoghphi10} for $n=3/2, 2$.
The Higgs mass effects are clearly important for large inflaton-Higgs coupling e.g., $g_{h\phi}=10^{-5}$, however as discussed in \sec{s.inflaton_decay} even for relatively small coupling e.g., $g_{h\phi}=10^{-10}$ for $n>3/2$ the kinematical suppression can be significant. This effect is can be seen in \fig{fig:rhoghphi10} with $n=2$.

It should also be pointed out that when the kinematic suppression of the inflaton decay width disappears at $a\!\gtrless\! a_{k}$ for $n\!\lessgtr\!3/2$, the averaged inflaton decay rate $\langle \Gamma_{\phi \rightarrow  hh}\rangle$ for massless and massive cases nearly merge. As mentioned above the convergence of the two cases is not perfect as for the massless reheating case the inflaton ``decays'' to four massless Higgs components. On the other hand, for the massive case (even if the mass contribution is tiny), during one half of the oscillation period (${\cal P}\!<\!0$ the electroweak symmetry broken phase), the decay final state is just one real Higgs particle $h_0$, while during the other half  (${\cal P}\!>\!0$ unbroken phase) the final state is made of four massive Higgs components $h_i$ degenerate in mass. Therefore, eventually, the averaged inflaton decay rate $\langle \Gamma_{\phi\to hh}\rangle$ for the massive case, in the limit when mass effects are negligible, is a factor 5/8 smaller than that of the massless case, see \fig{fig:rhoghphi5} and \fig{fig:rhoghphi10}.

\begin{figure}[t!]
\begin{center}
\includegraphics[width=0.5\linewidth]{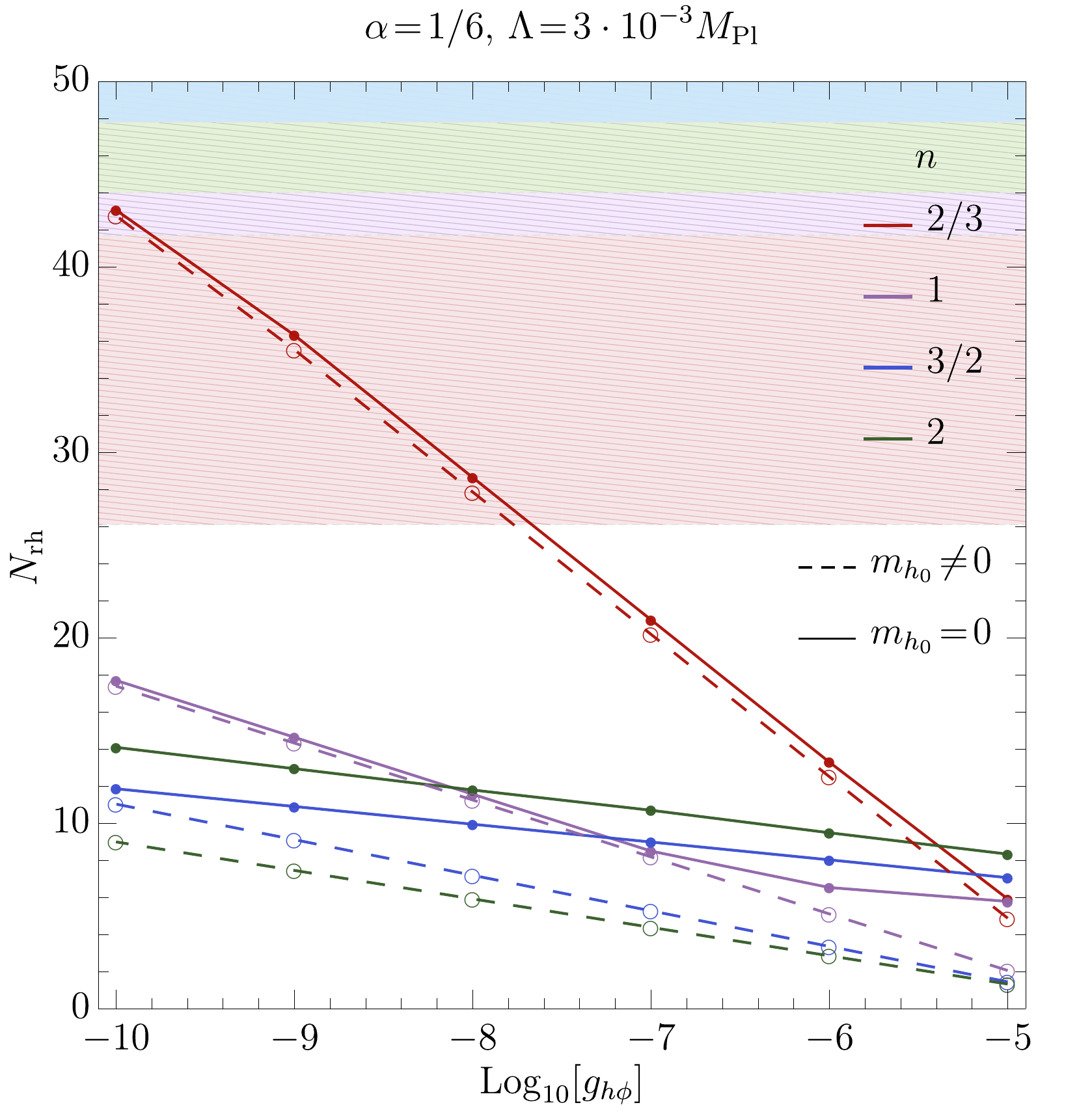}\!\!\!\!\!
\includegraphics[width=0.51\linewidth]{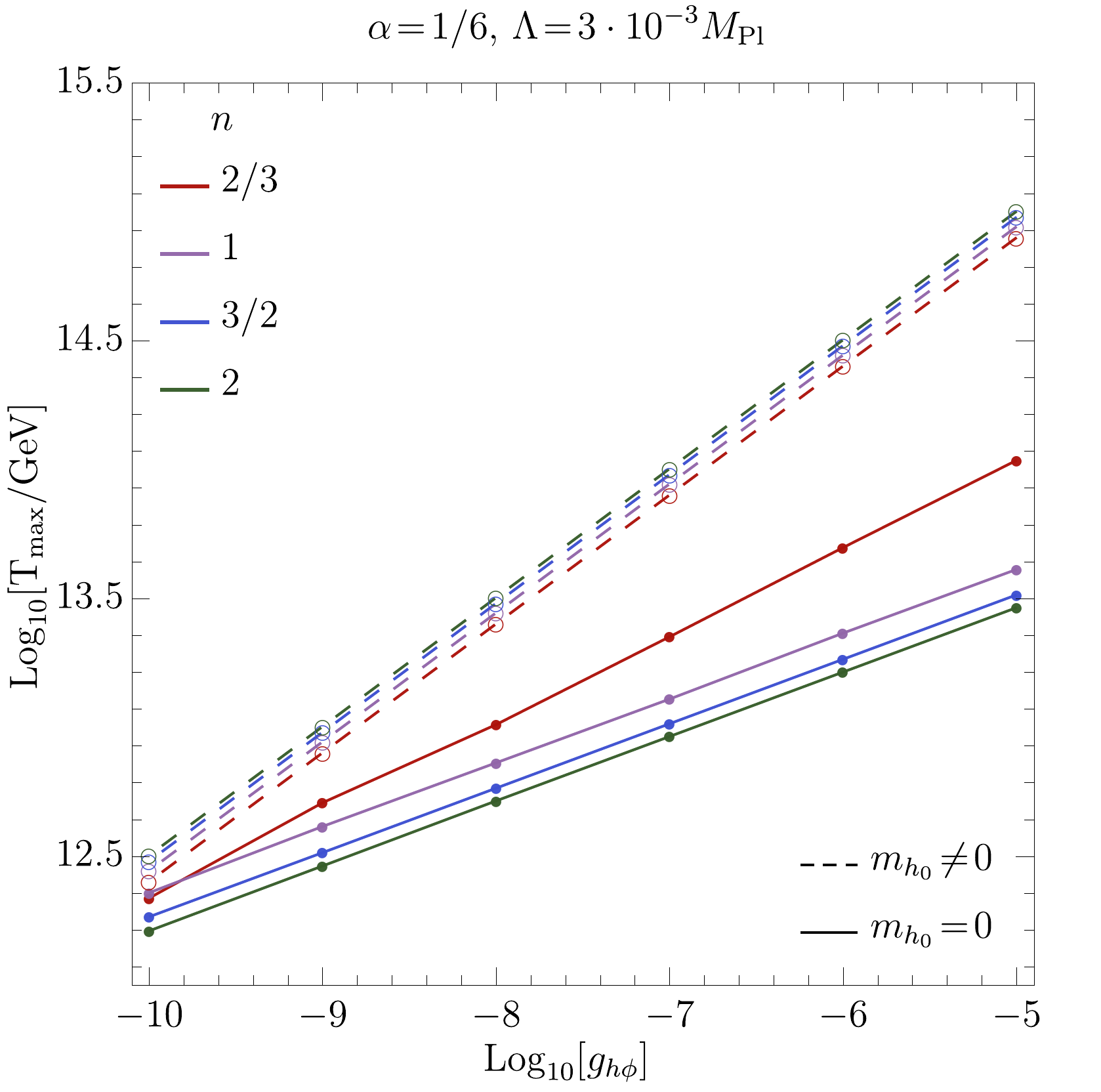}
\caption{Left panel: Relation between reheating numbers of e-folds $N_{\rm rh}$ and the value of the inflaton-Higgs coupling $g_{h \phi}$. Right panel: Relation between the maximal temperature, $T_{\rm max}$, obtained during reheating and the value of the inflaton-Higgs coupling $g_{h \phi}$. Filled (empty) dotes present pairs of $(g_{h \phi}, N_{\rm rh})$ obtained assuming the massive (massless) reheating scenario, whereas solid (dashed) lines are corresponding linear interpolations to the data. In the upper panel, colored regions show constraints on $N_{\rm rh}$ for different values of $n$, see Table \eqref{tab:Nrh}.}
\label{fig:NRH_ghphi}
\end{center}
\end{figure}
We have seen that the consequences of the non-zero SM Higgs mass during the reheating phase can be dramatic. Not only does it change the evolution of the radiation energy density and the temperature, but it also affects the duration of the reheating period, cf. the left panel of \fig{fig:NRH_ghphi}, where we have compared the number of e-folds for reheating phase, $N_{\rm rh}$, as a function of the inflaton-Higgs coupling, $g_{h \phi}$, in both the massless and massive cases. 
In the \textit{massive case}, reheating is not only delayed but also strongly suppressed. In the scenario with the massive Higgs field, the production of the SM radiation is always less efficient than in the \textit{massless case}. Therefore,  the thermal bath temperature, measured by $\rho_{\rm SM}$, in the \textit{massive scenario} is reduced compared to the \textit{massless case} during the reheating phase, shown in the right-panel of \fig{fig:NRH_ghphi} as a function of $g_{h\phi}$. It is worth noting that in the standard scenario with massless Higgs, the maximal temperature, $T_{\rm max}$, is typically attained shortly after the end of inflation. However, in the \textit{massless case} and for $n\!=\!2/3$ it is reached after a few e-folds of reheating when kinematic suppression becomes irrelevant and $\langle \Gamma_{\phi \rightarrow  hh} \rangle$ approaches $\langle \Gamma_{\phi \rightarrow  hh}^{(0)} \rangle$. 
The relation between the maximal temperature of the thermal bath and the value of the inflaton-Higgs coupling in both reheating scenarios is shown in the lower panel in \fig{fig:NRH_ghphi}. The largest discrepancy between the \textit{massless} and \textit{massive} cases is again observed for the $n\!=\!2$ curve. 
As the strength of the inflaton-Higgs interactions decreases, the role of the non-zero mass of the Higgs boson becomes less and less relevant for the dynamics of reheating, and as we can see in the left panel of Fig.~\ref{fig:NRH_ghphi}, all solid lines slowly approach the corresponding dashed lines. For the small value of the inflaton-Higgs coupling, e.g., $g_{h \phi}=10^{-10}$, the distinction between solid and the corresponding dashed lines is relatively small. However, the $n\!=\!2$ line is an exception to this rule, and even for $g_{h \phi} = 10^{-10}$ we observe a non-negligible deviation from the massless case. This is caused by the fact that for the $n\!=\!2$ case, the slope of $\rho_{\rm SM}$ as a function of $a$ is steeper in the \textit{massive scenario}, and thus it takes more e-folds of reheating to drop $\rho_\phi$ below $\rho_{\rm SM}$.

\section[DM-effects]{DM production during reheating}
\label{s.DM-effects}
In this section, we study implications of Higgs-boson-induced reheating discussed in the above section for the production of vector DM. 
As outlined in \sec{s.model} and specified in ~\eqref{eq:Lint}, the vector DM $X_\mu$ interacts directly with the inflaton field and the SM Higgs bosons through dim-5 ($\mathcal{C}_X^\phi m_X^2/\mpl \phi X_\mu X^\mu$) and dim-6 ($\mathcal{C}_{X}^{{\bm h}} m_X^2/\mpl^2 |{\bm h}|^2 X_\mu X^\mu$) operators, respectively. Moreover, the vector DM couples indirectly to the energy-momentum tensor of the inflaton and the SM through $s$-channel graviton exchange.
Furthermore, in the \textit{massive reheating} scenario, the Higgs portal operator $\mathcal{C}_{X}^{{\bm h}} m_X^2/\mpl^2 |{\bm h}|^2 X_\mu X^\mu$ expanded around $v_h$ provides a term $\propto v_h h_0 X_\mu X^\mu$, which accounts not only for the Higgs boson decays to DM pairs, but it also generates indirect DM-SM interactions mediated by the Higgs exchange. As already noted in \sec{s.model}, such processes can only occur in the one-half of the inflaton oscillations period, when the symmetry is broken and $v_h$ is non-zero.
To sum up, ignoring the inflaton-induced Higgs mass, i.e., in the \textit{massless reheating} scenario, the dark sector can be populated either due to gravitational interactions with the inflaton and SM particles or a result of direct higher-dimensional interactions with the $\phi$ and Higgs field. On the other hand, in the \textit{massive} scenario, there exist two additional DM production channels, i.e., direct decays of the SM Higgs boson $h_0$ and freeze-in from SM particles via s-channel $h_0$  exchange.

The DM dynamics is governed by the Boltzmann equation~\eqref{eq:beqn_x}, which can be recast in the following form in terms of the comoving number density~$N_X\equiv n_X a^3$:
\begin{align}
\frac{ d N_X}{ d a} &=\frac{a^2}{H} \Big[ \mathcal{D}_\phi+ \mathcal{S}_\phi + \mathcal{S}_{\rm{SM}} + \mathcal{D}_{h_0} \Big]. \label{eq:Xeq}
\end{align}
The first term on the r.h.s. of the above equation takes into account DM production through inflaton decays. The source term $\mathcal{S}_\phi$ describes the gain of DM number density due to gravitational interactions with the inflaton. In the scattering term, $\mathcal{S}_{\rm{SM}}$~ \eqref{eq:collision_decay}, the annihilation cross-section $\sigma_{XX \rightarrow {\rm  SM \, SM}}^{}$ includes contributions from the graviton exchange and the Higgs portal operator. Finally, the decay term $\mathcal{D}_{h_0}$, defined in \eq{eq:collision_decay}, accounts for Higgs boson decays into pairs of DM vectors.

All the terms on the r.h.s of \eqref{eq:Xeq} can be equally important for DM production. However, their origin is very different, and thus it is convenient to discuss each channel separately. In particular, the gravitational DM production, through the graviton exchange with the inflaton background field and the SM radiation bath, can be treated as an irreducible production mechanism that is always present, regardless of other DM interactions. Let us also emphasize that these two sources of gravitational production of DM are very different despite their deceptive resemblance. In this work, we treat $\phi$ as a homogenous, classical field, not a quantum particle. Within this framework, DM vectors are produced from the vacuum in the oscillating background of the inflaton field. To put it another way, the energy density of the $\phi$ field is transferred to the dark sector indirectly through gravity which couples to  $T_\phi ^{\mu \nu}$ and $T_X ^{\mu \nu}$. Note that in this case, DM particles are produced non-thermally since the inflaton field is not in thermal equilibrium. Thus, this kind of DM production mechanism is insensitive to the thermal history of the Universe and depends mainly on the initial value of the inflaton energy density and its evolution during the reheating period. Contrarily, the gravitational production from the SM particles realizes a standard freeze-in DM scenario, in which SM quantum states couple to $s$-channel virtual graviton which subsequently couples to pairs of DM species. In this case, SM species are assumed to be in thermal equilibrium, which implies a non-trivial dependence of the DM relic abundance on the evolution of the SM temperature $T$.

In what follows, we assume that DM production does not have a significant impact on the evolution of the first two Boltzmann equations, i.e., we adopt two assumptions: (i) the inflaton decays mainly to the SM, and (ii) DM particles are not in equilibrium with the SM thermal bath. Therefore, DM production does not substantially affect the evolution of the SM bath temperature and the Hubble rate. This requires a small DM branching ratio, i.e., $\langle \Gamma_{\phi \rightarrow XX} \rangle / \langle \Gamma_\phi \rangle  \ll 1$, and sufficiently weak interactions between the SM and DM. It is worth emphasizing that such suppression emerges naturally in our model with vector DM since the inflaton--DM as well as SM--DM interactions are sourced by the higher-dimensional terms~\eqref{eq:Lint}. Thus the above two assumptions seem to be well justified and robust within our model. Therefore, in the following, we solve the Boltzmann equation~\eqref{eq:Xeq} for DM evolution with the inflaton and SM energy densities as well as other related parameters obtained in the previous section.
 
\subsection{Inflaton induced gravitational DM production}
Let us start our discussion with the most generic production mechanism, namely the gravitational DM production in the background of the oscillating inflaton field. This scenario does not require any additional (besides gravitational) coupling of $X$ particles to the inflaton and thus can be treated as a kind of irreducible ``background'' to any DM production mechanism, which is present regardless of other DM interactions. In particular, the gravitational DM production from the $\phi$ field provides an unavoidable contribution to the relic DM abundance, which should be taken into account in any DM scenario. Recently, it was shown \cite{Clery:2022wib} that in the case of scalar and fermionic DM, pure gravitational production from the inflaton field can account for the observed abundance of DM particles, 
\begin{align}
\Omega_{X}^{\rm obs} h^2=0.1198 \pm 0.0012\,, \label{eq:OmegaX_obs}
\end{align} 
measured by the Planck Collaboration \citep{Aghanim:2018eyx}. In this work, we focus on spin-1 DM field, for which the source term $\mathcal{S}_\phi$ takes the following form,
\begin{align}
\mathcal{S}_\phi &\!=\! \frac{\rho_\phi^2}{8\pi\mpl^4}  \sum_{k=1}^\infty |\mathcal{P}_k^{2n}|^2 \bigg(\!1 -\frac{4m_X^2}{(k \omega)^2}+\frac{12m_X^4}{(k \omega)^4}\!\bigg) \sqrt{1\!-\! \frac{4 m_X^2}{(k \omega)^2}}, \label{eq:sphiX}
\end{align}
where we have expressed $V(\phi)$ using the envelope $\varphi$ and the quasi-periodic function $\mathcal{P}$ as
$V(\phi)\simeq \Lambda^4 (\varphi/M)^{2n} \mathcal{P}^{2n}(t)$ and decomposed $\mathcal{P}^{2n}(t)$ into the Fourier modes as
\begin{align}
\mathcal{P}^{2n}(t) = \sum_{k} \mathcal{P}_k^{2n} \,e^{-i k \omega t},
\end{align}
with
\begin{align}
\mathcal{P}_k^{2n}  = \frac{1}{\mathcal{T}(t_0)} \int_{t_0}^{t_0 + \mathcal{T}(t_0)} d t \mathcal{P}^{2n}(t)\, e^{i k \omega t},
\end{align}
and the inflaton mode energy/frequency $\omega=2\pi/{\cal T}$. For $n\neq 1$, the inflaton mode energy/frequency $\omega$ is time-dependant, where this dependency is the same as the inflaton effective mass time-dependance~\eqref{eq:m_phi}. Therefore, it is instructive to write the inflaton mode energy/frequency $\omega$ as
\begin{align}
\omega(a) = \omega_e \Big( \frac{a_e}{a}\Big)^{\frac{3(n-1)}{n+1}}\,,
\label{eq:omega}
\end{align}
where $\omega_e\!\equiv\! \omega(a_{e})$ is the mode frequency at the onset of reheating phase. 
The values of the $\mathcal{P}_k^{2n}$ coefficients decrease with $k$ for all considered values of $n$. Consequently, the $\sum_ k |\mathcal{P}_k^{2n}|^2$ sum quickly converges. Moreover, for $n\!=\!1$ the only non-zero Fourier coefficient is $\mathcal{P}_2^2=1/4$, whereas for $n\!=\!3/2$ ($n\!=\!2$) all even (odd) coefficients are zero. The numerical values of the $\sum_ k |\mathcal{P}_k^{2n}|^2$ sum for different values of $n$ are collected in Table~\ref{tab:PkSums}. For details see \app{s.graviational_production} where ${\cal S}_\phi\equiv {\cal D}^{(2)}_{\phi\to XX}$.

For the quadratic inflaton potential, i.e., $n\!=\!1$, the frequency $\omega=m_\phi = \sqrt{2} \Lambda^2/\mpl$ is time-independent during the reheating phase and the only non-zero Fourier coefficient for $|{\cal P}_k^{2n}|$ is for $k=2$. In this case, the source term \eq{eq:sphiX} simplifies as
\begin{align}
{\cal S}_\phi \overset{n=1}{=} \frac{1}{128 \pi} \frac{\rho_\phi^2}{\mpl^4} \bigg(\!1 -\frac{m_X^2}{m_\phi^2}+\frac{3m_X^4}{m_\phi^4}\!\bigg) \sqrt{1\!-\! \frac{m_X^2}{m_\phi^2}}.
\end{align} 
Note that the above expression receives time-dependence only through $\rho_\phi$. Thus, the rate of DM production is expected to be the largest at the onset of reheating. Similar behavior is observed for other values of $n$ for light DM (LDM) such that $m_X \!<\! \omega_e $. However, in generic case, i.e., $n\!\neq\!1$, the $S_\phi$ term has another source of time-dependence --- the frequency $\omega$, which increases (decreases) with time for $n\!<\!1$ ($n\!>\!1$).  For instance, for the $n\!=\!1$ case, the time-independence of $\omega$ implies that there exists a constant mass threshold during the whole period of reheating. Due to the fact that in this case, the only non-zero Fourier coefficient is $\mathcal{P}_2^{2}$, the phase space factor $\sqrt{1- (m_X/m_\phi)^2}$ requires $m_X \!<\!m_\phi$. This means that DM particles with masses exceeding the effective mass of the inflaton field cannot be gravitationally produced from the inflaton background. For the remaining values of $n$, we have infinitely many non-zero coefficients $\mathcal{P}_k^{2n}$ that contribute to the source term $\mathcal{S}_\phi$. In this case, we can compensate for the smallness of the frequency $\omega$ considering higher harmonics. By increasing the value of the denominator in the $\sqrt{1- (m_X/k \omega)^2}$ factor, we circumvent the standard kinematical suppression. However, in this case, $\mathcal{S}_\phi$ receives another suppression that comes from the Fourier coefficients, since $\mathcal{P}_k^{2n}$ rapidly decreases with increasing $k$.
Consequently, the main contribution to the $\mathcal{S}_\phi$ term comes from the first (minimal) non-zero harmonic $k_{\min}$, which for $n=2/3$ and $n=3/2$ is $k_{\min}=1$, whereas for $n=1$ and $n=2$ is $k_{\min}=2$.  

\begin{figure}[t!]
\begin{center}
\includegraphics[width=0.77\linewidth]{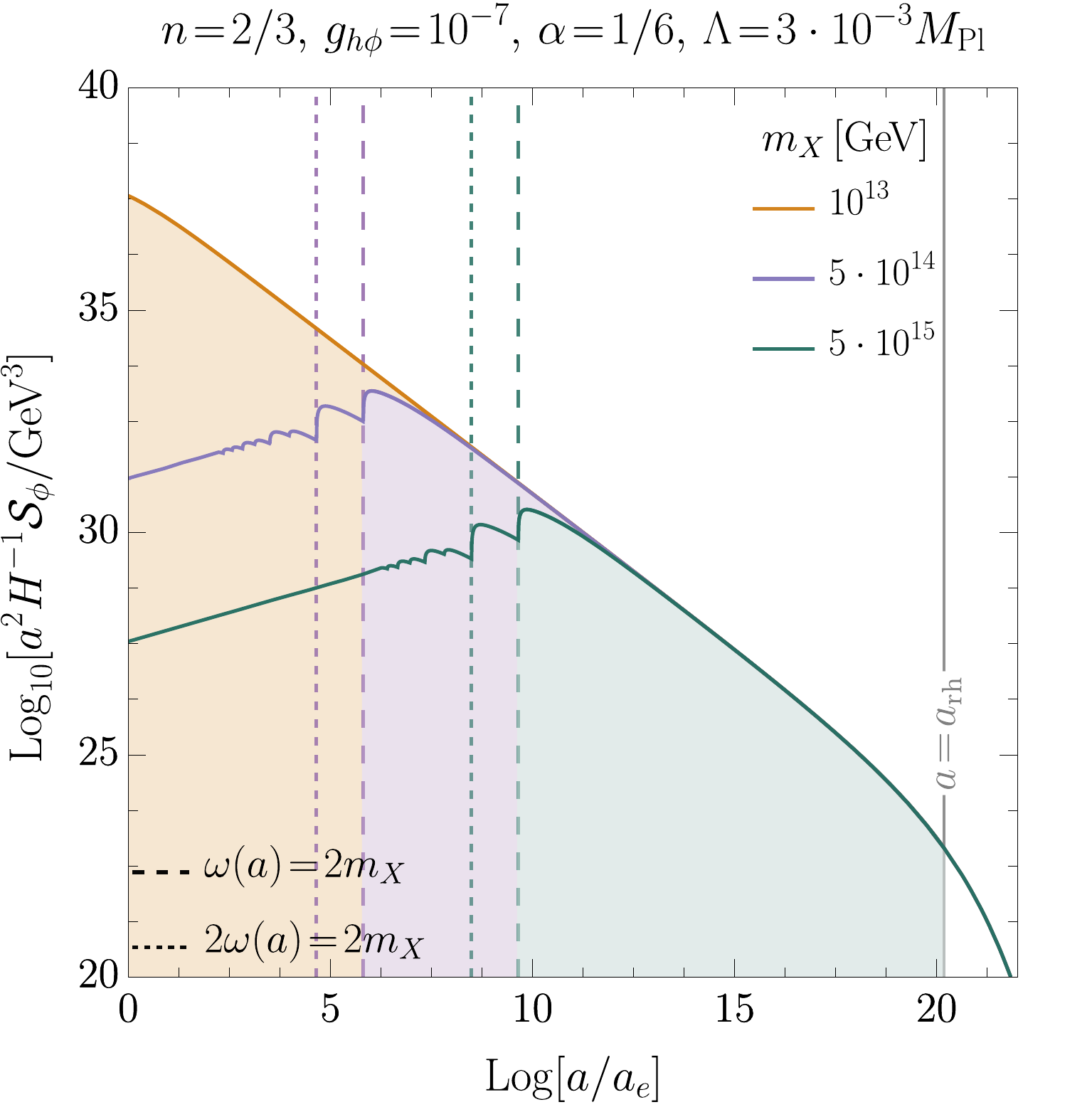}
\caption{The evolution of the comoving number density $dN_X/da=a^2 \mathcal{S}_\phi/ H$ as a function of the scale factor $a$ for $n\!=\!2/3$. For LDM, i.e., $2 m_X \! < \! \omega (a_e)$ (orange curve), the source term $\mathcal{S}_\phi$ reaches the maximal value at the beginning of reheating, i.e., $a=a_e$.  Whereas, for HDM, i.e., $2 m_X \! > \! \omega (a_e)$ (purple and green curves), the source term $\mathcal{S}_\phi$ is not maximized at $a=a_e$ due to the kinematical suppression, but instead grows with time until $\omega(a)\sim m_X$. }
\label{fig:sphin23_GP}
\end{center}
\end{figure}

We should also stress that for $n\!>\!1$ and for a given DM mass, the $m_X/\omega$ ratio is the smallest at the onset of reheating and increases with time. Thus, the DM production rate is the largest just after the end of inflation. Contrarily, for the $n<1$ case, e.g., $n\!=\!2/3$, the $m_X/\omega$ ratio decreases with time. Thus, in this case for heavy DM (HDM), i.e., $\omega_e <2m_X$, the source term $\mathcal{S}_\phi$ is the largest at some later moment $a_X \! > \! a_e$ when $\omega(a_X)\sim 2 m_X$. 
This behavior is manifestly shown in Fig.~\ref{fig:sphin23_GP}, where we have plotted the evolution of the $a^2 \mathcal{S}_\phi/ H^2$ terms as a function of the scale factor for LDM (orange curve) and HDM (purple and green curves) cases. For $n\! < \! 1$, the inflaton mode energy/frequency $\omega$ increases with time, see \eq{eq:omega}. 
Thus, even if the DM mass is larger than the mode frequency, i.e., $\omega<2m_X$ at the onset of reheating phase, the gradual increase in the inflaton mode energy $\omega$ can provide the phase for the production of DM before the end of reheating. This, in particular, means that the production of HDM is kinematically disfavoured at the onset of reheating, but if $2 m_X \!<\! \omega(a_{\rm{rh}})$ such particles can be nevertheless produced abundantly during the reheating period. It is useful to write explicitly the form of $a_X$, i.e., $k_{\min}\omega (a_X)=2m_X$, using \eq{eq:omega} as
\begin{align}
a_X &=  a_e \begin{dcases}\Big\lfloor\frac{2 m_X}{\omega_e } \Big\rfloor^{\frac{1+n}{3(1-n)}}, 	& n<1,	\\
1, & n\geq 1.
\end{dcases}		\label{eq:a_X}
\end{align} 
Above `floor' function $\lfloor\cdots\rfloor$ is defined as follows:
\beq
\lfloor x \rfloor =\begin{cases}
1, & x\leq1,	\\ 
x,	 & x>1.
\end{cases}  \label{eq:floor}
\eeq
In other words, $a_X$ indicates the lowest value of the scale factor for which the lowest non-zero harmonics $k_{\min}$ with a given $n$ can produce DM species with mass $m_X$. 

The number density of DM species produced gravitationally from the inflaton background can be obtained by solving the Boltzmann equation \eqref{eq:Xeq}. Keeping only the gravitationally produced DM from the inflaton field, i.e., $\mathcal{S}_\phi$ term, we get,
\begin{align}
n_X(a) = \frac{1}{a^3} \int_{a_e}^{a} d\tilde{a} \, \frac{\tilde{a}^2}{H(\tilde{a})} \mathcal{S}_\phi(\tilde{a}). \label{eq:nDMphi}
\end{align}
The comoving number density, $N_X$, approaches a constant limit around $a_f \simeq a_{\rm {rh}}$, which results from the fact that after the end of reheating, the inflaton is depleted and the contribution of $\rho_\phi$ to the total energy density quickly becomes negligible.  Thus, the present-day number density of DM particles, $n_X(a_0)$, can be well approximated by the solution of \eqref{eq:nDMphi} with the upper limit of the integral taken as $a=a_{\rm{rh}}$. Inserting solutions for $H(a)$ and $\rho_\phi(a)$, obtained in the previous section, we find
\begin{align}
n_X(a_0) &\simeq \frac{n+1}{6n-3}\frac{\sqrt{3}}{8\pi} \bigg( \frac{\sqrt{\rho_{\phi_e}  }}{\mpl}\bigg)^{\!3} \,\sum_{k=1}^\infty |\mathcal{P}_k^{2n}|^2  \nonumber \\
&\qquad\times \Big(\frac{a_e}{a_X}\Big)^{\!\frac{6n-3}{n+1}}\Big(\frac{a_e}{a_0}\Big)^{\!3}\bigg[1-\Big(\frac{a_X}{a_{\rm{rh}}}\Big)^{\frac{6n-3}{n+1}}\bigg]. \label{eq:nX1}
\end{align}
In the above equation, we have neglected the phase space factor proportional to $m_X^2/(k \omega)^2$. Moreover, we have also assumed that $N_X(a_{\rm rh}) \simeq N_X(a_0)$, which reflects the fact that the efficient DM gravitational production in the inflaton background is possible only during reheating. 

Now let us discuss the solution~\eqref{eq:nX1} in more detail. Firstly, we should emphasize that for $n\!>\! 1/2$, as employed in this work, the second factor in the squared bracket is always small, i.e., $(a_X/a_{\rm{rh}})^{6n-3/(n+1)} \! \ll \! 1$ since $a_X\leq a_{\rm rh}$, therefore this factor can neglected. Secondly,  for $n\! \geq\!1$ we should always take $a_X \!=\! a_e$, since in this case the inflaton mode energy $\omega$ is the largest at $a=a_e$, which implies that for $a>a_e$ the phase space suppression increases. Finally, for $n\!<\!1$, $a_X$ can be larger or equal $a_e$ depending on the relation between $2m_X$ and $\omega_e$. In particular, the for heavy DM, i.e., $2m_X>\omega$, the $a_X$ is related to the DM mass through the following relation 
\begin{align}
a_X &=  a_e \Big( \frac{2 m_X}{\omega_e }\Big)^{\frac{1+n}{3(1-n)}}, &n&<1.
\end{align} 
Taking into account all simplifications we can recast \eq{eq:nX1} as 
\begin{align} 
n_X(a_0) &\simeq  \frac{\sqrt{3}}{8\pi} \bigg( \frac{\sqrt{\rho_{\phi_e}  }}{\mpl}\bigg)^3  \Big( \frac{a_e}{a_0}\Big)^3 \frac{n+1}{6n-3} \sum_k |\mathcal{P}_k^{2n}|^{2}	\notag\\
&\qquad \times\begin{dcases}  \Big\lceil\frac{\omega_e }{2 m_X}\Big\rceil^{\!\frac{2 n-1}{1-n}} , & n\!<\! 1,	\\
1, & n\!\geq\!1. \label{eq:nX0sim}
\end{dcases}
\end{align}
In any successful model, the predicted value of a DM abundance today,
\begin{align}
\Omega_X= \frac{m_X n_X (a_0)}{\rho_c}, \label{eq:OmegaX}
\end{align}
should match with the observed value \eqref{eq:OmegaX_obs}. Above, $\rho_c =1.054 \times 10^{-5} h^2\; {\rm GeV}\, {\rm cm}^{-3}$  denotes the critical density.
Using \eq{eq:nX0sim} we can estimate the present-day amount of DM produced purely gravitationally from the inflaton field as
\begin{align}
&\Omega_X^{\rm{gr}, \phi}  = \frac{\sqrt{3}}{8\pi} \bigg( \frac{\sqrt{\rho_{\phi_e}  }}{\mpl}\bigg)^3  \Big( \frac{a_e}{a_{\rm{rh}}}\Big)^3 \frac{s_0}{s_{\rm rh}} \frac{1}{\rho_c} \nonumber \\ 
&\quad\times\frac{n+1}{6n-3} \sum_k |\mathcal{P}_k^{2n}|^{2}\begin{dcases}  m_X\Big\lceil\frac{\omega_e }{2 m_X}\Big\rceil^{\!\frac{2 n-1}{1-n}} , & n\!<\! 1,	\\
m_X, & n\!\geq\!1.,	
\end{dcases}	\label{Omega_phi}
\end{align}
where $s$ is the entropy-density, and $s_0\!=\! 2970\, {\rm cm}^{-3}$ denotes its present-day value and $s_{\rm rh}\!\equiv\!s(a_{\rm rh})$ is the entropy density at the end of reheating. In the above relation we have used the conservation of the entropy, i.e., $s_{\rm rh}/s(a_0) \simeq \frac{a_0^3}{a_{\rm{rh}}^3}$, neglecting the change of $g_{\star, S}$ in the period between $a_{\rm{rh}}$ and $a_0$. 

For $n\!=\! 2/3$ and heavy DM, i.e., $2m_X>\omega_e$ we note a very peculiar behaviour of $\Omega_X^{\rm{gr}, \phi} h^2$.  It turns out that in this case, the predicted abundance of DM particles does not depend on $m_X$, as the `ceiling' function in \eq{Omega_phi} results in $\omega_e/(2 m_X)$ and the $m_X$ factors cancel out. This mass-independence of $\Omega_X^{\rm{gr}, \phi} h^2$ for heavy DM is observed only if reheating lasts long enough such that $\omega(a)$ equals $2 m_X$ well before the end of reheating. Thus, the obtained approximation works only for DM particles with mass $\omega_e  \!\lesssim\! 2 m_X \! \lesssim \! \omega(a_{\rm{rh}})$. 
Note also that in other cases, for fixed value of $\Lambda$, $\alpha$, the present-day abundance of DM species depends only on the value of $n$, $m_X$ and thus on $a_i$ and $a_{\rm{rh}}$. Therefore, it is rather not sensitive to the evolution of the thermal bath but depends on the duration of the reheating period. 

\subsection{Gravitational freeze-in production}
Next, we discuss the second irreducible DM production mechanism, namely the gravitational freeze-in from the SM particles. Gravitational freeze-in is a universal phenomenon in which the SM particles annihilate to produce DM through the s-channel graviton exchange. This mechanism only requires gravitational interaction between the SM and DM. Similar to the mechanism discussed in the previous subsection, the gravitational freeze-in from SM is most efficient during the reheating period. However, in this case, DM species are produced in a quantum process from two SM particles in thermal equilibrium. Thus, the abundance of DM particles produced from SM quanta is sensitive to the thermal history of the Universe. Moreover, since the amplitude of such a process is proportional to $s/\mpl^2$, where $s$ is the Mandelstam variable, efficient DM production requires high energy or temperature of the SM annihilating particles. In principle, it was shown, e.g., ~\cite{Garny:2017kha,Tang:2017hvq}, that pure gravitational DM production from the SM sector can account for all the observed DM relic density \eqref{eq:OmegaX_obs} if the reheating efficiency and the thermal bath temperature are sufficiently high.   In contrast to the production in the inflaton background, the freeze-in DM production depends on the evolution of the radiation sector, or more precisely, on the evolution of the thermal bath temperature during the reheating phase. Thus, the efficiency of DM gravitational freeze-in production depends on the dynamics of reheating, which is determined by the scale of inflaton $\Lambda$, the inflaton--Higgs coupling $g_{h \phi}$, $n$ and the inflaton-induced mass of the Higgs boson.

For the gravitational freeze-in DM production, the Boltzmann equation simplifies as
\begin{align}
\frac{ d N_X}{ d a} =\frac{a^2}{H} \mathcal{S}_{\rm{SM}}=\frac{a^2}{H} \bar{n}_X^2\langle \sigma \lvert v \rvert \rangle_{XX \rightarrow h_{\mu \nu} \rightarrow \rm {SM \, SM }}^{},		\label{eq:Xprime_GP}
\end{align}
where we have kept only the $\mathcal{S}_{\rm{SM}}$ term on the r.h.s. of \eqref{eq:Xeq}. 
The thermally averaged cross-section $\langle \sigma \lvert v \rvert \rangle$ for any $2\to 2$ annihilation process can be calculated using the standard formula~\cite{Gelmini:2006pw},
\begin{widetext}
\vspace{-10pt}
\begin{align}
\langle \sigma \lvert v \rvert \rangle_{ii\to ff} &= \frac{(2 J_i+1)^2}{\bar{n}_i^2} \frac{T}{32 \pi^4}
 \int_{4m_{\rm max}^2}^\infty  d  s \; \sqrt{s} (s- 4 m_i^2) \, \sigma_{ii\to ff} (s) K_{1}\left( \frac{\sqrt{s}}{T}\right), \label{eq:tacs}
\end{align}
where $m_{\rm max} \equiv {\rm max} (m_i, m_f)$, and $\sigma_{ii\to ff}$ denotes the cross-section averaged over possible spin states of the DM particles $i$ in the initial state and SM particles $f$ in the final state.
For the gravitational freeze-in production of DM, there are four relevant annihilation processes, 
\begin{align}
\sigma_{XX \rightarrow h_{\mu \nu} \rightarrow h_i h_i} &= \frac{1}{9}\frac{1}{2}\frac{1}{ 480 \pi s^3\mpl^4} \frac{\sqrt{s- 4 \bar{m}_{h_i}^2}}{\sqrt{s- 4m_X^2}}  \bigg[ 2 \bar{m}_{h_i}^4\left(84 m_X^4 +68 m_X^2 s + 19 s^2 \right) 	\notag\\
& \hspace{5cm} - 2 \bar{m}_{h_i}^2 s (6m_X^2+s)(2m_X^2+7s)  + 3 s^2(6m_X^4+ 2m_X^2 s+ s^2) \bigg], \label{eq:GP_XXhh}\\
\sigma_{XX \rightarrow h_{\mu \nu} \rightarrow \bar{\psi}\psi} &=\frac{1}{9}\frac{1}{960 \pi s^2\mpl^4} \frac{\sqrt{s- 4 \bar{m}_{\psi}^2}}{\sqrt{s- 4m_X^2}} \!\bigg(\!1\!-\! \frac{4 \bar{m}_\psi^2}{s}\bigg)\!\bigg[\! 2 \bar{m}_{\psi}^2\left(84 m_X^4 +68 m_X^2 s + 19 s^2 \right)  \!+\!  s(48 m_X^4+ 56 m_X^2 s+ 13 s^2) \!\bigg],  \label{eq:GP_XXff}\\
\sigma_{XX \rightarrow h_{\mu \nu} \rightarrow V V} &= \frac{1}{9} \frac{1}{\delta_V} \frac{1}{480 \pi s^3 M_{\rm pl}^4}\frac{\sqrt{s- 4 \bar{m}_{V}^2}}{\sqrt{s- 4m_X^2}} \bigg[ \bar{m}_V^4 (504 m_X^4 + 408 m_X^2 s+ 114 s^2) + \bar{m}_V^2 (408 m_X^4 s + 536 m_X^2 s^2+ 118 s^3)  \nonumber \\
&\hspace{5cm} + 114m_X^4 s^2 + 118 m_X^2 s^3 + 29 s^4 \bigg],  \label{eq:GP_XXVV}\\
\sigma_{XX \rightarrow h_{\mu \nu} \rightarrow \gamma \gamma} &= \frac{1}{9} \frac{1}{2} \frac{1}{240 \pi \mpl^4} \frac{48 m_X^4 + 56 m_X^2 s+ 13 s^2}{\sqrt{s(s-4m_X^2)}}, 	
\end{align}
which account for gravitational annihilation of the $X$ species into SM Higgs bosons, fermions, and massive and massless vectors, respectively. Above $\delta_V\!=\!2$ for the identical gauge bosons in the final states, otherwise $\delta_V=1$. In the above annihilation cross sections, we have taken into account the non-trivial SM particle masses due to inflaton-induced electroweak vev $v_h$, where $\bar{m}$ denotes the time-averaged (over fast inflaton oscillations) mass. 
In the limit $m_{\rm{SM}} \rightarrow 0$, the corresponding thermally-averaged cross-sections are~\cite{Garny:2017kha},
\begin{align}
\langle \sigma |v| \rangle_{XX \rightarrow h_{\mu \nu} \rightarrow h_i h_i} &\!=\!  \frac{9}{32 \pi^4} \frac{1}{120 \pi \mpl^4} \frac{m_X^6 T^2}{\bar{n}_X^2}\bigg\{ 3 K_1^2\!\Big(\! \frac{m_X}{T}\!\Big) + 2 K_2^2\!\Big(\! \frac{m_X}{T}\!\Big) \bigg[ 1+4 \Big( \frac{T}{m_X}\Big)^2 \bigg]+ 4 \frac{T}{m_X} K_1\!\Big(\! \frac{m_X}{T}\!\Big) K_2\!\Big(\! \frac{m_X}{T}\!\Big) \bigg\},	\\
\langle \sigma |v| \rangle_{XX \rightarrow h_{\mu \nu} \rightarrow \bar\psi\psi} &\!=\! \frac{9}{32 \pi^4}  \frac{1}{90 \pi \mpl^4} \frac{m_X^6 T^2}{\bar{n}_X^2}\bigg\{ 11 K_1^2\!\Big(\! \frac{m_X}{T}\!\Big) +  K_2^2\!\Big(\! \frac{m_X}{T}\!\Big) \bigg[ 9+26 \Big( \frac{T}{m_X}\Big)^2 \bigg]+ 13 \frac{T}{m_X} K_1\!\Big(\! \frac{m_X}{T}\!\Big) K_2\!\Big(\! \frac{m_X}{T}\!\Big) \bigg\}, 	\\
\langle \sigma |v| \rangle_{XX \rightarrow h_{\mu \nu} \rightarrow \gamma \gamma} &\!=\! \frac{9}{32 \pi^4}  \frac{3}{135 \pi \mpl^4} \frac{m_X^6 T^2}{\bar{n}_X^2}\bigg\{ 11 K_1^2\!\Big(\! \frac{m_X}{T}\!\Big) +  K_2^2\!\Big(\! \frac{m_X}{T}\!\Big) \bigg[ 9+26 \Big( \frac{T}{m_X}\Big)^2 \bigg]+ 13 \frac{T}{m_X} K_1\!\Big(\! \frac{m_X}{T}\!\Big) K_2\!\Big(\! \frac{m_X}{T}\!\Big) \bigg\}.
\end{align}
\vspace{-10pt}
\end{widetext}
On the other hand, in the \textit{massive reheating scenario}, i.e., for $m_{\rm{SM}} \! \neq \!0$, the integral over $s$ in Eq.~\eqref{eq:tacs} cannot be obtained analytically because of the non-zero mass of the SM species, generated by their coupling to the inflaton or the Higgs doublet. In this case we integrate \eqref{eq:tacs} numerically using Eqs.~\eqref{eq:GP_XXhh}--\eqref{eq:GP_XXVV} and adopting the solution for $T(a)$ which we have found in the previous section. It is important to notice that even in the \textit{massive reheating scenario}, SM photons and gluons are still massless during reheating. Thus, one expects that in this case the freeze-in DM production is dominated by the annihilation of the massless SM gauge bosons, since other annihilation channels are Boltzmann suppressed if $m_{\rm{SM}}/T \!>\!1$. 
In the limit of asymptotically small or large DM mass the source term, $\mathcal{S}_{XX \rightarrow h_{\mu \nu} \rightarrow \rm{SM} \, \rm{SM}} =\bar{n}_X^2 \langle \sigma |v| \rangle_{XX \rightarrow h_{\mu \nu} \rightarrow \rm{SM} \,\rm{SM}} $, for massless SM particles behaves as
\begin{align}
\mathcal{S}_{XX \rightarrow h_{\mu \nu} \rightarrow \rm{SM} \, \rm{SM}} \!\simeq \!
\begin{dcases} 
\! \frac{A_{i}}{40 \pi^5} \frac{T^8}{\mpl^4} \,,	& m_X \! \ll \! T,  \\
\! \frac{B_{i}}{32\pi^4} \frac{m_X^5 T^3}{\mpl^4} e^{-2 m_X/T}  \,,	& m_X \! \gg \! T,
\end{dcases}
\end{align}
where $A_i \! = \!  3, 13, 26$ and $B_i \!=\! 3/16, 1, 2 $, for SM scalars, fermions and vectors, i.e., $i=0,1/2,1$, respectively. The asymptotic behavior of the $\mathcal{S}_{XX \rightarrow h_{\mu \nu} \rightarrow \rm{SM} \, \rm{SM}}$ term in the considered mass regimes indicates that efficient DM production occurs only if DM mass $m_X$ is smaller than the temperature of the thermal bath. The maximal temperature  $T= T_{\rm{max}}$ plays here an important role. Note that production of heavier, i.e., $m_X \! > \! T_{\rm{max}}$, DM species is exponentially suppressed, which means that the value of $T_{\rm{max}}$ sets the upper bound for DM particle mass that can be gravitationally produced from the thermal bath. 

To find a prediction for the relic abundance of DM vectors produced gravitationally, we should solve the Boltzmann equation \eq{eq:Xprime_GP} including all annihilation channels. 
Assuming that the integrand in \eqref{eq:Xprime_GP} is dominated by the contribution from early times, we can integrate this equation from $a=a_e$ up to some $a_{f}$, such that $a_f \! \ll \! a_0$. Since in the \textit{massive reheating scenario} we can enter regions where the inflaton-induced mass of the Higgs field can or cannot be neglected, the analytical solution of \eq{eq:Xprime_GP} is quite involved and not very illuminating.  However, in the \textit{massless reheating scenario}, we can obtain a rather simple formula for the present-day number density of DM species
\begin{align}
n_X^{{\rm gr}, {\rm SM}}(a_0) &\simeq  \frac{\sqrt{3} A}{40 \pi^5} \frac{1+n}{6n-3} \frac{(T_{\rm{rh}}^{(0)})^8}{\mpl^3 \sqrt{\rho_{\phi_e}  }} 	\notag\\
&\quad\times\Big(\frac{a_{\rm{rh}}}{a_e} \Big)^{\frac{6}{1+n}}\Big( \frac{a_e}{a_0} \Big)^3 \bigg[ \Big(\frac{a_{\rm rh}}{a_e} \Big)^{\frac{3(2n-1)}{n+1}} -1 \bigg],
\label{nX0}
\end{align}
where $A \equiv n_0 A_0 + n_{1/2} A_{1/2} + n_1 A_1/(40 \pi) = \mathcal{O}(10) $ for $n_{0,1/2,1}=4,45,12$ being the scalar, fermion and vector d.o.f. for the SM, respectively. Above $T_{\rm{rh}}^{(0)}\equiv T^{(0)}(a_{\rm{rh}})$ is the temperature at the end of reheating and can be obtained from \eq{eq:Tsol0}. Note that in the above expression we have neglected gravitational production after the end of reheating, assuming that the comoving number density becomes constant around $a \simeq a_{\rm{rh}}$. In principle, we have used the fact that the comoving number density of DM particles freezes-in, i.e., $N_X (a_{\rm{rh}}) \simeq N_X(a_0)$. 
Although this assumption seems robust, the full numerical solutions consider the contribution to $N_X$ from the late time after the end of reheating. Moreover, we have also included contributions from all annihilation channels in both considered reheating scenarios in the numerical analysis. Let us also emphasize that in the reheating scenario with the massive Higgs field, two sources suppress DM production, (i) the Boltzmann suppression that is present for the annihilation of massive SM particles, and (ii) the lower temperature of the thermal bath. Since the source term $\mathcal{S}_\phi$ for DM particles with mass $m_X < T_{\rm max}$ is proportional to $T^{8}$ the suppression in the bath temperature significantly affects the efficiency of the considered mechanism. 

The contribution to the total relic abundance \eqref{eq:OmegaX} from the gravitational scattering of SM particles in the \textit{massless reheating scenario} can be estimated as
\begin{align}
\Omega_X^{{\rm gr}, \rm{SM}^{(0)}} &\!\simeq\!\frac{\sqrt{3} A}{40 \pi^5} \frac{1+n}{6n-3} \frac{(T_{\rm{rh}}^{(0)})^8 m_X}{\mpl^3 \sqrt{\rho_{\phi_e}  }} \frac{s_0}{s_{\rm rh}\,\rho_c} \Big( \frac{a_{\rm rh}}{a_{e}} \Big)^{\frac{3n}{n+1}} ,	\label{eq:Omega_phi_massless}
\end{align}
for $n\!>\!1/2$, which is employed throughout this work. 
Now we can compare the contribution from the gravitational freeze-in production with the inflaton-induced gravitational production as
\begin{align}
\frac{\Omega_X^{{\rm gr}, \rm{SM}^{(0)}}}{\Omega_X^{{\rm gr}, \phi}} &=
\frac{A}{5 \pi^4} \frac{(T_{\rm{rh}}^{(0)})^8}{\rho_{\phi_e}  ^2} \Big( \frac{a_{\rm{rh}}}{a_e} \Big)^{\!\frac{3(2n+1)}{n+1}} \non \\
&\quad\times \frac{1}{\sum_{k}|\mathcal{P}_k^{2n}|^2}\begin{dcases} \Big\lceil\frac{\omega_e }{2 m_X}\Big\rceil^{\!\frac{1-2n}{1-n}} , & n\!<\! 1,	\\
1, & n\!\geq\!1.
\end{dcases}
\end{align}
The above estimation for the $\Omega_X^{{\rm gr}, \rm{SM}^{(0)}}  / \Omega_X^{{\rm gr}, \phi} $ ratio indicates that for light DM with mass $m_X \! < \!  \omega_e $ the inflaton induced gravitational production dominates over the contribution from the SM freeze-in mechanism. As we have highlighted in the previous subsection, for $n\! \geq \! 1$ the initial value of the inflaton's frequency, $\omega_e $, sets the upper bound for the mass of DM species that can be efficiently produced from the inflaton. However, the maximal temperature of the thermal bath may exceed the initial value of $\omega$. In such a case, the DM production from the inflaton field is kinematically suppressed, however, for $m_X \! < \! T_{\rm {max}}$, the DM particles can still be produced from the thermal bath. Moreover, for $n\!<\! 1$, the inflaton induced gravitational production always dominates over the thermal DM production. In this case, it turns out that for heavy DM particles with mass $m_X \in (\omega_e , \omega(a_{\rm rh}))$ and for generic values of model parameters ($\Lambda, \alpha$ and $g_{h \phi}$), the inflaton mode energy at the end of reheating $\omega(a_{\rm rh})$ exceeds $T_{\rm max}$. 

\begin{figure}[t!]
\begin{center}
\includegraphics[width=0.77\linewidth]{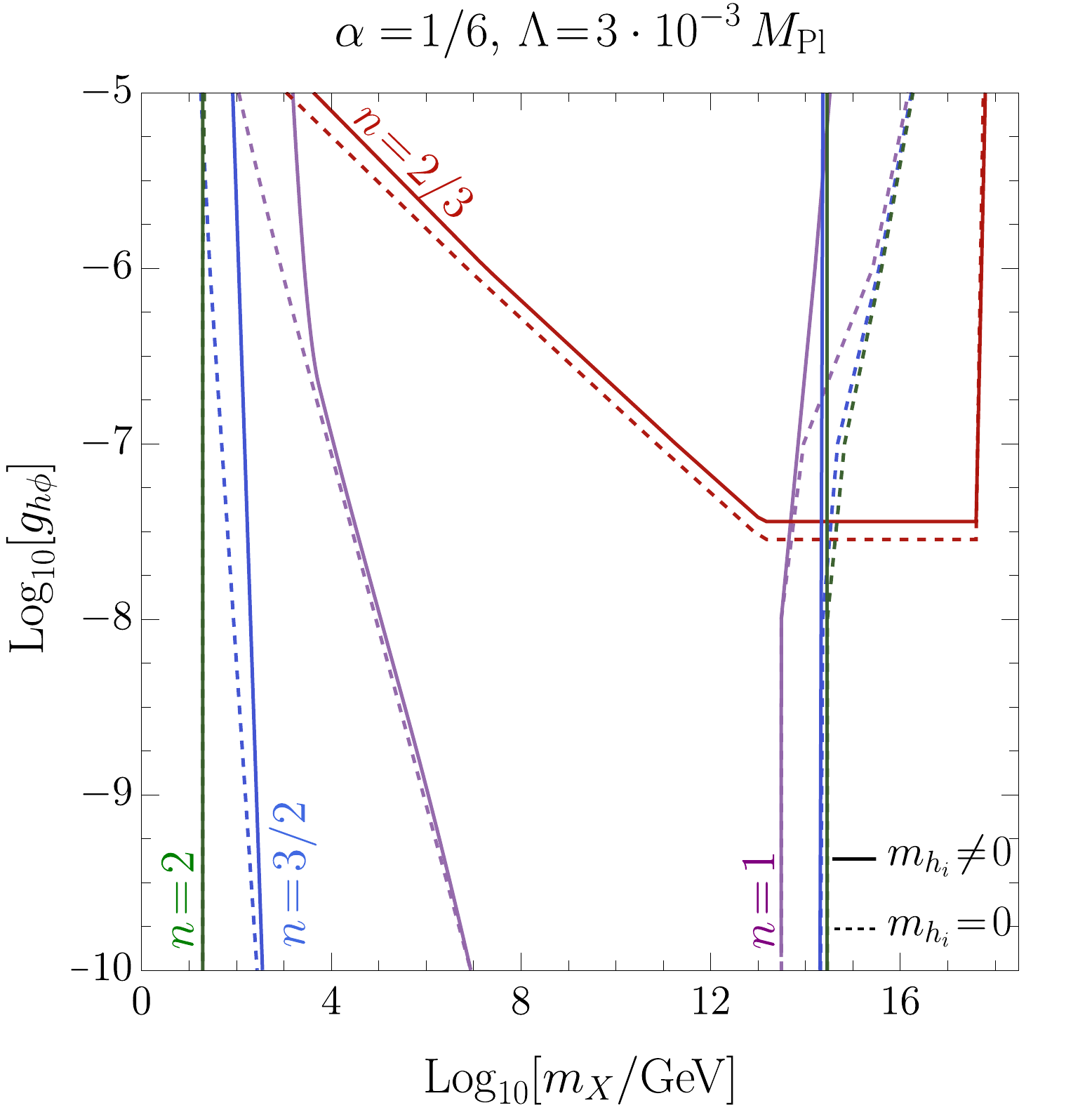}
\caption{ The relation between the inflaton-Higgs coupling, $g_{h \phi}$, and the mass of DM species, $m_X$, that predicts the observed DM relic abundance \eqref{eq:OmegaX_obs} from purely gravitational interactions. Two contributions, i.e., from the inflaton field and the SM thermal bath, are included. Solid (dashed) curves are for the \textit{massive (massless) reheating scenario,} respectively. Predicted abundance exceeds the observed value $\Omega_X^{\rm{obs}}h^2$ in the region above colored curves (in other words in-between the curves).}
\label{fig:ghphimx_GP}
\end{center}
\end{figure}

The relation between the inflaton-Higgs coupling, $g_{h \phi}$, and the mass of DM species, $m_X$, for the considered values of $n$ (red: $n\!=\!2/3$, purple: $n\!=1$, blue: $n\!=\!3/2$ and green $n\!=\!2$) satisfying the $\Omega_X^{{\rm gr}, \phi} h^2 + \Omega_X^{{\rm gr}, \rm{SM}^{(0)}} h^2\!=\! \Omega_X^{\rm obs} h^2$ constraint is presented in Fig.~\ref{fig:ghphimx_GP}. Solid (dashed) curves are for the \textit{massive (massless) reheating scenario,} respectively. In the region above (below) each curve, the DM is overabundant (underabundant). In particular, we note from this figure, that for all considered values of $n$ and a fixed $\ghp$, there is a vast mass region for which DM particles are overproduced. This region is bounded from below (small $\mx$) by the position of the colored curves, which show a relation between $g_{h \phi}-m_X$ for which the amount of DM vectors produced gravitationally from the inflaton field matches the observed abundance\,\footnote{Here the contribution from the SM gravitational freeze-in is negligible.}. 
For small values of $g_{h \phi} \lesssim 10^{-8}$, the overproduction mass region is bounded from above also by the inflaton induced gravitational production. In this case, SM bath temperature is not high enough to produce DM pairs through graviton exchange. The upper limit for $\mx$ is set by the kinematics of the contribution from the lowest Fourier mode in \eq{Omega_phi}. On the other hand, increasing the value of $g_{h \phi}$ enhances the efficiency of reheating, hence the temperature of the SM radiation bath is higher and the annihilation through graviton exchange opens out contributing to DM production. As a result, the limiting curves bend outward excluding larger regions. 
This behavior is observed for $n\!\geq \! 1$ around $g_{h \phi} \sim 10^{-8}$ in the \textit{massless reheating scenario.}
In this region of large $\mx$ the inflaton contribution is eliminated by the kinematics.
In the \textit{massive reheating scenario}, the kinematical suppression reduces the radiation energy density such that even for large values of $g_{h \phi}$ the temperature is not high enough to produce DM particles in SM gravitational freeze-in. In this case, we observe that only the $n\!=\!1$ curve (purple, solid line) deviates from the mass threshold $m_X = m_\phi$, though this deviation is much milder than in \textit{massless reheating scenario} (purple dashed line).

We shall emphasize that the result shown in \fig{fig:ghphimx_GP} is strong and conservative, i.e., if any other DM production mechanism was added, the allowed region could only shrink, in other words, the region excluded in Fig.~\ref{fig:ghphimx_GP} is always excluded by the gravitational freeze-in and/or inflaton induced gravitational production.
Let us also stress two important features of Fig.~\ref{fig:ghphimx_GP}. First of all, for the quartic inflaton potential, i.e.,  $n\!= \! 2$, the predicted relic abundance of DM particles produced gravitationally in the inflaton background does not depend on the value of $g_{h \phi}$. To put it another way, the amount of DM vectors produced gravitationally from the inflaton field is insensitive to the efficiency of the reheating period. 
From \eq{eq:Omega_phi_massless} we can see that for $n\!=\!2$, the relic abundance of light DM particles,
\begin{align}
\Omega_X^{{\rm gr}, \phi}  \overset{ n=2}{=} &\frac{\sqrt{3}}{32 \pi} \Big( \frac{30}{\pi^2 g_\star}\Big)^{\!1/4} \frac{m_X s_0}{\rho_c}  \frac{\rho_{\phi_e}  ^{3/4}}{\mpl^3}  \sum_k |\mathcal{P}_k^4|^{2},
\end{align}
depends only on the initial value of the inflaton energy density, $\rho_{\phi_e}  $, which is fixed for our benchmark values of $\alpha$ and $\Lambda$. Moreover, it turns out that the above formula is also valid in the \textit{massive reheating scenario}. From the above formula, one can determine the value of $m_X$ for which the predicted relic abundance matches the observed value. For the benchmark values of $\alpha, \Lambda$ we obtain $m_X \simeq 3.71 \rm{GeV}$, which agrees with the numerical result, cf. Fig.\eqref{fig:ghphimx_GP}. Secondly, we note that for the $n\!=\! 2/3$ case for $g_{h \phi} \sim 10^{-7.5}$ the produced amount of DM becomes mass-independent. This unusual behavior is observed for DM vectors with mass in the range $(\omega_e , \omega(a_{\rm rh}))$ and was discussed in detail in the previous subsection. 

In the following part of this work, we use the results obtained in this section and treat them as a constraint on DM production due to gravitational DM overabundance. Strictly speaking, for fixed $\alpha$, $\Lambda$, $n$ and for a given $g_{h \phi}$ there exists a mass region in which DM species are overproduced, purely through gravity. Consequently, we exclude this mass region while discussing other DM production mechanisms.

\subsection{Inflaton decay}
Once we have discussed irreducible mechanisms of DM production, we can investigate the possibility of direct inflaton-DM interactions. The lowest-dimensional direct coupling between the inflaton and vector DM $X$ fields arises at a dim-5 level and allows DM vectors to be pair produced from the inflaton decay. In this case, the vector DM pair is created from the vacuum in a non-thermal quantum process in the presence of the oscillating inflaton field. In the Boltzmann equation~\eqref{eq:Xeq} the decay term, $\mathcal{D}_\phi$, is given by
\begin{align}
\mathcal{D}_\phi &= \frac{ |\mathcal{C}_X^\phi|^2}{32 \pi}  \bigg(\frac{\rho_\phi}{\Lambda^4}\bigg)^{\!\!1/n}\, \sum_{k=1}^\infty   \lvert \mathcal{P}_k \rvert^2 (k \omega)^4	\notag\\
&\qquad \times\Big( 1 -\frac{4m_X^2}{(k \omega)^2}+\frac{12m_X^4}{(k \omega)^4}\Big) \sqrt{1\!-\! \frac{4 m_X^2}{(k \omega)^2}}.		\label{eq:Dphi}
\end{align} 
For detailed derivation of the above result see \app{s.graviational_production}, where $\mathcal{D}_\phi\equiv \mathcal{D}^{(1)}_{\phi\to XX}$ in \eq{eq:D1phiXX}.
In the above expression, one might want to include the time variation of $m_X$ that originates from dim-5 effective interaction  \eqref{eq:Lint}. Note, however, that the inflaton induced DM mass is $C_X^\phi (\phi /\mpl)m_X^2$, which is the largest for maximal value of the inflaton field, i.e., $\phi\simeq \phi_e\sim \mpl$ up to ${\cal O}(1)$ $\alpha$-dependent corrections. Therefore, the inflaton-induced DM mass is always smaller than the bare DM mass for $C_X^\phi \!\lesssim \!1$, hence the time dependence of $m_X$ can be dropped and $m_X$ is not subjected to the short time-scale oscillations.

Let us also stress, that for the quadratic inflaton potential, i.e., $n\!=\!1$, the inflaton mode energy/frequency $\omega=m_\phi$ becomes time-independent and the source term \eqref{eq:Dphi} simplifies as follows,
\begin{align}
D_\phi \!\overset{n=1}{=}\! \frac{\lvert C_X^\phi \rvert^2}{128\pi}  \frac{\rho_\phi\, m_\phi^4}{\Lambda^4} \Big(\! 1 -\frac{4m_X^2}{m_\phi^2}+\frac{12m_X^4}{m_\phi^4}\!\Big) \!\sqrt{\!1\!-\! \frac{4 m_X^2}{m_\phi^2}}.
\end{align}
In the general case, the mass threshold evolves in time during the reheating phase. As we have already discussed above, for $n\! <\! 1$, the inflaton mode energy $\omega$ increases with time during reheating, allowing heavy DM particles ($m_X>\omega_e$) to be pair-produced from the inflaton decay. On the other hand, the inflaton mode energy $\omega$ decreases during the reheating period for $n \! > \! 1$, which means that the production of DM above $m_X>\omega_e/2$ is kinematically suppressed. Even though we do not observe a sharp threshold limit for DM mass, the creation of $X$ particles with mass $2 m_X \! > \! \omega$ is exponentially suppressed by the $\mathcal{P}_k$ values for higher modes $k>1$. 

Furthermore, it is also clear that DM can be efficiently produced through $\phi$ decays only when the energy density of the inflaton field still makes a relevant contribution to the total energy density. After the end of reheating phase, i.e., the equality of inflaton--radiation of energy densities, the inflaton energy density eventually drops to a negligible amount, and DM production freezes. The drop of the inflaton energy density depends on the value of $n$ and can be either very rapid for $n\!\geq \!1 $ or relatively slow for $n \!<\! 1$. This, in turn, implies that DM particles are efficiently produced through inflaton decay only during the reheating phase for $n \geq 1 $. In contrast, for $n\!=\!2/3$, the $\phi \rightarrow XX$ channel is not immediately switched off after the end of reheating.
Hence, in this case, the DM production is somewhat extended to a few e-folds after the equality of inflaton--radiation energy densities, which might be relevant only for the very heavy DM species. However, in the following analytical discussion, we neglect this subtlety assuming that the efficient DM production happens only during the reheating phase. Consequently, the present number density of LDM particles can be estimated as 
\begin{align}
n_X&(a_0) \!=\! \frac{\sqrt{3} \pi n^4 \Lambda^6}{8 \mpl^3}  |\mathcal{C}_X^\phi|^2 \frac{\Gamma\!\left( \frac{n+1}{2n}\right)^4}{\Gamma\!\left( \frac{1}{n}\right)^4} 
\Big(\! \frac{\rho_{\phi_e}  }{\Lambda^4}\!\Big)^{\!\!\frac{3n-2}{2n}}\!  \sum_k k^4 |\mathcal{P}_k|^2 	\notag\\
&\times\!\!\begin{dcases}
\frac{n+1}{6n-9} \frac{a_e^3}{a_0^3} \Big(\frac{a_e}{a_X}\Big)^{\!\!\frac{6n-9}{n+1}} \bigg[1- \Big( \frac{a_X}{a_{\rm{rh}}}\Big)^{\!\frac{6n-9}{n+1}}\bigg], & n\!\neq \! 3/2, \\
\frac{a_e^3}{a_0^3} \ln\!\Big( \frac{a_{\rm{rh}}}{a_e}\Big), & n\! = \! 3/2. \label{eq:nXID}
\end{dcases}
\end{align}
Note that in the above expression we have used $a_X$, which for $n \! \geq \!1$ coincides with the initial value of the scale factor, i.e., $a_e \! = \! a_X$. On the other hand, if  $n \! \leq \!1$, $a_X$ is determined by the condition $\omega (a_X) \!=\! 2 m_X$ and is given by \eq{eq:a_X}. 

Moreover, let us emphasize that the first term in the above square bracket dominates for $n\! < \! 3/2$, whereas, for $n\! > \! 3/2$, $(a_e/a_{\rm{rh}})^{(6n-9)/(n+1)} \! \ll \! 1$. Neglecting subdominant terms, we find the following prediction for the present abundance of DM species produced from the inflaton decay,
\begin{align}
\Omega_X^{\phi} &\simeq \frac{\sqrt{3} \pi n^4 m_X \Lambda^6}{8 \rho_c\mpl^3}  |\mathcal{C}_X^\phi|^2 \frac{\Gamma\!\left( \frac{n+1}{2n}\right)^4}{\Gamma\!\left( \frac{1}{n}\right)^4} 
\Big(\! \frac{\rho_{\phi_e}  }{\Lambda^4}\!\Big)^{\!\!\frac{3n-2}{2n}}\!  \sum_k k^4 |\mathcal{P}_k|^2 \non \\
&\times \frac{a_e^3}{a_{\rm{rh}}^3}\frac{s_0}{s_{\rm{rh}}} 
\begin{dcases}
\frac{n+1}{6n-9},  & n\! > \! 3/2, \\
\ln\!\Big( \frac{a_{\rm{rh}}}{a_e}\Big), & n\! = \! 3/2, \\
\frac{n+1}{9-6n} \Big( \frac{a_{\rm{rh}}}{a_e}\Big)^{\!\frac{9-6n}{n+1}} , & n\!<\!3/2.
\end{dcases}
\end{align}
It is instructive to compare the vector DM production through the direct inflaton decay to the inflaton induced gravitational production of DM, i.e.,
\begin{align}
\frac{\Omega_X^{\phi} }{\Omega_X^{{\rm gr}, \phi} } &= |\mathcal{C}_X^\phi|^2 \pi^2 n^4  \frac{\Gamma\!\left( \frac{n+1}{2n}\right)^4}{\Gamma\!\left( \frac{1}{n}\right)^4} \bigg( \frac{\Lambda^4}{\rho_{\phi_e}  }\bigg)^{\!1/n} \frac{\sum_k k^4 |\mathcal{P}_k|^2}{\sum_k |\mathcal{P}_k^{2n}|^2 } \non \\
&\times 
\begin{dcases}
\frac{2n-1}{2n-3},  & n\! > \! 3/2, \\
\frac{12}{5}\ln \!\Big( \frac{a_{\rm{rh}}}{a_e}\Big), & n\! = \! 3/2, \\
\frac{2n-1}{3-2n}\Big\lceil \frac{\omega_e }{m_X}\Big\rceil \Big( \frac{a_{\rm{rh}}}{a_e}\Big)^{\!\frac{9-6n}{n+1}} , & n\!<\!3/2.
\end{dcases}
\end{align}
One can conclude that for $n\! \geq  \! 3/2$ the direct and indirect (through the graviton exchange) production of DM from the inflaton are approximately the same order for $\mathcal{C}_X^\phi \! \sim \! 1$. Furthermore, in this case, both processes are relatively insensitive to the dynamics of the reheating period, e.g., its duration. On the other hand, the direct decay of the inflaton to vector DM, $\phi \rightarrow XX$, is enhanced relative to the gravitational production by the $(a_{\rm rh}/a_e)^{(9-6n)/(n+1)}$ ratio for $n\! \leq \! 3/2$. 
\begin{figure*}[t!]
\begin{center}
\includegraphics[width=0.37\linewidth]{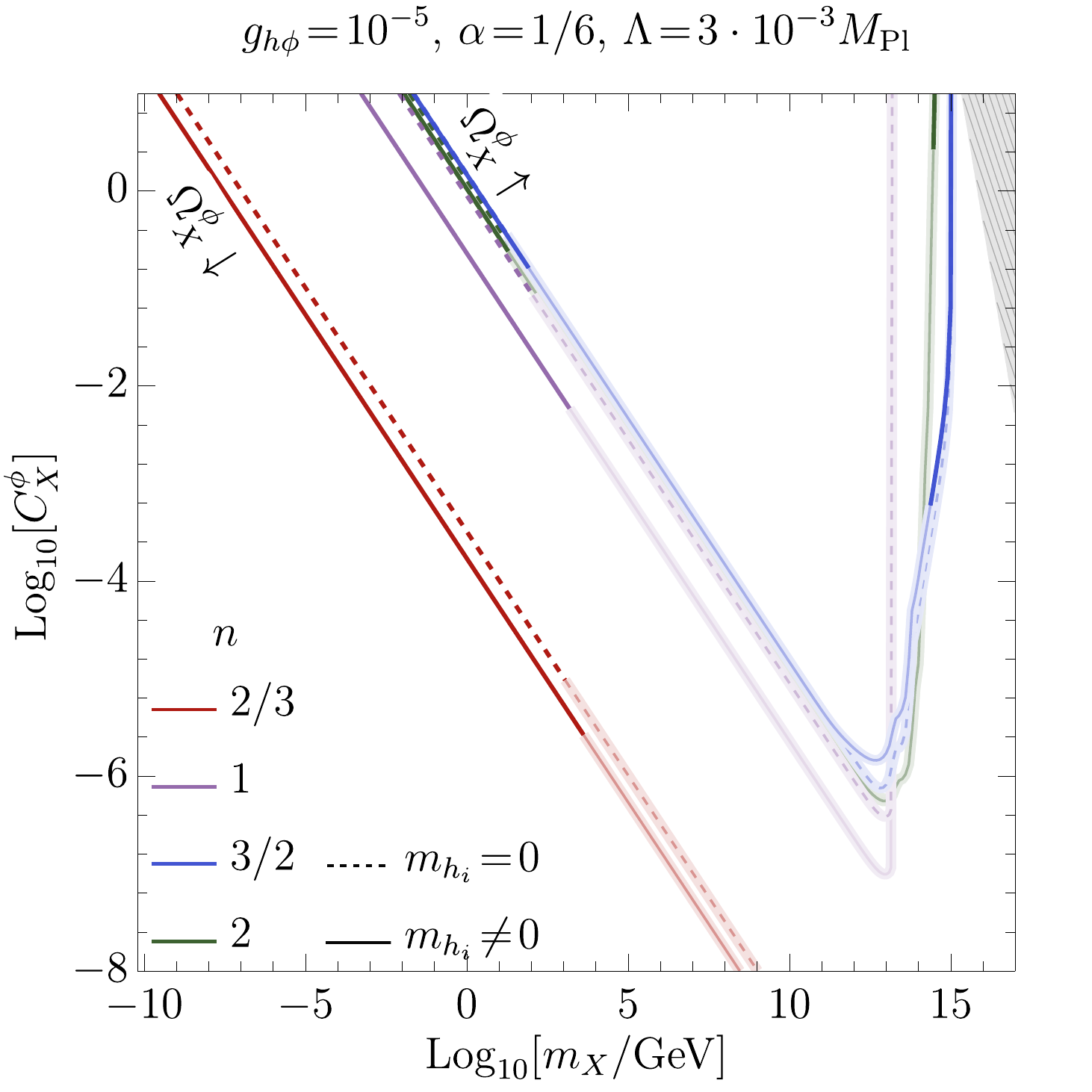}\!\!\!
\includegraphics[width=0.37\linewidth]{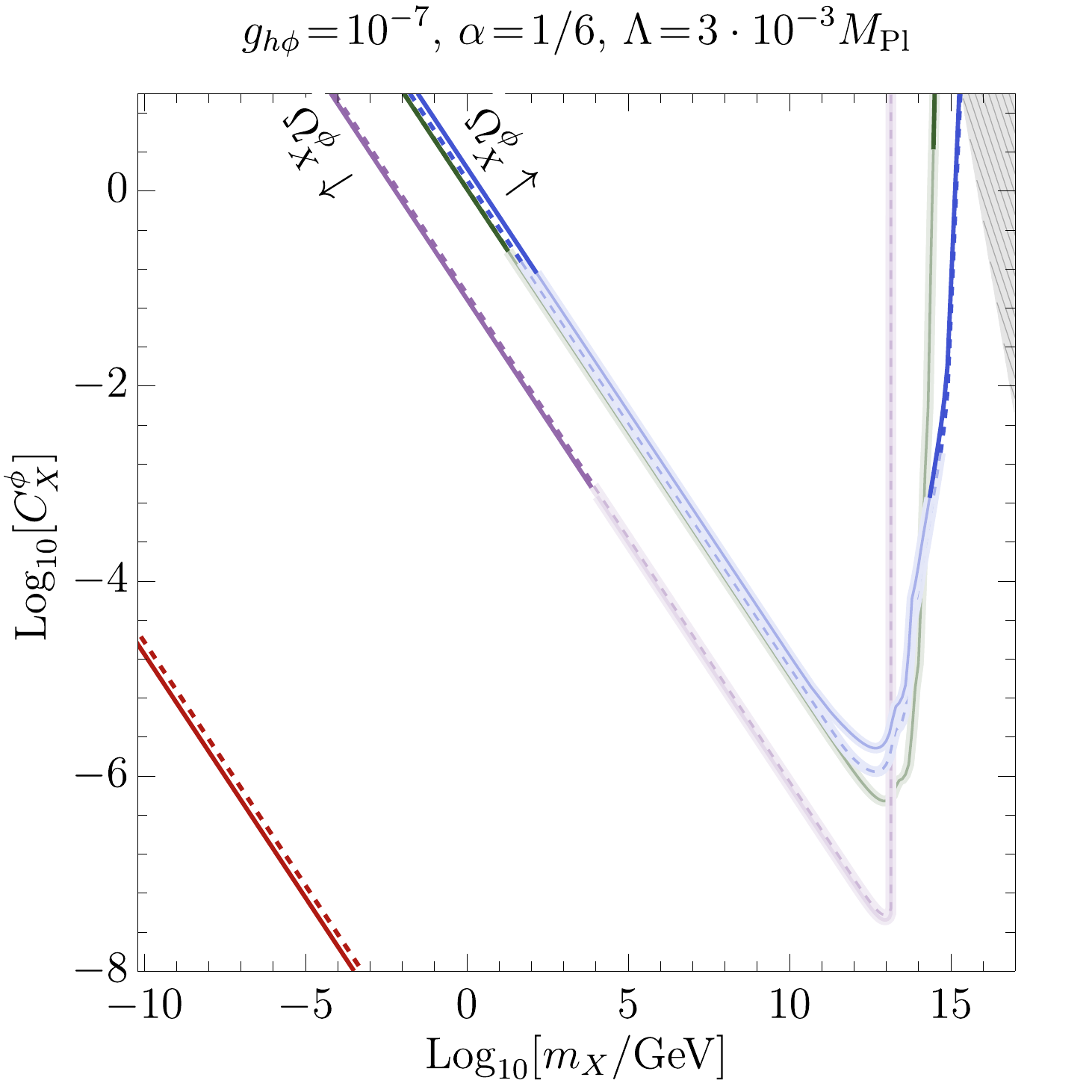}
	\caption{The relation between the Wilson coefficient $\mathcal{C}_X^{\phi}$ and $\mx$ for \mbox{$\ghp\!=\!10^{-5}$} (left panel) and 
	$\ghp\!=\!10^{-7}$ (right panel) that is consistent with the requirement of observed DM abundance \eqref{eq:OmegaX_obs}. Solid (dashed) curves present relation between $\mathcal{C}_X^\phi$ and $m_X$ for massive (massless) SM Higgs bosons. In faded regions, DM particles are overproduced purely due the gravitational production.}
	\label{fig:phiXXFinalScan}
	\end{center}
\end{figure*}

In Fig.~\ref{fig:phiXXFinalScan} we have shown relation between the Wilson coefficient $\mathcal{C}_X^\phi$ and the vector DM mass $m_X$ that gives the correct relic abundance for given values of $g_{h \phi}$ and $n$.
In the region above (below) the $\mathcal{C}_X^\phi-m_X$ curves the predicted relic density is overproduced (underproduced) in comparison to the observed value of  $\Omega_X^{\rm{obs}}h^2$.  In the gray hatched region, the value of the Wilson coefficient $\mathcal{C}_X^{\phi}$ exceeds the perturbativity limit \eqref{eq:cxp_lim}. In the faded region of the curves, DM particles are overproduced purely by gravitational interactions. Comparing the relation between $\mathcal{C}_X^\phi$ and $m_X$ for two considered values of $g_{h \phi}$, we see that for $n\!>1 \!$ the predicted amount of vector DM weakly depends on the details of the reheating dynamics. Very different behavior is observed for the $n\!=\! 2/3$ curves. In this case, DM particles are easily overproduced by decays of the inflaton, and fine-tuning is required to give the correct relation between $\mathcal{C}_X^\phi$ and $m_X$. As we have already noticed above, the $\Omega_X^\phi/\Omega_X^{\rm{gr}, \phi}$ ratio is proportional to $(a_{\rm{rh}}/a_e)^3$ which can be very large for the $n\!=\! 2/3$ case, due to the long duration of the reheating period. That is why one needs extremely small values of the Wilson coefficient $\mathcal{C}_X^\phi$ to suppress this effect. Moreover, in this case, DM particles are gradually produced during the whole period of reheating, and the production rate increases with time. 
Thus, for the $n\!=\!2/3$ scenario (in general for $n\!<\!1$), the production of DM vectors is prolonged in time, and simultaneously enhanced. For $n\!\geq\!1$ the reheating number of e-folds $N_{\rm rh}$ is typically much smaller than for $n\!=\!2/3$. Thus, DM particles are produced for a shorter period. Moreover, the direct inflaton decay to DM, $\phi \rightarrow XX$, is most efficient at the onset of reheating for $n\!\geq\!1$. In this case, the comoving number density of DM species approaches a constant limit quickly after the start of the reheating phase. Consequently, for $n\!>\!1$,  DM production from the inflaton decay becomes almost independent of the duration of the reheating period.

Finally, let us also mention that DM production from the inflaton decay can be regarded as a non-thermal production mechanism, which does not depend on the temperature of the thermal bath. That is why we do not observe a significant difference in the relation between $\mathcal{C}_X^\phi$ and $m_X$ in the \textit{massless} and \textit{massive} reheating scenarios. Rather small deviations are observed only for the $n\!=\!1$ and $n\!=\!2/3$ curves and result from the fact that in these two scenarios, DM relic abundance is sensitive to the number of e-folds during reheating phase. Since in the \textit{massless scenario} reheating ends earlier, DM particles are created for a shorter period. Thus, for a given mass of DM species, one needs a slightly larger value of $\mathcal{C}_X^\phi$ to reproduce observed DM relic abundance in this case. 


\subsection{Higgs boson decay}
Let us now discuss the DM production mechanism that emerges from the inflaton-induced EW symmetry breaking during the reheating phase. 
As it was already pointed out before, in the \textit{massive reheating scenario} the Higgs boson develops a time-dependent vev due to its coupling to the oscillating inflaton. The DM--Higgs effective contact interaction term, $\mathcal{C}_X^{{\bm h}}/2\,(m_X^2/\mpl^2) \lvert {\bm h} \rvert^2\, X_{\mu}X^{\mu} $, expanded around $v_h$ in the EW symmetry broken phase provides a cubic term, $\cxh/2 \, (m_X/\mpl)^2 v_h h_0 X_{\mu}X^{\mu}$, that accounts for the direct Higgs field decay to the vector DM. 
The resulting time-averaged Higgs decay width $\langle \Gamma_{h_0}^{XX} \rangle\!\equiv\! \langle \Gamma_{h_0\to XX} \rangle$ is given by
\begin{align}
\langle \Gamma_{h_0}^{XX} \rangle  \!= \!\frac{\lvert \cxh \rvert^2}{128\pi} \bigg\langle \!\frac{ v_h^2 m_{h_0}^3}{\mpl^4}\Big(\! 1 -\frac{4m_X^2}{m_{h_0}^2}+\frac{12m_X^4}{m_{h_0}^4}\!\Big)\!\sqrt{1- \frac{4 m_X^2}{m_{h_0}^2}}\!\bigg\rangle.		\label{eq:Gamma_h_XX}
\end{align}
Let us emphasize that the $h_0$ decays to DM vectors can only occur during one-half of the inflaton oscillations period, i.e., when $\mathcal{P}\! <\!0$ and the Higgs boson acquire a non-zero vev~\eqref{eq:Higgs_vev}. Moreover, we should note that such interactions are only allowed during the reheating phase. After the end of reheating, $v_h$ rapidly drops as $v_h \propto |\phi|$, and the $h_0 \rightarrow XX$ channel is switched off as the Higgs boson goes into the EW symmetric phase. Thus, DM vectors are produced from the SM Higgs field in the period between $a_e$ and $a_{\rm{rh}}$.
After that, the comoving DM number density $N_X$ produced in the Higgs decays becomes constant, i.e.,
\begin{align}
N_X(a_0) \simeq N_X(a_{\rm{rh}}) = \int_{a_0}^{a_{\rm rh}} da \frac{a^2}{H(a)} \mathcal{D}_{h_0}, \label{eq:NXHiggs}
\end{align}
where
\begin{align}
\mathcal{D}_{h_0} &\equiv \bar{n}_{h_0} \langle \Gamma_{h_0}^{XX} \rangle_{\rm{th}}= \langle \Gamma_{h_0}^{XX} \rangle \frac{K_1(m_{h_0}/T)}{K_2(m_{h_0}/T)},
\end{align}
and $\langle \Gamma_{h_0}^{XX} \rangle_{\rm th}$ is the thermally-averaged Higgs decay width.
We should point out that the amount of DM produced from the Higgs decay strongly depends on the relation between Higgs mass $m_{h_0}$ and SM bath temperature $T$. The equilibrium number density of the Higgs particles, $\bar{n}_{h_0}$, is exponentially suppressed in the non-relativistic limit, i.e., $m_{h_0} \! \gg \! T$. However, since the inflaton-induced Higgs boson mass typically decreases more rapidly than the temperature $T$, therefore the DM production rate grows with time. Thus, the main contribution to the integral in Eq.~\eqref{eq:NXHiggs} comes from the moment at which the Higgs mass becomes comparable with the thermal bath temperature, i.e., $m_{h_0}\!\sim\! T$.

The present-day relic density of the DM species produced from the Higgs decay, $\Omega_X^{\bm h} h^2$, can be calculated analogously to the mechanisms discussed above.  The contribution of $\Omega_X^{\bm{h}} h^2$ to the total DM abundance is presented in Fig.~\ref{fig:FinalScan} in dashed colored curves, which we discuss in the following subsection.

\subsection{Higgs portal freeze-in production}
Let us now explore a standard freeze-in scenario in which DM particles with negligible initial density are gradually produced from the SM thermal bath in the early universe. In particular, we focus on a scenario in which  DM vectors are pair-produced from Higgs particles via the effective dim-6 contact operator $\cxh\, m_X^2/(2\mpl^2) X_\mu X^\mu |{\bm h}|^2$ \citep{Kolb:2017jvz}. As we have already pointed out above, in the \textit{massive reheating scenario}, the Higgs portal generates the 3-point vertex $h_0 \rightarrow XX$, which not only accounts for the direct Higgs decays but also enables massive SM particles to annihilate into DM vectors through the $ {\rm SM\, SM} \rightarrow h_0 \rightarrow XX$ channel. It should be noted that such interactions might occur during reheating only in the broken phase when $v_h \!\neq \!0$. Furthermore, we assume that DM interactions with the SM sector are feeble such that vector DM can neither reach thermal equilibrium with themselves nor with the SM thermal bath due to the Planck mass suppress effective operator. As a result, the $XX \rightarrow {\rm SM\, SM}$ channel is almost completely turned off, and thus, DM particles, produced in the reverse process, accumulate as the Universe expands, i.e., the freeze-in mechanism.
In the rest part of this section, we will discuss solutions to the Boltzmann equation \eqref{eq:Xeq}, assuming that DM species are only produced from the annihilation of SM particles, neglecting contribution from the inflaton decay, i.e., keeping $\mathcal{C}_X^\phi \!=\!0$. Additionally, we impose gravitational constraints, obtained in the first two subsections, on the final results. 

The cross-section that accounts for all considered diagrams of the type $XX \rightarrow {\rm SM\, SM}$ can be written as a sum of four contributions
\begin{align}
\sigma_{XX \rightarrow {\rm SM} \, {\rm SM}} &= \sigma_{XX \rightarrow h_0 h_0} + (n_0-1)  \sigma_{XX \rightarrow h_j h_j} \non \\
&+ n_{1/2}\, \sigma_{XX \rightarrow \bar\psi\psi} + n_1 \sigma_{XX \rightarrow VV}.
\end{align}
Note that the $\sigma_{XX \rightarrow h_0 h_0}$ term contains contributions from the  $ X X \rightarrow h_0 h_0$ process, s-channel Higgs exchange $ X X \!\rightarrow \!h_0^\star\! \rightarrow\! h_0 h_0$ and the interference term. The second term in the above expression, $ \sigma_{XX \rightarrow h_j h_j}$, accounts for the contact diagram $ X X \rightarrow h_j  h_j$ with $j \! \neq\! 0$. While $\sigma_{XX  \rightarrow \bar\psi\psi}$ and $\sigma_{XX \rightarrow VV}$ describe the DM annihilation into massive SM fermions and vectors through the Higgs exchange, respectively (see \fig{fig:feyn_SM_DM} second line). 
Note, however, that we neglect the interference terms between the Higgs portal and graviton exchange ${\rm SM\, SM}\to XX$ processes. As discussed in the previous section, we treat the graviton-mediated freeze-in production as an independent mechanism as it only depends on the gravitational interactions. Whereas, the Higgs portal interaction is non-minimal and requires a UV model where the DM--Higgs effective operator is generated.
Let us start our discussion with the contact Higgs-DM interaction which is present in both considered reheating scenarios.  The spin-averaged annihilation cross-section for the $XX \rightarrow h_i h_i$ diagram takes the following form
\begin{align}
\sigma_{XX \rightarrow h_i h_i} &\!=\! \frac{1}{9} \frac{\lvert \mathcal{C}_X^{{\bm h}} \rvert^2}{128 \pi} \frac{s}{\mpl^4}\bigg(\!1- \frac{4 m_X^2}{s}+\frac{12 m_X^4}{s^2} \bigg)\sqrt{\frac{s- 4 \bar{m}_{h_i}^2}{s- 4 m_X^2}}.
\end{align}
Note that for $m_{h_i} \neq 0$ the integral over $s$ in Eq.~\eqref{eq:tacs} cannot be calculated analytically. 

In the {\it massless reheating scenario}, we get the following thermally-averaged cross-section,
\begin{align}
\langle \sigma \lvert v \rvert \rangle_{XX \rightarrow  h h} & \overset{m_{h_i} =0 }{=} \frac{3 \lvert \mathcal{C}_X^{{\bm h}} \rvert^2 }{64 \pi^5\mpl^4} \frac{m_X^6 T^2}{\bar{n}_X^2} \bigg[\frac{4T^2}{m_X^2} K_2^2\!\Big(\! \frac{m_X}{T}\!\Big)		\notag\\
&+K_1^2\!\Big(\! \frac{m_X}{T}\!\Big)+ \frac{2T}{m_X} K_1\!\Big(\! \frac{m_X}{T}\!\Big) K_2\!\Big(\! \frac{m_X}{T}\!\Big) \bigg].
\end{align}
In the above expression we have factored out a factor of 4, which arises from the fact that for temperatures higher than the temperature of electro-weak symmetry breaking, we have $\sum_{i=0}^3 \langle \sigma \lvert v \rvert \rangle_{XX \rightarrow h_i h_i} = 4 \langle \sigma \lvert v \rvert \rangle_{XX \rightarrow h_0 h_0} \equiv \langle \sigma \lvert v \rvert \rangle_{XX \rightarrow h h}$.
In this case, the source term,\mbox{ $\mathcal{S}_{XX \rightarrow  hh} \equiv \langle \sigma \lvert v \rvert \rangle_{XX \rightarrow h h} \bar{n}_X^2 $}, can be approximated as
\begin{align}
\mathcal{S}_{XX \rightarrow  hh} \! \overset{m_{h_i} =0 }{ \simeq }\!
\frac{3 \lvert \mathcal{C}_X^{{\bm h}} \rvert^2}{128 \pi^4\mpl^4} 
\!\begin{dcases}
\!\frac{32}{\pi} T^8, &  m_X\! \ll\! T,	 \\ 
 \!m_X^5  T^3\, e^{-2 m_X/T} , &  m_X \!\gg\! T.
\end{dcases}	\label{eq:STerm0XXHH}
\end{align}
From the above formula, it is clear that efficient DM production can only occur in the early Universe when the temperature of the thermal bath exceeds the DM mass.  Moreover, this mechanism is most efficient at the maximal temperature $T_{\rm max}$, which, in the standard (massless Higgs boson) scenario, is obtained shortly after the end of inflation, see \eq{eq:Tmax0}. 

\begin{figure}[t!]
\begin{center}
\includegraphics[width=0.5\linewidth]{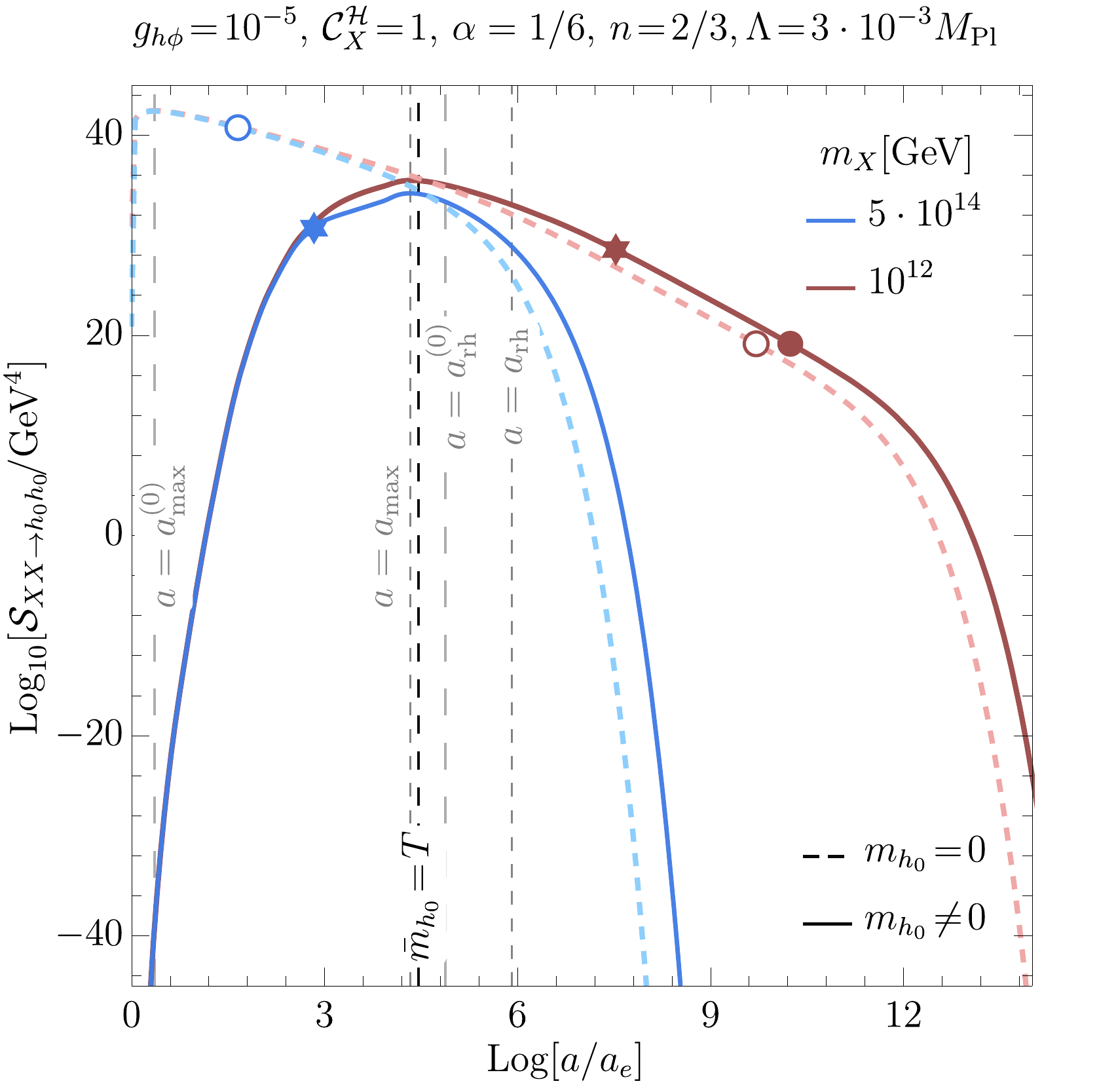}\!\!\!
\includegraphics[width=0.5\linewidth]{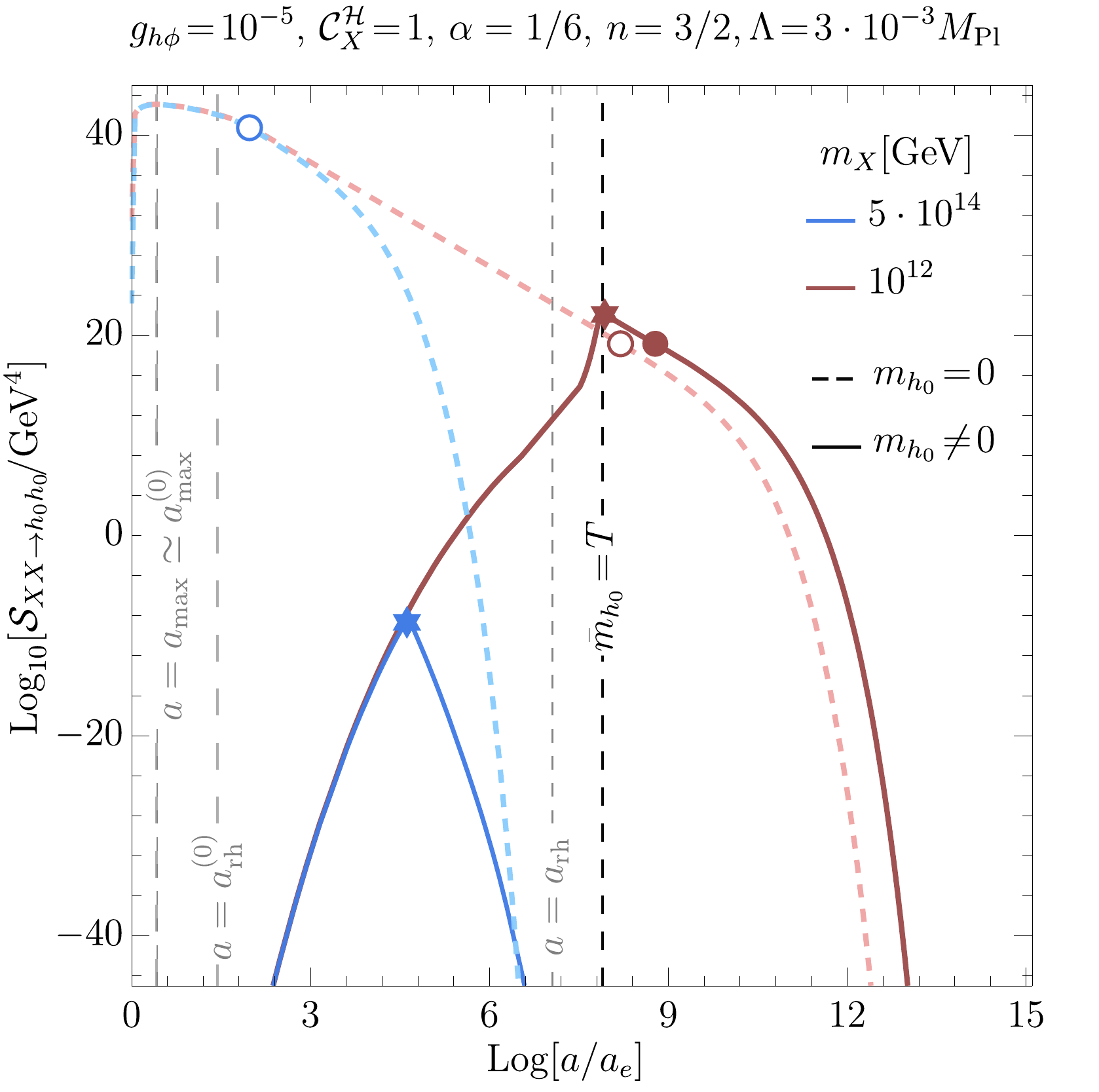}
	\caption{ Evolution of the source term $\mathcal{S}_{XX \rightarrow h_0 h_0} (a)$ for the massive (solid) and massless (dashed) Higgs field for $n=2/3$ (left panel) and $n=3/2$ (right panel). In both panels blue (red) curves are for $m_X\!=\! 5 \cdot 10^{14} \, {\rm GeV}$ ($m_X\!=\!  10^{12} \, {\rm GeV}$), other parameters are specified above each plot. Dashed gray lines $a\!=\!a_{\rm max}$ ($a\!=\!a_{\rm max}^{(0)}$) indicate the moment of time at which $T$ is maximal in the massive (massless) reheating scenario, while black dashed lines signalize $\bar{m}_{h_0}(a)\!=\!T(a)$.
Filled (empty) dots indicate $m_X=T(a)$ equality, for the massless (massive case), respectively, whereas filled stars represent the moment at which $m_{h_0}(a)\!=\!m_X$.}
	\label{fig:STXXh0h0}
	\end{center}
\end{figure}

The evolution of the source term $\mathcal{S}_{XX \rightarrow h_0 h_0}$ as a function of the scale factor $a$ for two values of $n$, i.e., $n\!=\!2/3$ (left panel) and $n\!=\!3/2$ (right panel), and for two DM masses $m_X \!= \! 5 \cdot 10^{14} \,\rm{GeV}$ and $m_X \!= \! 10^{12} \,\rm{GeV}$ is shown in Fig.~\ref{fig:STXXh0h0}. As we can see from these two figures,  in the \textit{massless scenario} (dashed curves), the source term has a maximum at $a\!=\!a_{\rm max}^{(0)}$, i.e., at the moment at which the temperature is the highest. Note that we have used DM mass such that $m_X \! < \! T_{\rm max}^{(0)}$. Following approximation \eqref{eq:STerm0XXHH}, $S_{XX \rightarrow  hh}$ evolves independently of the DM mass at the onset of the reheating period, as long as $m_X \!<\! T_{\rm max}^{(0)}$.  When the thermal bath temperature drops below the DM mass $m_X$, the source term becomes exponentially suppressed and DM particles are no longer efficiently produced. The temperature-DM mass equality (empty circles) can happen both during reheating (e.g., for $m_X\!=\! 5 \cdot 10^{14}\, {\rm GeV}$ and $n\!=\!2/3$) and after (e.g., for $m_X\!=\!  10^{12}\, {\rm GeV}$ and $n\!=\!2/3$ or $n\!=\!3/2$). In both cases, the comoving number density of DM species $X$ becomes constant quickly after this moment. For the sake of simplicity, in the following analytical approximations, we assume that the main contribution to the comoving number density of vector DM comes from the reheating period. However, in numerical calculations, we also take into account the evolution of the source term after the end of reheating.

\begin{figure}[t!]
\begin{center}
\includegraphics[width=0.5\linewidth]{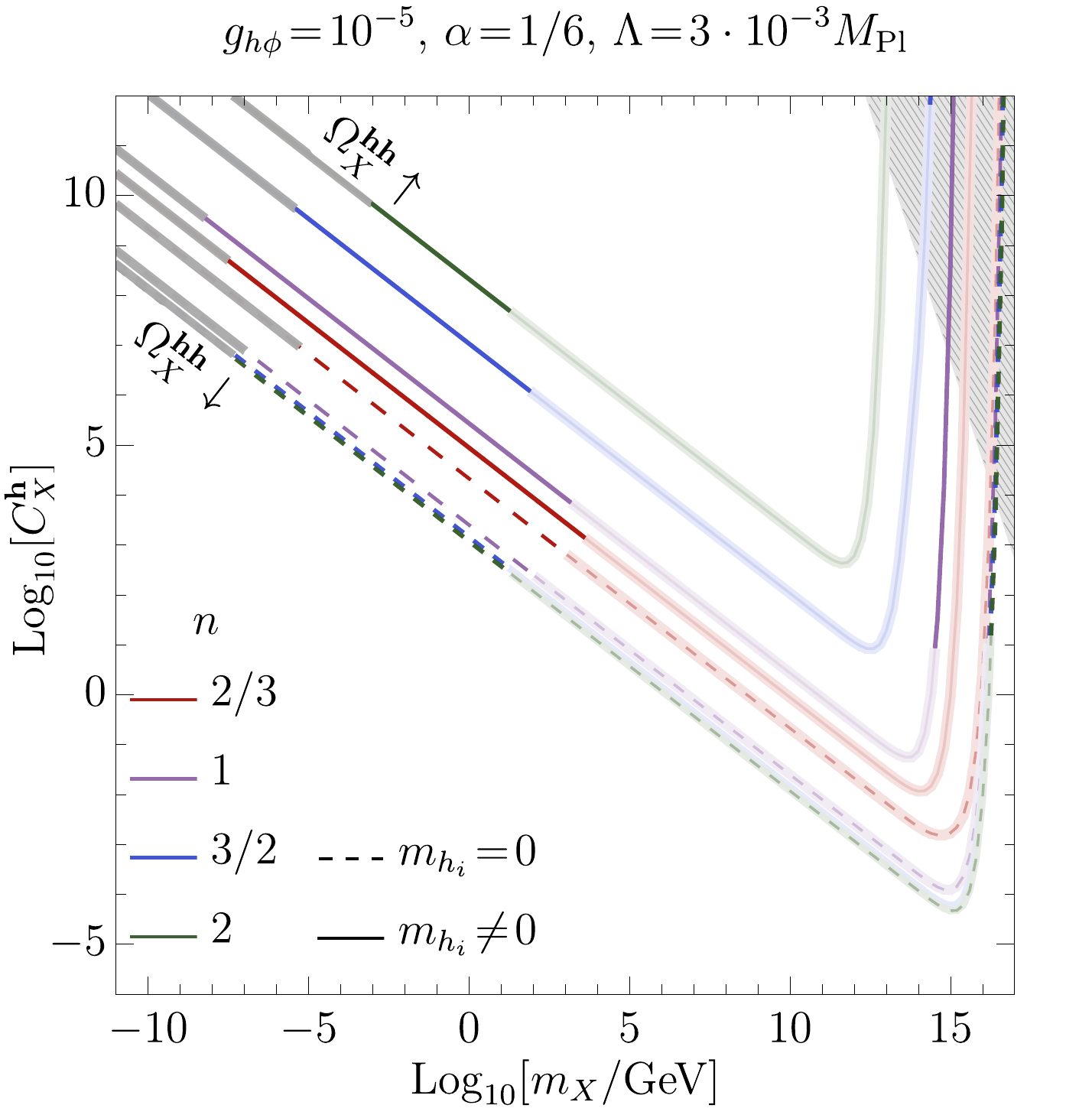}\!\!\!
\includegraphics[width=0.5\linewidth]{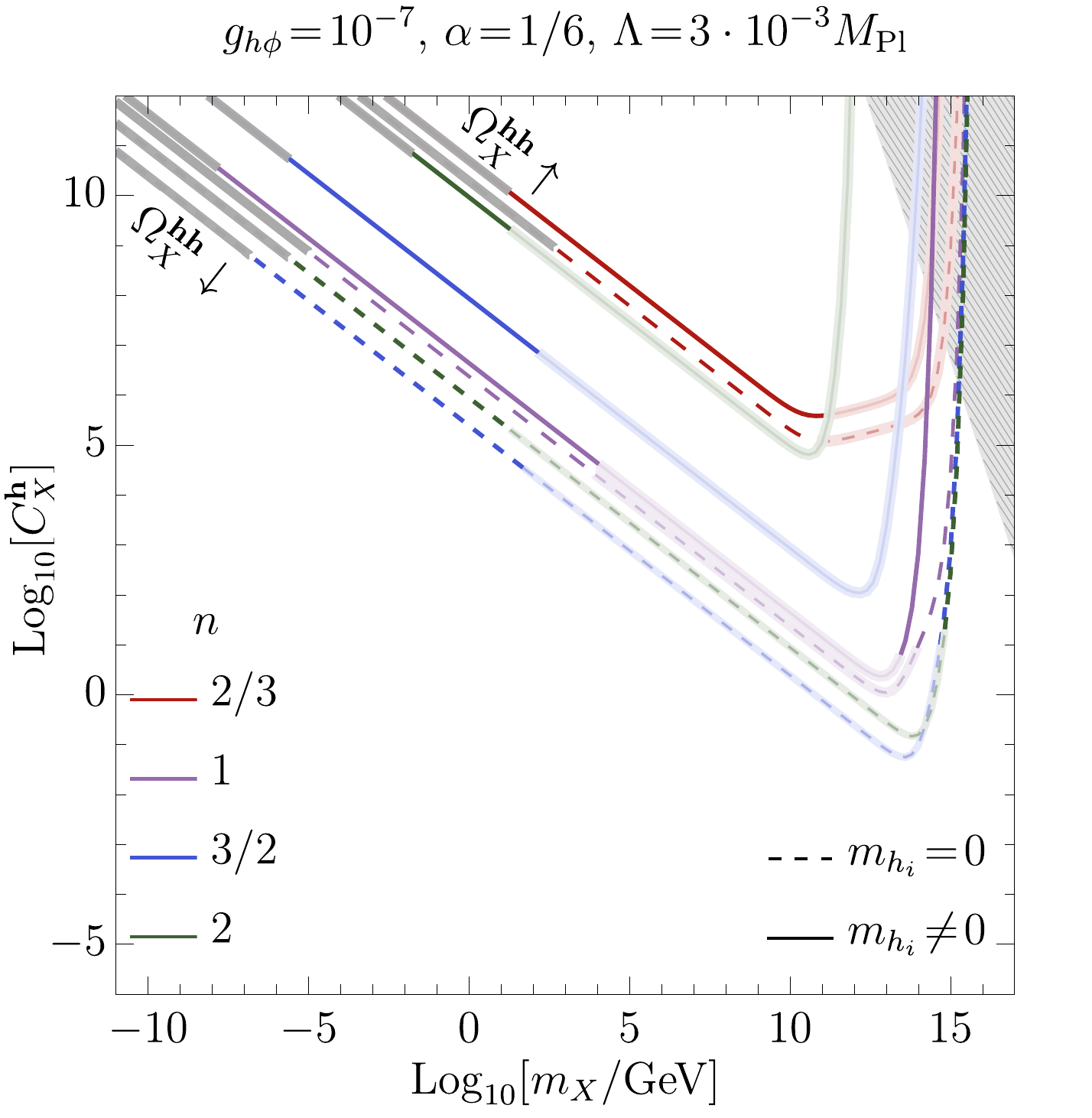}
	\caption{The Wilson coefficient $\mathcal{C}_X^{{\bm h}}$ as a function of $\mx$ for $\ghp=10^{-5}$ (left panel) and 
	$\ghp=10^{-7}$ (right panel) consistent with the requirement of observed DM abundance. Solid (dashed) lines satisfy $\Omega_X^{\rm obs} h^2$ constraint assuming that the Higgs field is massive (massless) during reheating. In gray hatched region $\mathcal{C}_X^{{\bm h}}$ exceeds the perturbativity limit \eqref{cxp_lim}. Solid (dashed) lines are constrained by the corresponding faded regions due to gravitational production. }
	\label{fig:CXHmx XXh0h0}
	\end{center}
\end{figure}

We assume that the comoving number density of vector DM, approaches a constant value around the end of reheating phase, which implies the following formula for the present DM number density,
\begin{align}
n_X^{{\bm h}{\bm h}} (a_0) &\! \overset{m_{h}\!=\! 0}{\simeq} \frac{\sqrt{3}(n+1) |\mathcal{C}_X^{{\bm h}}|^2}{4 \pi^5(2n-1)} \frac{(T_{\rm{rh}}^{(0)})^8}{\mpl^3 \sqrt{\rho_{\phi_e}  }} \non \\
&\quad\times \Big( \frac{a_{\rm{rh}}}{a_e} \Big)^{\!\frac{3(1-n)}{1+n}} \Big( \frac{a_{\rm rh}}{a_0} \Big)^3  \bigg[ \Big(\frac{a_{\rm rh}}{a_e} \Big)^{\!\frac{3(2n-1)}{n+1}} -1 \bigg].
\end{align}
The above result holds for vector DM mass $m_X \! < \! T_{\rm max}^{(0)}$. Note the similarity between the above equation and \eq{nX0}, obtained from the gravitational annihilation. If $\mathcal{C}_X^{{\bm h}} \simeq 1$, the strength of these two interactions is comparable, and thus by adjusting the value of the Wilson coefficient, one can control the production of DM. In particular, we can estimate the value of the $\mathcal{C}_X^{{\bm h}}$ coupling for which the gravitational and the Higgs portal DM production are equal, i.e.
\begin{align}
 \frac{n^{{\bm h}{\bm h}}_X (a_0)}{n^{{\rm gr}, \rm{SM}}_X (a_0)}=1 \implies \mathcal{C}_X^{{\bm h}} \lvert_{\rm{eq}} = \sqrt{\frac{A}{30}},
\end{align}
where $A\!\sim\!\op(10)$ factor. 
Thus, for a DM vector with mass $m_X \! < \! T_{\rm max}^{(0)}$, if $\mathcal{C}_X^{{\bm h}}$ exceeds $\mathcal{C}_X^{{\bm h}} \lvert_{\rm{eq}} $ the contact diagram ${\bm h}{\bm h} \rightarrow XX$ dominates over the gravitational annihilation of SM particles, while if $\mathcal{C}_X^{{\bm h}} \!<\!\mathcal{C}_X^{{\bm h}} \lvert_{\rm{eq}} $, the contribution from the Higgs portal is negligible.
The relic density of DM particles produced through the Higgs portal is given by
\begin{align}
\Omega_X^{{\bm h}{\bm h}} &\simeq \frac{\sqrt{3}(n+1) |\mathcal{C}_X^{{\bm h}}|^2}{4 \pi^5(2n-1)} \frac{(T_{\rm{rh}}^{(0)})^8}{\mpl^3 \sqrt{\rho_{\phi_e}  }}\frac{s_0}{s_{\rm rh}\,\rho_c} \Big( \frac{a_{\rm rh}}{a_{e}} \Big)^{\!\frac{3n}{n+1}},
\end{align}
for $n\!>\!1/2$.
For a given DM mass $m_X$ one can find the value of the Wilson coefficient $ |\mathcal{C}_X^{{\bm h}}|^2$ such that $\Omega_X^{{\bm h}{\bm h}}  \!= \! \Omega_X^{\rm{obs}}$. The relation between $ |\mathcal{C}_X^{{\bm h}}|^2$ and $m_X$ that gives the observed DM relic density is plotted in Fig.~\ref{fig:CXHmx XXh0h0}.

In this case of freeze-in from the {\it massive reheating scenario}, we should note two non-trivial aspects of the non-zero inflaton-induced Higgs boson mass. First, as demonstrated in the previous section, the inflaton-induced Higgs mass significantly affects the dynamics of reheating, modifying the evolution of the thermal bath temperature and suppressing the energy accumulated in the radiation sector. 
Secondly, at the onset of the reheating period, $m_h$ typically exceeds the SM bath temperature~$T$. In this case, even if $m_X \! < \! T_{\rm{max}}$, one expects similar exponential suppression in the source term as was observed for $m_X \!\gg\! T_{\rm max}$. Such suppression results from the lower limit of the $s$-integral in Eq.\eqref{eq:tacs}, which for $m_X \! < \! m_h$ is equal $4 \bar{m}_{h_i}$. Since the Higgs mass term decreases with time, the suppression of $\mathcal{S}_{XX \rightarrow hh}$ is the strongest at the beginning of reheating. This, in turn, results in a significant reduction of DM production compared to the \textit{massless reheating scenario}, in which the maximum temperature and thus the most efficient production is reached just after the onset of reheating phase. In the \textit{massive reheating scenario}, depending on the value of $g_{h \phi}$ and $n$, the maximal temperature is obtained at the very beginning of reheating or slightly later.  

Let us first consider the case for which the maximum temperature is obtained shortly after the start of reheating, which happens for $n\!=\!3/2$, cf. left panel of Fig.~\ref{fig:STXXh0h0}. At that moment, the averaged Higgs mass $\bar{m}_{h_0}$ exceeds $T$, and DM particles cannot be efficiently produced due to the exponential suppression in the source term. If the mass of DM species is larger than the reheating temperature $T_{{\rm rh}}$ (solid blue curve in \fig{fig:STXXh0h0}),  the source term $\mathcal{S}_{XX \rightarrow h_0 h_0}$ rapidly increases until the moment at which $\bar{m}_{h_0}$ drops below $m_X$ (blue star) and then decreases as $e^{-m_X/T}$. For DM particles with mass smaller than $T_{\rm rh}$ (solid red curve), $\mathcal{S}_{XX \rightarrow h_0 h_0}$ reaches its maximum value at the moment when $m_{h_0}$ crosses $T$ (red star), which in this case occurs right after the end of reheating. 
After that, the source term decreases as $\propto T^{8}$ until the moment at which DM particles become non-relativistic (red, filled circle). On the other hand, the dependence of $\mathcal{S}_{XX \rightarrow h_0 h_0}$ on the scale factor $a$ looks slightly different if the maximum temperature is not reached at the very beginning of reheating, cf. right panel of Fig.~\ref{fig:STXXh0h0}.  For $n\!=\!2/3$ and $m_X\!>\! T_{\rm max}$ (blue solid curve), the source term grows exponentially until $\bar{m}_{h_0}$ crosses $m_X$ (blue star). Then, $\mathcal{S}_{XX \rightarrow h_0 h_0}$ evolves as a power-low, $\mathcal{S}_{XX \rightarrow h_0 h_0} \propto T^{8}$, up to the point at which $T\!=\!T_{\rm max}$. After that, the source term receives strong exponential suppression. For DM particles with a mass smaller than $T_{\rm max}$ (solid, red curve),  the peak of the $\mathcal{S}_{XX \rightarrow h_0 h_0}$ term is reached at $T\!\sim\!m_{h}$. Then, it decreases as $T^8$ until the temperature drops below $m_X$ (red, solid circle). The source term receives an exponential Boltzmann suppression from that moment, and the DM production again becomes inefficient. 

The predicated abundance of DM species in the \textit{massive reheating scenario} can be estimated numerically, analogously to the \textit{massless reheating} case discussed above. The final results are presented in Fig.~\ref{fig:CXHmx XXh0h0}, which shows the relation between $m_X$ and $\mathcal{C}_X^{{\bm h}}$ required to achieve the observed DM relic density in both considered scenarios. For a given value of $g_{h \phi}$ and $n$, colored solid (dashed) curves give the correct abundance in the \textit{massive (massless)} Higgs reheating model. In the gray hatched region in the upper right corner, the value of the Wilson coefficient exceeds the perturbativity limit $(\mpl / m_X)^2$. Several curves are also constrained by the condition $ \mathcal{C}_X^{{\bm h}} \! <\! (\mpl/T_{\rm max})^2$ (the thick grey part of the lines). Moreover, the faded parts of the curves show regions of DM masses where vector DM is overproduced due to pure gravitational interactions. 

Note that for a fixed DM mass, a larger value of the $\mathcal{C}_X^{{\bm h}}$ coupling is needed to obtain the observed DM relic abundance in the \textit{massive reheating scenario} in comparison to the \textit{massless reheating scenario}. Hence, all dashed curves lie below the corresponding solid ones. This behavior results from the fact that in the \textit{massive reheating scenario} DM production is Boltzmann suppressed and occurs in the lower temperatures than in \textit{massless reheating scenario}. Furthermore, decreasing the value of the inflaton-Higgs coupling leads to a cutdown of the allowed parameter space in both considered scenarios. For weaker inflaton-Higgs coupling, the efficiency of the reheating process is reduced, which implies a lower temperature of the thermal bath. Thereby all the curves move up towards a larger value of $\mathcal{C}_X^{{\bm h}}$. Moreover, in the \textit{massless reheating scenario}, for a given $m_X$ the largest (smallest) value of $\mathcal{C}_X^{{\bm h}}$ is required for the $n\!=\! 2/3$ $(n\!=\!2)$ cosmology to produce the correct amount of DM abundance. In contrast, in the \textit{massive reheating scenario}, we observe different hierarchies in the position of the solid curves. Finally, in both considered scenarios, for DM particles with mass $m_X \!<\! T_{\rm max}$, the maximal value of the source term ${\cal S}_{XX\rightarrow h h}$ does not depend on $m_X$. This, in turn, means that the relic abundance $\Omega_X^{{\bm h}{\bm h}} h^2$ of DM species is proportional to the vector DM mass for $m_X \!<\! T_{\rm max}$.  Contrarily, for $m_X \!>\! T_{\rm max}$, the Boltzmann suppression of the source term causes a  turnover of the curves around $m_X \simeq T_{\rm max}$. Consequently, a larger value of the Wilson parameter $\mathcal{C}_X^{{\bm h}}$ is required to produce heavy DM particles.

Finally, we move on to the last DM production mechanism discussed in this work, i.e., the freeze-in production from massive SM particles mediated by a Higgs particle $h_0$~\footnote{The unitary gauge is adopted here.}.  
The spin-averaged cross-section for the $XX \rightarrow h_0 h_0$, $XX \rightarrow \bar\psi\psi$, $X X \rightarrow VV$ processes are given by
\begin{widetext}
\vspace{-10pt}
\begin{align}
\sigma_{XX \rightarrow h_0 h_0} &= \frac{1}{9} \frac{\lvert \mathcal{C}_X^{{\bm h}} \rvert^2}{128 \pi} \frac{s}{\mpl^4}\bigg(\!1- \frac{4 m_X^2}{s}+\frac{12 m_X^4}{s^2} \bigg)\sqrt{\frac{s- 4 \bar{m}_{h_0}^2}{s- 4 m_X^2}}
\frac{(s+ 2 \bar{m}_{h_0}^2)^2}{(s- \bar{m}_{h_0})^2 +\bar{ m}_{h_0}^2 \bar{\Gamma}_{h_0}^2},\\
\sigma_{XX \rightarrow \bar\psi\psi} &=  \frac{1}{9} N_c^\psi \frac{\lvert \mathcal{C}_X^{{\bm h}} \rvert^2}{32 \pi} \frac{s}{\mpl^4}\bigg(\!1- \frac{4 m_X^2}{s}+\frac{12 m_X^4}{s^2} \bigg)\sqrt{\frac{s- 4 \bar{m}_{\psi}^2}{s- 4 m_X^2}}\frac{\bar{m}_\psi^2(s-4 \bar{m}_\psi^2)}{(s- \bar{m}_{h_0})^2 + \bar{m}_{h_0}^2 \bar{\Gamma}_{h_0}^2} ,\\
\sigma_{XX \rightarrow VV } &=  \frac{1}{9} \frac{\lvert \mathcal{C}_X^{{\bm h}} \rvert^2}{64 \pi\, \delta_V} \frac{s}{\mpl^4}\bigg(\!1- \frac{4 m_X^2}{s}+\frac{12 m_X^4}{s^2} \bigg)\bigg(\!1- \frac{4 m_V^2}{s}+\frac{12 m_V^4}{s^2} \bigg)\sqrt{\frac{s- 4 \bar{m}_{V}^2}{s- 4 m_X^2}} \frac{s^2}{(s- \bar{m}_{h_0})^2 + \bar{m}_{h_0}^2 \bar{\Gamma}_{h_0}^2},
\end{align}
\vspace{-10pt}
\end{widetext}
where $V=W^\pm,Z$ with $\delta_W=1$, $\delta_Z=2$ for the $W^\pm$ and $Z$ bosons, respectively. 
The time-averaged total Higgs decay width $\bar{\Gamma}_{h_0}$ can be written as a sum of three contributions,
\begin{align}
\bar{\Gamma}_{h_0}= \bar{\Gamma}_{h_0 \rightarrow \bar\psi\psi} +\bar{\Gamma}_{h_0 \rightarrow VV} +  \bar{\Gamma}_{h_0 \rightarrow XX}, 
\end{align}
where $\bar \Gamma_{h_0 \rightarrow XX }$ is given in \eq{eq:Gamma_h_XX}, and 
\begin{align}
\bar{\Gamma}_{h_0 \rightarrow \bar\psi\psi} &=\sum_{\psi} \frac{N_c^\psi}{8 \pi} \Big(\frac{\bar{m}_\psi}{\bar{v}_h} \Big)^2 \bar{m}_{h_0} \bigg(1- \frac{4 \bar{m}_\psi^2}{\bar{m}_{h_0}^2}\bigg)^{3/2},		\\
\bar{\Gamma}_{h_0 \rightarrow VV} &\!=\! \sum_{V}\frac{1}{16 \pi\,\delta_V} \frac{\bar{m}_{h_0}^3}{\bar{v}_h^2} \Big(\! 1 -\frac{4\bar{m}_V^2}{m_{h_0}^2}+\frac{12\bar{m}_V^4}{m_{h_0}^4}\!\Big)\sqrt{1-\frac{4\bar{m}_V^2}{\bar{m}_{h_0}^2}} .
\end{align}
Note that $\sigma_{XX \rightarrow h_0 h_0}$ contains contributions from the contact interaction $XX \rightarrow h_0 h_0$, the indirect interaction  $XX \rightarrow h_0^\star \rightarrow h_0 h_0$, as well as interference between these two diagrams. Moreover, we should stress that we do not consider any RGE running effects in our calculations and adopt the EW value for the Higgs quartic couplings, i.e., $\lambda_h\simeq 0.13$. During the reheating phase, SM fermions and massive vector bosons acquire a non-zero mass via a standard Higgs mechanism, 
\begin{align}
m_{\rm \tiny SM} &= m_{\rm \tiny SM}^{\rm \tiny EW} \frac{v_h}{v_{\rm \tiny EW}} =\frac{m_{\rm \tiny SM}^{\rm \tiny EW}}{v_{\rm \tiny EW}} 
\begin{dcases}
0, & \phi(t) \! > \! 0,\\
\sqrt{\frac{\lvert \mu_h^2(\phi) \rvert}{\lambda_h}},  & \phi(t) \!<\! 0,
\end{dcases}
\end{align} 
where $v_{\tiny{\rm  EW}} = 246\; {\rm GeV}$, and $m_{\rm \tiny SM}^{\rm \tiny EW}$ denotes the electro-weak mass of the SM species. After the end of reheating, when the inflaton field value reduces to a negligible amount, masses of SM particles decrease very rapidly, and DM production through the ${\rm SM} \,{\rm SM} \rightarrow h_0^\star \rightarrow XX $ channel is terminated. 

Let us also emphasize that for the EW value of the Higgs quartic coupling, a mass hierarchy in the SM sector is time-independent, and it turns out that for all considered values of $n$, $\bar{m}_V \!<\! \bar{m}_t \!<\! \bar{m}_{h_0} $. The ratio of the masses is thus clearly different from that observed today. At this point, we should notice that the mass of the SM Higgs particle is somehow fixed by the inflaton dynamics and does not depend on $\lambda_h$. On the other hand, masses of other SM species are sensitive both to the inflaton potential form and the value of $\lambda_h$. Thus, the observed EW mass hierarchy for the SM particles can be recovered by running the value of $\lambda_h$ during reheating. 

\begin{figure}[t!]
\begin{center}
\includegraphics[width=0.5\linewidth]{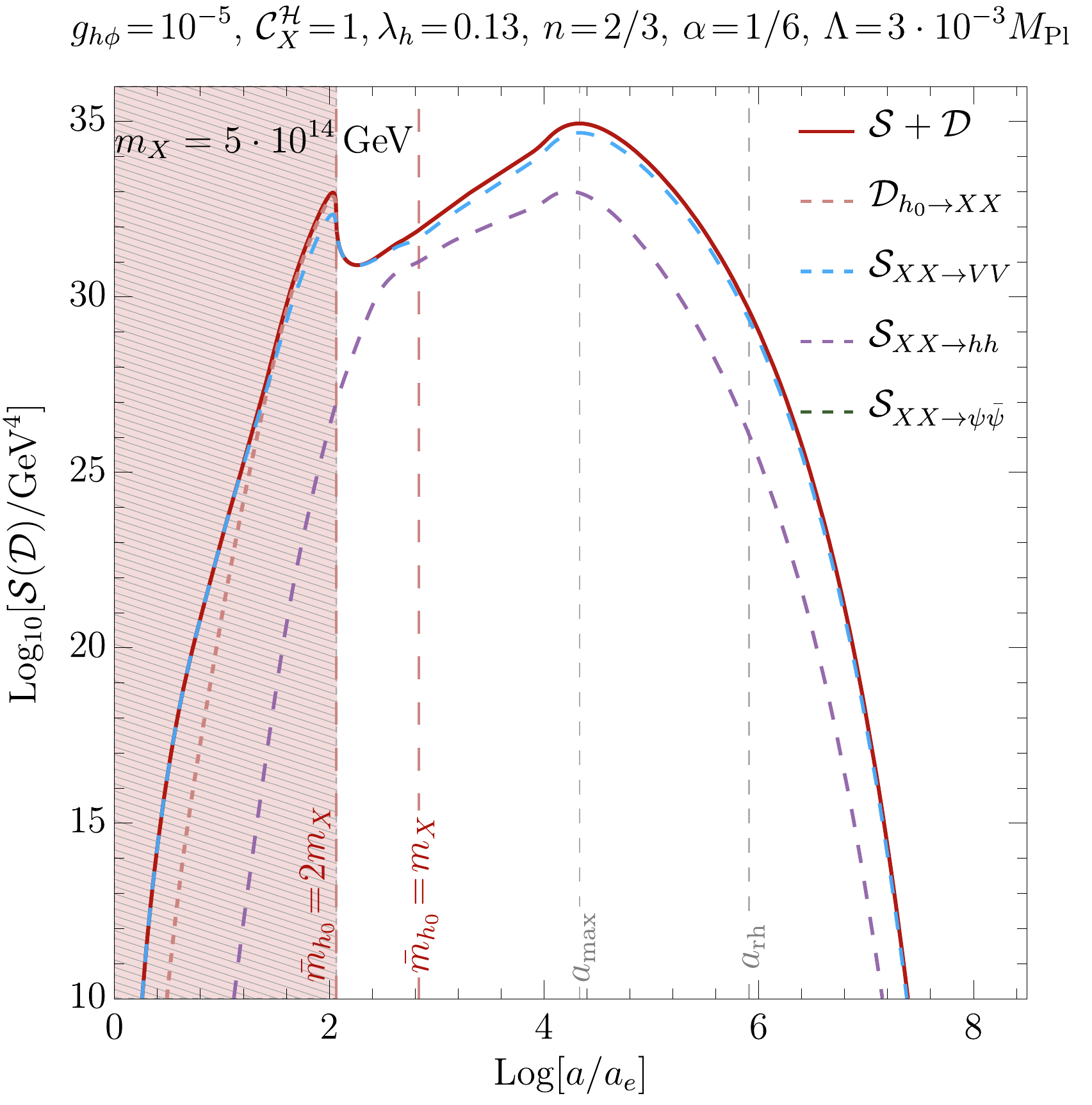}\!\!\!
\includegraphics[width=0.5\linewidth]{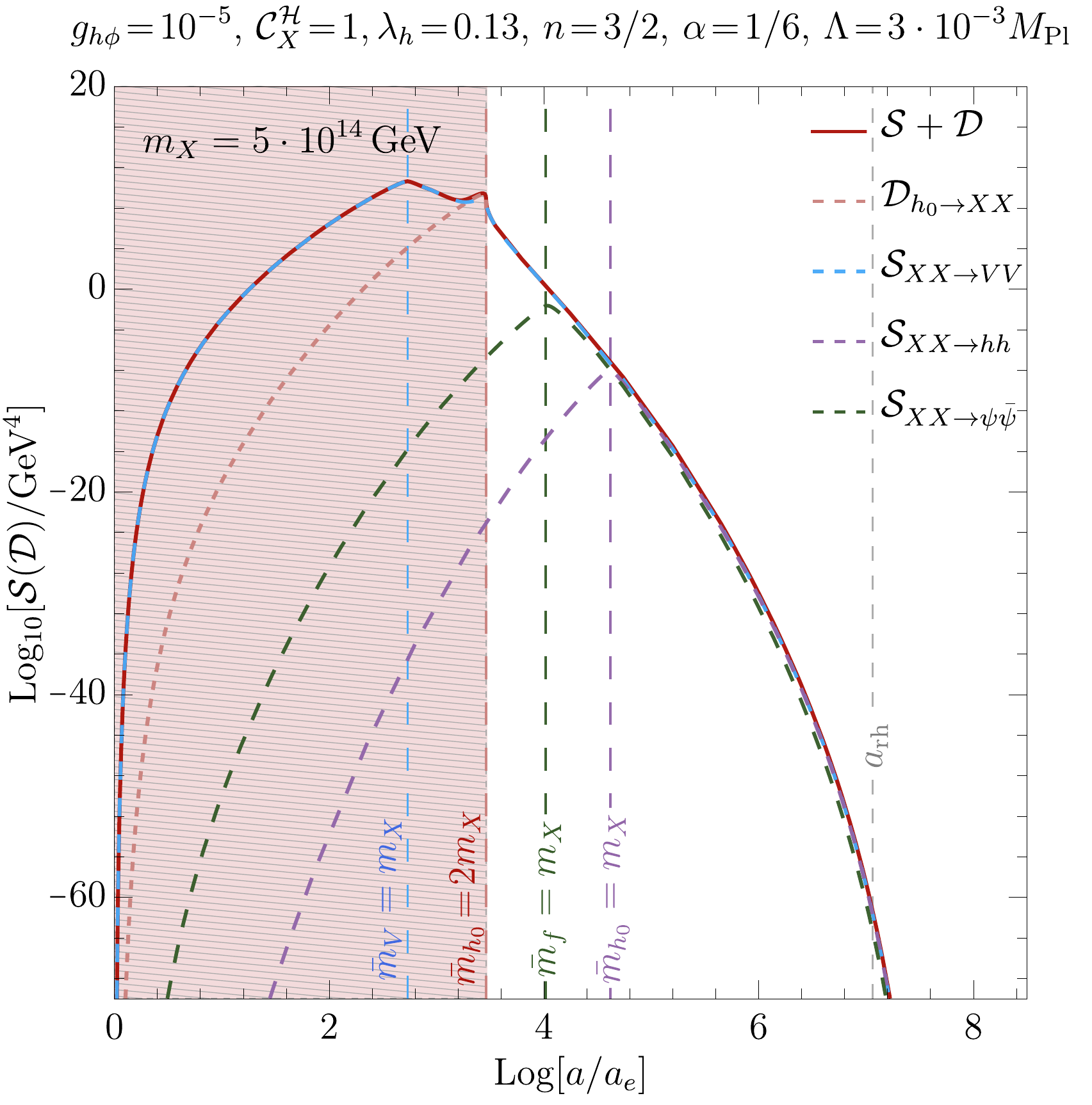}
	\caption{Evolution of the decay term, $\mathcal{D}_{h_0 \rightarrow XX}$, and various source terms, i.e., $\mathcal{S}_{XX \rightarrow VV}$, $\mathcal{S}_{XX \rightarrow \bar\psi\psi}$ and $\mathcal{S}_{XX \rightarrow  hh}$ as a function of the scale factor $a$ for $n\!=\!2/3$ (left panel) and $n\!=\!3/2$ (right panel). The Higgs boson decays are only allowed in the pale red, dashed region.}
	\label{fig:DSTerms}
	\end{center}
\end{figure}

In Fig.~\ref{fig:DSTerms} we compare the evolution of the scattering $\mathcal{S}_{XX \rightarrow VV}$, $\mathcal{S}_{XX \rightarrow \bar\psi\psi}$, $\mathcal{S}_{XX \rightarrow  hh}$, and decay $\mathcal{D}_{h_0 \rightarrow XX}$ terms for two values of $n$, i.e., $n=2/3$ (left panel) and $3/2$ (right panel) with fixed DM mass $m_X =5  \cdot 10^{14} \; {\rm GeV}$. At the beginning of reheating for $n=2/3$, DM particles are produced mainly by the Higgs decay and annihilation of SM vectors. When the mass of the Higgs field drops below DM mass threshold, the total source term, $\mathcal{S} + \mathcal{D}$, becomes dominated by the $\mathcal{S}_{XX \rightarrow  VV}$ term. Slightly after that moment, $\bar{m}_{h_0}$ equalizes with $m_X$ and $\mathcal{S}_{XX \rightarrow h h}$ become of the same order as $\mathcal{S}_{XX \rightarrow VV}$. Then, we restore the standard evolution of the source term, i.e., $\mathcal{S} + \mathcal{D}$, peaks at $a=a_{\rm max}$ when the temperature of the thermal bath reaches its maximum value $T_{\rm max}$ and then decreases exponentially. We also should emphasize that the source term that accounts for the $\bar\psi\psi$ annihilations into DM pairs is always subdominant since the cross-section is proportional to the fermion mass. The total source term, $\mathcal{S} + \mathcal{D}$, and thus DM production is dominated by the $h h \rightarrow XX$ and $VV \rightarrow XX$ annihilations, while the contribution from $h_0 \rightarrow XX$ decays is negligible. It turns out that this observation is also valid for lighter DM particles. For $n=3/2$, the total source term is dominated by the contribution from the $\mathcal{S}_{XX \rightarrow VV}$ term. The decay term, $\mathcal{D}_{h_0 \rightarrow XX}$, is initially subdominant and becomes of the same order as $\mathcal{S}_{XX \rightarrow VV}$ near the mass threshold. All source terms grow until the mass of SM particles that annihilate into DM vectors drops below $m_X$. After that, they receive exponential suppression.

\begin{figure*}[t!]
\begin{center}
\includegraphics[width=0.37\linewidth]{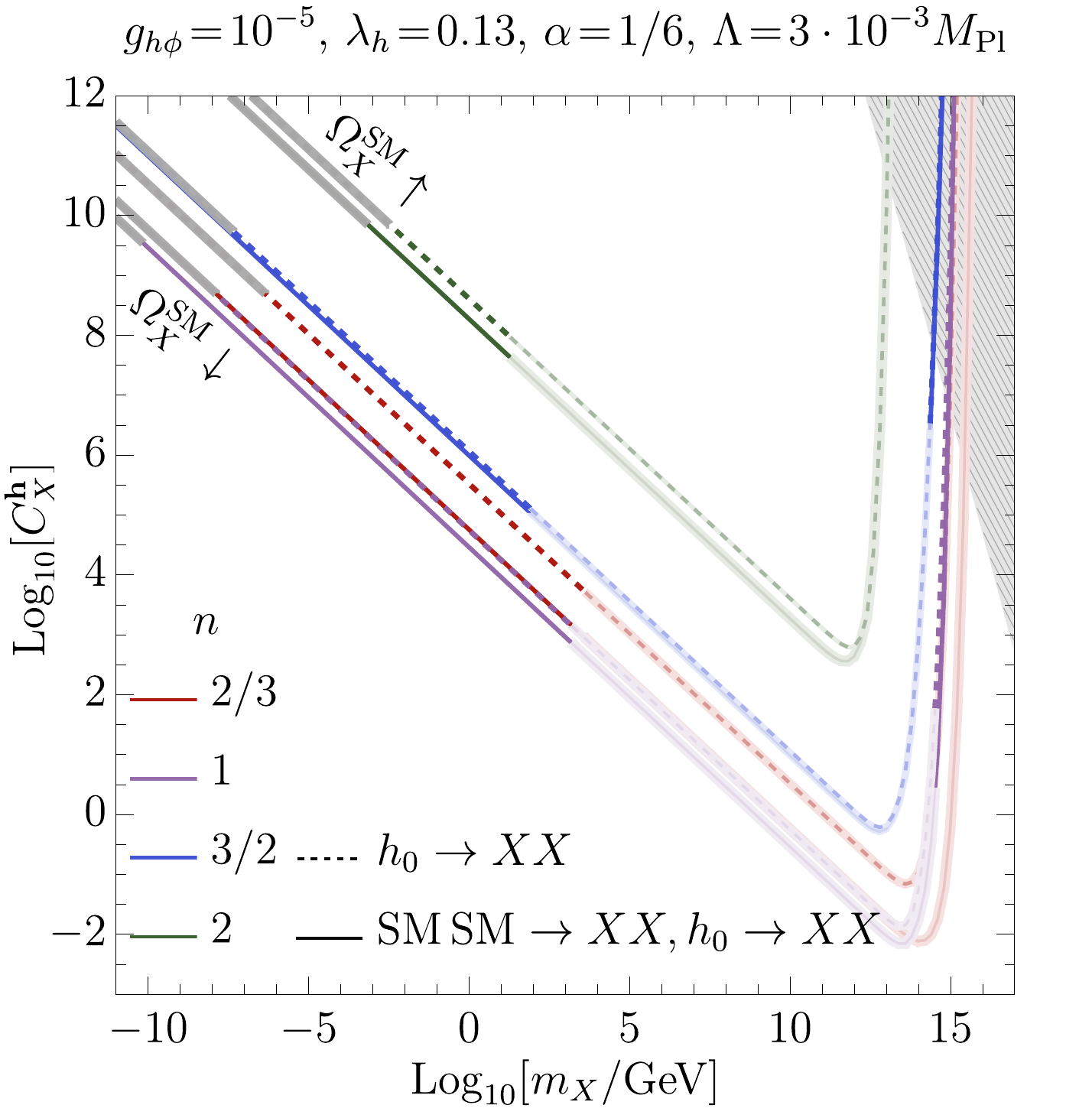}\!\!\!
\includegraphics[width=0.37\linewidth]{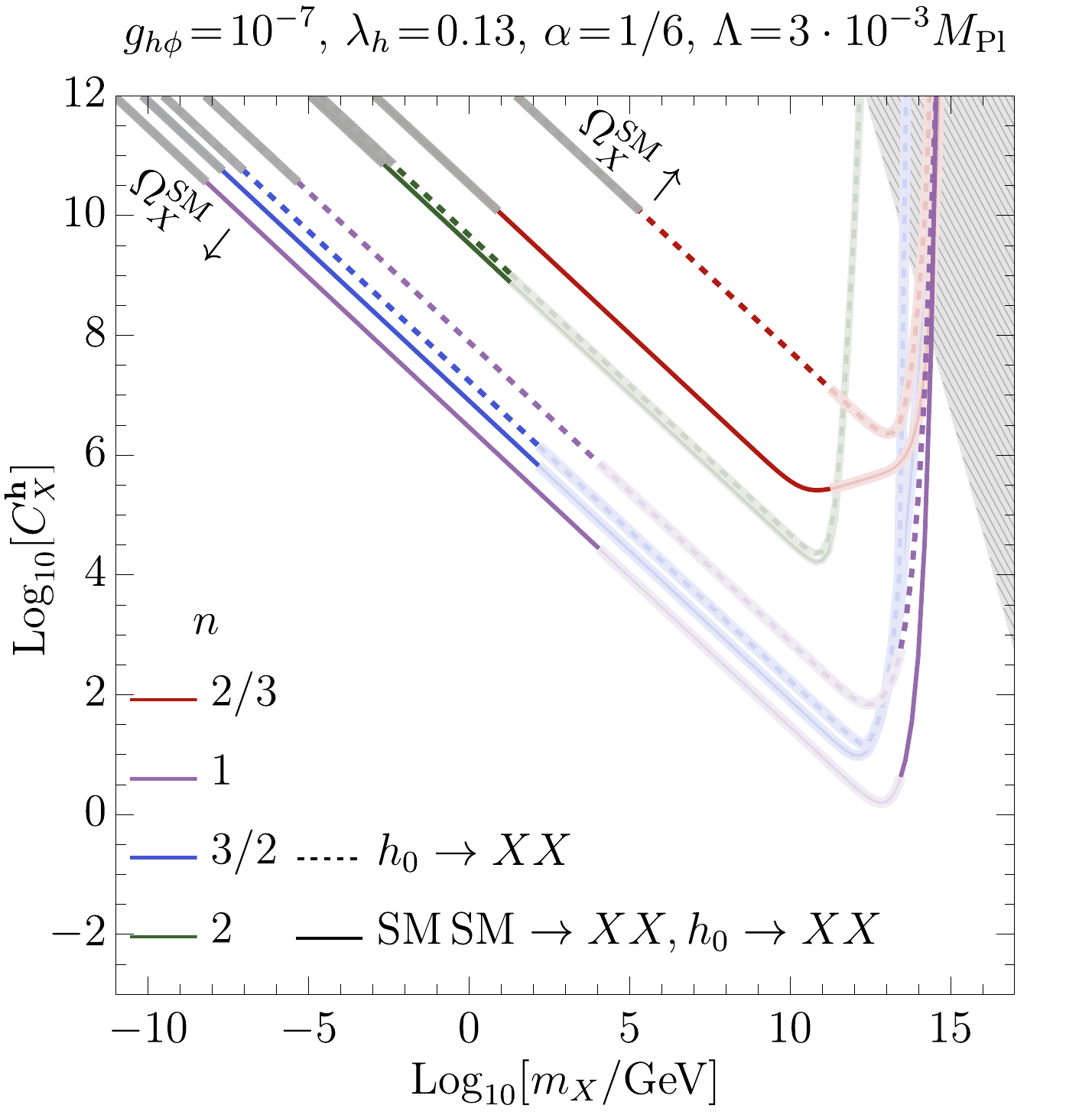}
	\caption{The Wilson coefficient $\mathcal{C}_X^{{\bm h}}$ as a function of $\mx$ for $\ghp=10^{-5}$ (left panel) and 
	$\ghp=10^{-7}$ (right panel) consistent with the requirement of observed DM abundance \eqref{eq:OmegaX_obs}. Dashed lines show the contribution of the Higgs decay to the vector DM, $h_0\to XX$, to the total DM relic abundance. In the gray hatched region $\cxp$ exceeds the perturbative limit. In these plots the faded colored parts of the curves are constrained by the pure gravitational production of DM, whereas the gray part of the curves is excluded by the condition $ \mathcal{C}_X^{{\bm h}} \! <\! (\mpl/T_{\rm max})^2$. }
	\label{fig:FinalScan}
	\end{center}
\end{figure*}
The full scan of the $\mathcal{C}_{X}^{{\bm h}} - m_X$ plane for a given $n$ and two values of the Higgs-inflaton coupling is shown in Fig.~\ref{fig:FinalScan}. Each curve determines the value of $\mathcal{C}_{X}^{{\bm h}}$ such that for a given DM mass, the total relic abundance is consistent with the observed value $\Omega_X^{\rm{obs}}h^2$. Note that here we consider only the \textit{massive reheating scenario} and take into account all interactions responsible for the DM production from the SM sector. Dashed curves present the contribution of the decay term, $\mathcal{D}_{h_0 \rightarrow XX}$, to the total DM relic density. The gray hatched region (top-right corner) $\cxp$ exceeds the perturbative limit, whereas the gray part of the curves, is excluded by requiring $ \mathcal{C}_X^{{\bm h}} \! <\! (\mpl/T_{\rm max})^2$. Moreover, the faded colored parts of the curves are constrained by the pure gravitational production of DM. To highlight the result in these plots, let us focus on the $n=1$ case with $\ghp=10^{-5}$ (left panel, purple curve). For this case the allowed vector DM mass, which produced the observed DM relic, is in the range $m_X\!\in\!(10^{-10}, 10^{3})\gev$ for the values $\cxh\!\in\!(10^{9}, 10^{3})$. Note that a smaller vector DM mass requires a larger value of $\cxh$. We point out that large values of $\cxh$ are consistent with the effective field theory since we parametrized the dim-6 effective operator in \eq{eq:Linthphi} with $1/\mpl^2$. Therefore, the real cutoff scale of new physics (NP) should be understood as $\Lambda_{\rm NP} \!\sim\!\mpl/\sqrt{\cxh}$.

\section[summary]{Summary \label{s.conclusions}}
We have investigated reheating dynamics in a system of an inflaton field~$\phi$, the SM Higgs doublet~${\bm h}$ and an Abelian vector DM~$X_\mu$. We have employed a cubic interaction of the inflaton and the Higgs boson of the form $\ghp \mpl \phi |{\bm h}|^2$ to facilitate the reheating process. To specify inflation dynamics, the $\alpha$-attractor T-model potential for the inflaton field has been adopted, which provides a relatively flat potential for large field values suitable for inflation and a monomial potential of the form $\phi^{2n}$ during the reheating phase. Our analysis is valid for generic values of $n$, however, for numerical illustrations we have considered $n=2/3,1,3/2,2$. 

One of the novel aspects discussed in this work is the inflaton-induced mass term of the Higgs boson that generates a non-trivial phase-space suppression of the reheating efficiency. As a result, the reheating period is extended, and the maximal temperature of the SM thermal bath is reduced due to the non-trivial Higgs mass, which we have dubbed as {\it massive reheating} scenario. The inflaton--Higgs coupling leads, due to oscillations of the inflaton, to periodic transitions between phases of broken and unbroken electroweak gauge symmetry. The consequences of the oscillations have been studied in detail. It turned out that Higgs mass effects are, in general, substantial and must be taken into account while investigating the reheating period in the presence of the inflaton--Higgs interactions.

As a dark matter candidate, a massive Abelian vector boson, $X_\mu$, has been adopted. Various production mechanisms of $X_\mu$ have been discussed; (i) gravitational production from the inflaton background during the reheating phase, (ii) gravitational freeze-in from the SM radiation, (iii) inflaton decay through dim-5 effective operators, and (iv) Higgs portal SM freeze-in (and Higgs decays) utilizing dim-6 effective operator.
First of all, the gravitational production of DM is treated as a background/irreducible relic from (i) and (ii) and has been adopted to exclude ranges of DM mass $\mx$ (for given values of other parameters) corresponding to the pure gravitational overproduction of DM.
Then each DM production scenario had been separately discussed, and relevant coupling constants (as a function of DM mass $\mx$) required by the observed relic abundance were calculated. For instance, for the inflaton decay (iii) or Higgs portal production mechanism (iv), the abundance condition implies a constraint on the Wilson coefficients $\cxp$ or $\cxh$, respectively. Effects implied by the Higgs mass were illustrated by comparing the massless and massive Higgs cases. 
Then SM annihilations into DM have been considered, and parameters consistent with the abundance have been determined.

It is worth emphasizing the result presented in Fig.~\ref{fig:ghphimx_GP} are generic for the gravitational production of DM from the background inflaton field and SM freeze-in through graviton exchange. In this figure the excluded conservative regions in the space $(\ghp,\mx)$ are shown for various values of $n$. Note that the $\ghp$ parameter controls the efficiency of reheating and it is hence directly related to the SM bath temperature as well as the duration of reheating. Such gravitational production of DM is present in any model where reheating is non-instantaneous and, to a large extent, unavoidable. In our model, we consider the simplest way to reheat, i.e., by the cubic interaction between the inflaton and SM Higgs fields, $\ghp \phi |\boldmath{h}|^2$. Note, however, that even in the limit of vanishing inflaton-Higgs interaction, the figure is useful, then two-dimensional regions reduce to excluded ranges of DM masses. Furthermore, it is straightforward to extend these results for gravitational production of DM to other models of reheating.

\section*{Acknowledgments}
The work of B.G. and A.S. is supported in part by the National Science Centre (Poland) as a research project, decisions no 2017/25/B/ST2/00191 and 2020/37/B/ST2/02746.
\onecolumngrid
\appendix
\section[constraints]{Constraints on reheating from inflation	\label{s.constraints}}
In this appendix, we outline the procedure for obtaining constraints on the model parameters from inflationary dynamics. 
For the $\alpha$-attractor T-model of inflation, we have three parameters of the model~\eqref{eq:inf_pot}, namely $n$, $\alpha$, and $\Lambda$. As mentioned above the inflation scale $\Lambda$ is uniquely constrained by tensor-to-scalar (power-spectrum) ratio~$r$ and amplitude of the scalar power-spectrum $A_s$ from the CMB measurements. Furthermore, as we see below the parameter $\alpha$ is also constrained by the tensor-to-scalar ratio~$r$, see also a recently dedicated analysis~\cite{Ellis:2021kad}.

We start with revising some of the necessary inflationary parameters and their corresponding values from the CMB measurements. The tensor-to-scalar ratio is defined as
\beq
r \equiv \frac{\Delta_{t}^{2}(k)}{\Delta_{s}^{2}(k)},
\eeq
where $\Delta_{t}^{2}(k)$ and $\Delta_{s}^{2}(k)$ are the dimensionless tensor and scalar power spectra, respectively~\cite{Baumann:2009ds}. In the following we use $k_\star=0.05\,{\rm Mpc}^{-1}$ as pivot scale for the CMB observations by Planck~\cite{Planck:2018jri}. The amplitude of scalar power-spectrum measured by Planck at $k=k_\star$ is
\beq
\Delta_{s}^{2}(k_\star)=2.1 \times 10^{-9}.
\eeq
whereas the upper-bound on tensor-to-scalar ratio $r$ is, 
\beq
r\leq 0.032,
\eeq
at 95\% C.L. by the Planck~\cite{Planck:2018jri} and BICEP/{\it Keck}~\cite{BICEP:2021xfz} combined analysis~\cite{Tristram:2021tvh}.
Hence, from the above equations, the amplitude of the tensor power-spectrum is constrained to be,
\beq
\Delta_{t}^{2}(k_\star)\leq 6.7 \times 10^{-11}.	\label{eq:tensor_con}
\eeq
The explicit form of tensor power spectrum for $k=k_\star$ is given by~\cite{Baumann:2009ds},
\beq
\Delta_{t}^{2}(k_\star)=\frac{2}{\pi^{2}} \frac{H_{\star}^{2}}{\mpl^{2}},
\eeq
where $H_{\star}\equiv H(a_\star)$ and $a_\star$ marks the moment of horizon crossing of the mode $k_\star$ during inflation. Therefore, from \eq{eq:tensor_con} we set an upper-bound on the Hubble scale during inflation as 
\beq
H_\star\simeq H_I\leq 4.4 \times 10^{13}\gev.	\label{eq:Hinf_con}
\eeq
Since, the Hubble rate during inflation is given by, 
\beq
H_\star^2 \simeq  \frac{V(\phi_\star)}{3\mpl^2} \approx \frac{\Lambda^4}{3\mpl^2},
\eeq 
where $\phi_\star \equiv \phi(a_\star)$. In consequence, the upper-bound on the Hubble scale can be recast on the scale of inflation $\Lambda$ as
\beq
\Lambda \lesssim 1.4 \times 10^{16}\gev.	\label{eq:Lamb_con}
\eeq

Next, we calculate the number of e-folds during inflation $N_\star$, defined as
\beq
N_{\star}=\frac{1}{\mpl} \int_{\phi_{e}}^{\phi_{\star}} \frac{d \phi}{\sqrt{2 \epsilon_{V}(\phi)}},	\label{eq:Nstar}
\eeq
where $\epsilon_V(\phi)$ is given in \eqref{eq:epsilonV}, and $\phi_e$ refers to the value of $\phi$ at the end of inflation as in \eq{eq:varphie}, and $\phi_\star = \phi(k_\star/H_\star)$.
We obtain $\phi_\star$ from measurement of the spectral tilt~$n_s$~\eqref{eq:r_ns} by the Planck satellite for the modes $k_\star$, i.e.,
\beq
n_s=0.9649\pm 0.0042.
\eeq
Employing the above result for $n_s$ and using \eq{eq:r_ns}, we can calculate the value of $\phi_\star$, which then can be used to get the corresponding value for the number of e-folds during inflation~$N_\star$ from \eq{eq:Nstar}. For example, employing $1~\sigma$ level constraint for $n_s$ leads to range for the number of e-folds during inflation $N_{\star}\in [50, 65]$.

In the following, we calculate constraint on the number e-fold during reheating $N_{\rm rh}$ due to inflationary observables, in particular $N_\star$, and due to the BBN lower-bound on the reheating temperature $T_{\rm rh}\gtrsim 1\mev$~\cite{deSalas:2015glj}. Note that the comoving scale $k_\star$ relates the quantities during inflation at $a_\star$, when the modes exit the horizon, with the late universe when they re-enter the horizon at the pivot scale. In particular, one can evolve the scale factor starting from the mode crossing during inflation to the late time universe with different epochs. For instance, we can write
\beq
\frac{k_\star}{H_\star}=\frac{a_\star}{a_e}\frac{a_e}{a_{\rm rh}}\frac{a_{\rm rh}}{a_{\rm eq}}a_{\rm eq},
\eeq
where $a_{\rm eq}$ is the scale factor at the matter-radiation equality. Taking the logarithm of the above expression gives, 
\beq
\ln\!\bigg(\frac{k_\star}{H_\star}\bigg)=-N_\star - N_{\rm rh} - N_{\rm rd} +\ln(a_{\rm eq}),	\label{eq:kstar}
\eeq
where $N_\star\!\equiv\!\ln(a_e/a_\star)$, $N_{\rm rh}\!\equiv\!\ln(a_{\rm rh}/a_e)$, $N_{\rm rd}\!\equiv\!\ln(a_{\rm eq}/a_{\rm rh})$ are the number of e-folds during inflation, reheating, and radiation-dominated epochs. The above expression relates different epochs in the history of the universe. 

Assuming a generic equation of state during the reheating phase, one can calculate the $N_{\rm rh}$ in terms of temperature at reheating and energy density at the end of inflation as follows.  From \eq{eq:rhophia} we get the energy density at the end of reheating as
\beq
\rho_{\rm rh}=\rho_{e}\left(\frac{a_{e}}{a_{\rm rh }}\right)^{3\left(1+\bar w \right)},	\label{eq:rho_rh}
\eeq
where averaged equation of state during reheating phase $\bar w=(n-1)/(n+1)$ is defined in \eq{eq:w}, $\rho_e\equiv 3V(\phi_e)/2$ is the inflaton energy density at the end of inflation, and $\rho_{\rm rh }$ is the energy density at the end of reheating defined as
\beq
\rho_{\rm rh}\equiv\rho_\phi(a_{\rm rh})=\rho_{\rm SM}(a_{\rm rh})=\frac{\pi^{2}g_{\star}^{\rm rh}}{30}  T_{\rm rh}^{4}, 	\label{eq:rho_rh_def}
\eeq
with $g_{\star}^{\rm rh}\equiv g_{\star}(T_{\rm rh})$ being the effective number of relativistic d.o.f. at the reheating temperature $T_{\rm rh}$. Assuming no extra relativistic d.o.f. beyond the SM $g_{\star}(T\gtrsim 100\gev)\simeq 106.75$ and $g_{\star}(T\sim 1\mev)\simeq 10.75$, whereas in between, i.e., for $1\mev \lesssim T\lesssim 100\gev$. the $g_{\star}$ varies depending of number of active relativistic d.o.f..
We can write the number of e-folds during reheating phase from \eqref{eq:rho_rh} and \eqref{eq:rho_rh_def} as 
\beq
N_{\rm rh} \equiv \ln\! \bigg(\frac{a_{\rm rh}}{a_{e}}\bigg)=\frac{1}{3\left(1+\bar w\right)} \ln\! \bigg(\frac{45}{\pi^{2}g_{\star}^{\rm rh}}\frac{V(\phi_{e})}{T_{\rm rh}^{4}}\bigg), 	\label{eq:Nrh}
\eeq
where the energy density at the end of inflation $\rho_e=3 V(\phi_e)/2$ defined as $\epsilon_V=1$. Employing the entropy conservation during the radiation-dominated epoch, one can relate the reheating temperature $T_{\rm rh}$ with the temperature at the matter-radiation equality $T_{\rm eq}$ as 
\beq
T_{\rm rh}=\bigg(\frac{g_{\star s}^{\rm eq}}{g_{\star s}^{\rm rh}}\bigg)^{\!1/3}\bigg(\frac{a^{\rm eq}}{a^{\rm rh}}\bigg)T_{\rm eq},
\eeq
where $g_{\star s}^{i}$ is the effective number of relativistic d.o.f. contributing to entropy density at temperature $T_i$. Now we can rewrite \eq{eq:Nrh} as
\begin{align}
N_{\rm rh}&=\frac{1}{3\left(1+\bar w\right)} \ln\! \bigg(\frac{45}{\pi^{2}g_{\star}^{\rm rh}}\frac{V(\phi_{e})}{T_{\rm eq}^{4}} \Big(\frac{g_{\star s}^{\rm rh}}{g_{\star s}^{\rm eq}}\Big)^{\!4/3}\Big(\frac{a^{\rm rh}}{a^{\rm eq}}\Big)^4\bigg), 	\notag\\
&\simeq\frac{4}{3\left(1+\bar w\right)} \bigg[\frac14\ln\! \Big(\frac{45}{\pi^{2}g_{\star}^{\rm rh}}\Big)+\ln\! \Big(\frac{\Lambda}{T_{\rm eq}}\Big) + \frac13\ln\!\Big(\frac{g_{\star s}^{\rm rh}}{g_{\star s}^{\rm eq}}\Big)-N_{\rm rd}\bigg],
\end{align}
where in the last step we use $V(\phi_e)\approx \Lambda^4$. We use \eq{eq:kstar} to express $N_{\rm rd}$ in the above equation, after solving for $N_{\rm rh}$, we obtain,
\begin{align}
N_{\rm rh}&\simeq\frac{-4}{1-3\bar w} \bigg[\frac14\ln\! \Big(\frac{45}{\pi^{2}g_{\star}^{\rm rh}}\Big) + \frac13\ln\!\Big(\frac{g_{\star s}^{\rm rh}}{g_{\star s}^{\rm eq}}\Big) +\ln(a_{\rm eq})	+\ln\! \Big(\frac{\Lambda}{H_{\star}}\Big)+ \ln\! \Big(\frac{k_\star}{T_{\rm eq}}\Big)+N_{\star}\bigg].
\end{align}
Note this relation is only valid for an effective equation of state during reheating $\bar w\neq1/3$ since for $\bar w=1/3$ expansion rate of the Universe is the same as that of the radiation-dominated universe. 
Assuming the reheating temperature $T_{\rm rh}\gtrsim 1\mev$, we can approximate the above relation as
\begin{align}
N_{\rm rh}&\approx\frac{4}{1-3\bar w} \bigg[61.6-\ln\! \Big(\frac{\Lambda}{H_{\star}}\Big)-N_{\star}\bigg],
\end{align}
where we have used $g_{\star}^{\rm rh}=g_{\star s}^{\rm rh}=106.75$, $g_{\star}^{\rm eq}=3.94$, $a_{\rm eq}\simeq 1/3400$, $T_{\rm eq}\simeq 0.8\,{\rm eV}$ and $k_\star=0.05\, {\rm Mpc^{-1}}$. 
For $\bar w\neq1/3$ the above relation sets a constraint on the duration of reheating from inflationary observables along with the constraint on the reheating temperature $T_{\rm rh}\gtrsim1\mev$ from BBN as
\beq
N_{\rm rh} \lesssim \frac{4}{3\left(1+\bar w\right)} \bigg[6.7+\ln\! \Big(\frac{\Lambda}{1\gev}\Big)\bigg].	\label{eq:Nrh_BBN}
\eeq
Using the upper-bound on $\Lambda\lesssim 1.4\times 10^{16}\gev$, we get the following upper-bound on $N_{\rm rh}\lesssim 59/(1+\bar w)$. For $\bar w=1/3$, i.e., $n=2$, one finds $N_{\rm rh}\lesssim 44$. We summarize the constraints on $N_{\rm rh}$ in \tab{tab:Nrh}.
\begin{table}[t!]
$
\begin{array}{|c|c|c|c|}
\hline
\alpha & n & N_{\rm rh}[n_s\!:\!1\sigma] & N_{\rm rh}[n_s\!:\!2\sigma]\\
\hline
 1/6 & 2/3 & 13.8 & 26.1 \\
 1/6 & 1 & 22.0 & 41.7 \\
 1/6 & 3/2 & 48.0 & 47.8\\
 1/6 & 3 & 38.4 & 38.4\\
 1 & 2/3 & 15.2 & 27.5\\
 1 & 1 & 23.4 & 43.1\\
 1 & 3/2 & 47.8 & 47.7\\
 1 & 3 & 38.2 & 38.0\\
 \hline
\end{array}
$
\caption{Constraints on $N_{\rm rh}$ for different values of $\alpha$ and $n$. For $n=2$ the constraint is only from the lower limit on reheating temperature from BBN, which is $N_{\rm rh}\lesssim 44$.}
\label{tab:Nrh}
\end{table}

\section[RDMproduction]{Inflaton induced gravitational production} 
\label{s.graviational_production}
In this appendix, we calculate the energy gain of the SM and DM sectors due to the inflaton field. In particular, we calculate the collision terms that account for interactions between these three sectors. We focus on cubic interactions of the inflaton with the SM Higgs field and vector DM $X$ as well as indirect interactions through gravity. This appendix is supplementary to \sec{s.reheating} and \sec{s.DM-effects}.

During the reheating phase, we assume the inflaton field as a classical, homogeneous background that coherently oscillates in time. The character of these oscillations depends on the form of the inflaton potential during the reheating phase. Following Refs.~\cite{Shtanov:1994ce, Garcia:2020wiy, Ichikawa:2008iq, Clery:2021bwz}, we parametrize the evolution of the inflaton field $\phi(t)$ by a slowly time-varying envelope $\varphi(t)$ and a fast oscillating quasi-periodic function $\mathcal{P}(t)$, i.e., $\phi(t) = \varphi(t) \,\mathcal{P}(t)$.  The envelope function $\varphi(t)$ is given by \eq{eq:envelope_sol}, whereas the fast oscillating function $\mathcal{P}(t)$ is decomposed into Fourier modes with oscillation frequency~$\omega$.

The inflaton field is a given by (appropriate EoM) external time-dependent field which enters the Lagrangian implying various interesting phenomena.
In this framework, SM or DM particles are produced in quantum processes from the vacuum in the classical inflaton background. Thus, we consider the quantum transition from the vacuum, i.e., $\ket{\alpha_0}= \ket{0}$ to two-particle final states. 
The S-matrix element is defined as
\begin{align}
S_{\beta \alpha} = \Big\langle\beta_0 \Big|T\Big[\exp{\!\Big(\!- i \int d^{4}x\, \mathcal{L}_{\rm int}\Big)}\Big]\Big|\alpha_0\Big\rangle,
\end{align}
where the interaction Lagrangian relevant for the inflaton induced interactions is,
\beq
\begin{aligned}
\mathcal{L}_{\rm int} =&\, g_{h \phi} \mpl \phi |{\bm h}|^2 +\frac12 \mathcal{C}_X^\phi \frac{m_X^2}{\mpl} \phi X_\mu X^\mu + \frac{h^{\mu \nu}}{\mpl} \left( T_{\mu \nu}^\phi \!+\!  T_{\mu \nu}^{\rm{SM}} \!+\! T_{\mu \nu}^{\rm{DM}} \right).
\end{aligned}
\eeq
Above $T_{\mu \nu}^\phi$ denotes the energy-momentum tensor for the inflaton field,
\begin{align}
T_{\mu \nu}^\phi = \partial_\mu \phi \partial_\nu \phi - \eta_{\mu \nu} \Big[\frac{1}{2}\partial^\alpha \phi \partial_\alpha \phi - V(\phi)\Big],
\end{align}
while the energy-momentum tensor for real scalar (spin-0) and vector (spin-1) fields take the forms
\begin{align}
T_{\mu \nu}^S&\!=\! \partial_\mu S \partial_\nu S - \frac{\eta_{\mu \nu}}{2} \Big[\partial^\alpha S \partial_\alpha S - m_S^2 S^2\Big],\\
T_{\mu \nu}^{V} &\!=\! - \eta^{\alpha \beta} X_{\mu \alpha} X_{\nu \beta} + m_V^2 X_\mu X_\nu - \eta_{\mu \nu} \Big[\!- \frac{1}{4} \eta^{\rho \sigma} \eta^{\alpha \beta} X_{\rho \alpha} X_{\sigma \beta} + \frac{1}{2} m_V^2 \eta^{\rho \sigma} X_\rho X_\sigma \Big].
\end{align}

The next step is to quantize the graviton $h_{\mu \nu}$, vector DM $X$ and scalar field $S$ fields in terms of creation and annihilation operators as 
\begin{align}
\hat{h}_{\mu \nu} (x) &\!=\! \sum_{\lambda= ++, --} \int d \Pi_p \Big(\epsilon_{\mu \nu}^{\lambda}(p) \hat{a}_\lambda (p) e^{- i p x} + {\rm h.c.} \Big), \\
\hat{X}(x) &\!=\!\sum_{\sigma= L, \pm}\int d \Pi_p \Big(\epsilon_{\mu}^{\sigma}(p) \hat{a}_\sigma (p) e^{- i p x} + {\rm h.c.} \Big), \\
\hat{S}(x) &\!=\! \int d \Pi_p \Big(\hat{a} (p) e^{- i p x} +  {\rm h.c.} \Big),
\end{align}
where the phase-space is $d \Pi_p \equiv \frac{d^3 p }{(2 \pi)^3 \sqrt{2 p_0}}$. 
The polarization vectors for the spin-2 massless field satisfy the following relation:
\begin{align}
\sum_{\lambda= ++, --}\!\!\! \epsilon_{\mu \nu}^\lambda(p) \epsilon_{\alpha \beta}^{\lambda *} (p) &= \frac{1}{2}\Big[\eta_{\mu \alpha}\eta_{\nu \beta}+ \eta_{\mu \beta}\eta_{\nu \alpha}- \eta_{\mu \nu}\eta_{\alpha \beta}\Big]\equiv P_{\mu \nu \alpha \beta},
\end{align}
whereas, for massive vector field we have 
\begin{align}
\sum_{\sigma= L, \pm} \epsilon_\mu^\sigma (p) \epsilon_\nu^{\sigma *}(p ) = - \eta_{\mu \nu} + \frac{p_\mu p_\nu}{m_X^2}.
\end{align}
The annihilation and creation operators satisfy the following commutation relations:
\begin{align}
[\hat{a}_\upsilon (p),\hat{a}^\dagger_{\upsilon^\prime}(q) ]&=(2 \pi)^3 \delta^{(3)} (p-q) \delta_{\upsilon, \upsilon^\prime}.
\end{align}
\subsection{S-matrix for the SM and DM production}

In this subsection, we derive the S-matrix for various processes in the presence of the inflaton background field. 
At the leading order, we calculate the S-matrix for the Higgs boson and vector DM production with single interaction of the interaction Lagrangian, i.e.,
\begin{align}
S_{\beta_0 \alpha_0}^{(1)} = -i\int d^{4}x\, \Big\langle\beta_0 \Big|T\Big[\mathcal{L}_{\rm int}(x)\Big]\Big|\alpha_0\Big\rangle,
\end{align}
These processes would mimic the ``decay'' of the inflaton to the SM Higgs boson and vector DM. 
In the case of SM Higgs boson, the lowest-order non-zero element of the S-matrix is given by
\begin{align}
S_{\phi \rightarrow h_i h_i }^{(1)} &=-\frac{i}{2} g_{h \phi} \mpl \bra{\beta_0} \int dt \int d^3 x \phi(t) \hat{h}_i \hat{h}_i \ket{0}, \non \\
& = -\frac{i}{2}  g_{h \phi} \mpl \varphi(t) \sum_{k=-\infty}^{\infty} \mathcal{P}_k \int {\rm d}t \, e^{- i k \omega t}  \int {\rm d}^3 x  \int d \Pi_p\int d \Pi_q 
 \bra{0} a(p_1) a(p_2) a^{\dagger}(p) a^{\dagger}(q) \ket{0} 
\sqrt{4 p_1^0 p_2^0} \,e^{ i p \cdot x} e^{i q \cdot x} , 	\non \\ 
&=  -i g_{h \phi} \mpl \varphi(t) \sum_{k} \mathcal{P}_k  (2 \pi)^4 \delta (k \omega - p_1^0 -p_2^0) \delta^{(3)}(\vec{p}_1+\vec{p}_2).
\end{align}
Analogously, for the DM production channel, we find
\begin{align}
S_{\phi \rightarrow XX }^{\sigma \sigma^\prime (1)} &= -i \mathcal{C}_X^\phi \frac{m_X^2}{\mpl} \varphi(t) g^{\mu \nu} \epsilon_\mu^{\sigma^\prime \star}(p_1) \epsilon_\nu^{\sigma \star}(p_2)   \sum_{k} \mathcal{P}_k  (2 \pi)^4 \delta (k \omega - p_1^0 -p_2^0) \delta^{(3)}(\vec{p}_1+\vec{p}_2).
\end{align}

Next, we calculate the S-matrix for the SM and DM production from the inflaton background through the gravitational interactions. 
To take into account such processes mediated by virtual gravitons, we have to expand the $S$ matrix up to the second-order, i.e.,
\begin{align}
S_{\beta_0 \alpha_0}^{(2)} \!=\!\frac{(-i)^2}{2!} \!\int d^4 x \int \!d^4 y \Big\langle\beta_0 \Big|T\Big[\mathcal{L}_{\rm{int}}(x)\mathcal{L}_{\rm{int}}(y)\Big]\Big|\alpha_0\Big\rangle.
\end{align}
Such processes mimic the ``annihilation'' of the inflaton field to a two-particle final state mediated by a graviton. In this case, we can rewrite the above expression as 
\begin{align}
S_{\beta_0 \alpha_0}^{(2)} &=\frac{(-i)^2}{2!} \frac{1}{\mpl^2}\int d^4 x \int d^4 y \Big\langle\beta_0 \Big|T\Big[h^{\mu \nu}(x)h^{\alpha \beta}(y)\Big(T_{\mu \nu}^\phi(x) T_{\alpha \beta}^{F}(y) + T_{\mu \nu}^{F}(x) T_{\alpha \beta}^{\phi}(y) \Big)\Big]\Big|\alpha_0\Big\rangle,		\label{eq:s-matrix_grav}
\end{align}
where $T_{\mu \nu}^{F}$ denotes the energy-momentum tensor for a generic field $F$ in the final state.
 
Let us first consider gravitational mediation from a quantum vacuum in the presence of the inflaton background to a pair of scalar $S$ particles. We obtain the following expression for the $h^{\mu \nu}(x) h^{\alpha \beta}(y) T_{\mu \nu}^\phi (x) T_{\alpha \beta}^{S}(y)$ term of the S-matrix:
\begin{align}
S_{\phi\to S S}^{(2)} &\supset  \frac{(-i)^2}{2!} \frac{1}{\mpl^2} \int d^4 x \int d^4 y \sqrt{4 p_1^0 p_2^0} \Big\langle 0 \Big| \hat{a}(p_1) \hat{a}(p_2) T[h^{\mu \nu}(x)h^{\alpha \beta}(y)  T_{\mu \nu}^\phi (x) T_{\alpha \beta}^{S} (y) ] \Big|0\Big\rangle, \nonumber \\
&= \frac{(-i)^2}{2!} \frac{1}{\mpl^2}\sum_{\lambda, \lambda^\prime} \int d^4 x \int d^4 y \int d \Pi_p\int d \Pi_q\int d \Pi_{k_1 }\int d \Pi_{k_2} T_{\mu \nu}^\phi (x) 
\Big\langle 0 \Big|\big( \epsilon^{\mu \nu, \lambda}(k_1) \hat{a}_\lambda (k_1) e^{- i k_1 x} + {\rm h.c.} \big) \nonumber \\ 
&\quad \times\big(\epsilon^{\alpha \beta, \lambda^\prime}(k_2) \hat{a}_{\lambda^\prime} (k_2) e^{- i k_2 y} + {\rm h.c.} \big)\Big|0\Big\rangle\Big\langle 0 \Big|\hat{a}(p_1) \hat{a}(p_2) \Big[\left(i p_\alpha \hat{a}^\dagger (p) e^{i p y}  + {\rm h.c.} \right) \left(i q_\beta \hat{a}^\dagger (q) e^{i q y} + {\rm h.c.} \right) \nonumber \\
&\quad-\frac{\eta_{\alpha \beta} }{2} \left( i p^\rho \hat{a}^\dagger (p) e^{i p y}  + {\rm h.c.} \right)\left( i q_\rho \hat{a}^\dagger (q)e^{i q y} + {\rm h.c.} \right)+ \eta_{\alpha \beta} \frac{m_{S}^2}{2} \left( \hat{a}(p) e^{ i p y} + {\rm h.c.}\right)\left(  \hat{a}(q) e^{ i q y}+ {\rm h.c.} \right)\Big] \Big|0\Big\rangle,
\end{align}
where $p_{1}, p_2$ are the final state momenta.
Contraction of graviton operators leads to the following propagator:
\begin{align}
D^{\mu \nu \alpha \beta}(x-y) &= \int \frac{d^4 \tilde{q}}{(2 \pi)^4} \frac{i}{\tilde{q}^2 + i \epsilon} e^{-i (x-y) \tilde{q}} P^{\mu \nu \alpha \beta},
\end{align}
such that
\begin{align}
T_{\mu \nu}^\phi P^{\mu \nu \alpha \beta} &= \partial^\alpha \phi \partial^\beta \phi  - \eta^{\alpha \beta} V(\phi).
\end{align}
After some tedious but straightforward computations, we get
\begin{align}
S_{\phi\to S S}^{(2)} \supset \frac{(-i)^2}{\mpl^2} \int d^4 x \int d^4 y& \int \frac{d^4 \tilde{q}}{(2 \pi)^4} \frac{i}{\tilde{q}^2 + i \epsilon} e^{-i (x-y) \tilde{q}} \non \\
&\times\Big(\!- (\partial^\alpha \phi \, {p_1}_\alpha ) (\partial^\beta \phi \,  {p_2}_\beta ) -  V(\phi) p_{1, \alpha} p_2^\alpha - 2m_{S}^2 V(\phi)  + \frac{1}{2} \partial^\alpha \phi \partial_\alpha \phi (p_1 \cdot p_2+m_{S}^2) \Big).
\end{align}
Using the fact that the inflaton field is a homogeneous background, i.e., $\phi = \phi(t)$, we can write the kinematics as
\begin{align}
p_1 \cdot p_2 = \frac{s}{2}-m_{S}^2, && p_1^0= p_2^0 = \frac{\sqrt{s}}{2}, \label{eq:assumptions}
\end{align}  
where $s\equiv (p_1+p_2)^2$ is the Mandelstam variable.
Now, we can further simplify the above result as
\begin{align}
S_{\phi\to S S}^{(2)} \supset -\frac{1}{2}\frac{(-i)^2 }{\mpl^2} \int d^4 x \int d^4 y \int \frac{d^4 \tilde{q}}{(2 \pi)^4} \frac{i}{\tilde{q}^2 + i \epsilon}  V(\phi)\Big( s+ 2m_{S}^2\Big) e^{-i (x-y) \tilde{q}} e^{+i (p_1+p_2) y}. \label{eq:S1}
\end{align}
It is convenient to Fourier transform the inflaton potential $V(\phi)$ instead of the inflaton field $\phi$, i.e., 
\begin{align}
V(\phi) &\simeq \Lambda^4 \left(\frac{\varphi}{M}\right)^{2n} \mathcal{P}^{2n}(t) = \rho_\phi \,\mathcal{P}^{2n}(t)\,,  &\mathcal{P}^{2n}(t) &= \sum_{k} \mathcal{P}^{2n}_k e^{- i  k \omega t}. \label{eq:decomposition}
\end{align}
Therefore,
\begin{align}
S_{\phi\to S S}^{(2)} \supset \frac{1}{2}\frac{\rho_\phi}{\mpl^2} \sum_k \mathcal{P}^{2n}_k \Big(1 + \frac{2 m_{S}^2}{s} \Big) \delta(k\omega - p_1^0 - p_2^0) \delta^{(3)}(\vec{p}_1+ \vec{p}_2).
\end{align}
The same result emerges from the $h^{\mu \nu}(x) h^{\alpha \beta}(y) T_{\mu \nu}^{S}(x) T^\phi_{\alpha \beta}$ term in the S-matrix~\eq{eq:s-matrix_grav}. Therefore the final result for the S-matrix element describing the gravitational production of massive scalars reads
\begin{align}
S_{\phi\to S S}^{(2)}=\frac{\rho_\phi}{\mpl^2} \sum_k \mathcal{P}^{2n}_k \Big(1 + \frac{2 m_{S}^2}{s} \Big) \delta(k\omega - p_1^0 - p_2^0) \delta^{(3)}(\vec{p}_1+ \vec{p}_2). \label{eq: S_00SS}
\end{align}

Similarly, for the final state vector $V$ particles the $h^{\mu \nu}(x) h^{\alpha \beta}(y) T_{\mu \nu}^\phi (x) T_{\alpha \beta}^{V} (y)$ term of the S-matrix takes the form
\begin{align}
S_{\phi\to VV}^{(2)} &\supset\frac{(-i)^2}{2!} \frac{2}{\mpl^2} \int d^4 x  \int d^4 y  \int \frac{d^4 \tilde{q}}{(2 \pi)^4} \frac{i}{\tilde{q}^2 + i \epsilon} e^{-i (x-y) \tilde{q}} e^{i (p_1+p_2)y}\nonumber \\
&\qquad \times \Big[\underbrace{- \frac{\sqrt{s}}{2}\dot{\phi}^2 \Big((\epsilon_2^* \cdot p_1) (\epsilon_1^*)^0+(\epsilon_1^* \cdot p_2) (\epsilon_2^*)^0 - \sqrt{s} (\epsilon_1^*)^0 (\epsilon_2^*)^0 \Big) + \frac{\dot{\phi}^2}{2}(\epsilon_2^* \cdot p_1) (\epsilon_1^* \cdot p_2) + V(\phi) m_V^2 \epsilon_1^* \cdot \epsilon_2^*}_{{\cal A}\eqref{eq:S2_phiVV}}\Big].	\label{eq:S2_phiVV}
\end{align}
We choose the frame in which the spin-1 particles move along the $z$-direction, i.e.,
\begin{align}
p_1^\mu &= \Big(\frac{\sqrt{s}}{2}, 0, 0, p_z\Big),  &p_2^\mu = \Big(\frac{\sqrt{s}}{2}, 0, 0, -p_z\Big).
\end{align} 
In this case, polarization vectors can be written as 
\begin{align}
&(\epsilon_{1,(2)}^L)^\mu = \left(\pm \frac{p_z}{m_V}, 0, 0, \frac{\sqrt{s}}{2 m_V} \right),
&(\epsilon_{1,(2)}^{\pm})^\mu = \frac{1}{\sqrt{2}} (0, 1, \pm i, 0).
\end{align}
For the transverse modes, the expression in the square bracket in \eq{eq:S2_phiVV} reduces to
\begin{align}
{\cal A}^{\pm}\eqref{eq:S2_phiVV}&=-V(\phi) m_V^2,
\end{align}
whereas for the two longitudinally-polarized modes, we have
\begin{align}
{\cal A}^L\eqref{eq:S2_phiVV}&= - V(\phi) \Big( \frac{s}{2} - m_V^2 \Big).
\end{align}
Thus, the total S-matrix element accounting for the gravitational production of spin-1 particles is given by
\begin{align}
S_{\phi\to VV}^{(2)} = \frac{\rho_\phi}{\mpl^2}\sum_k \mathcal{P}_k^{2n} \mathcal{M}_{V}(k) (2 \pi)^4 \delta(k \omega - p_1^0 - p_2^0) \delta^{(3)} ( \vec{p}_1 + \vec{p}_2 ),
\end{align}
where the matrix elements for the longitudinal and transverse polarizations are
\begin{align}
\mathcal{M}_{V}^L(k)  &\equiv  1- \frac{2m_V^2}{(k \omega)^2},
&\mathcal{M}_{V}^\pm (k) &\equiv  \frac{2m_V^2}{(k \omega)^2}.
\end{align}

\subsection{Production of the SM and DM from the inflaton}

Next, we can calculate the probability $P$ for the production of two particles with momenta $p_1$ and $p_2$ in the presence of the inflaton background field
\begin{align}
P(p_1, p_2) &= \frac{\lvert  S \rvert^2}{\braket{0|0} \braket{p_1|p_1} \braket{p_2|p_2}}, \label{eq:trans_prob}
\end{align}
where $\braket{0|0}=1$ and $\braket{p_{i}|p_{i}}=(2 \pi)^3 2 p_{i}^0 \delta^{(3)}(0)$ for $i=1,2$. Whereas $S$ is the S-matrix for a given process, which for our model can be:
\bit
\item[(1)] direct production of final state pair of particles $f$, 
\begin{align}
\big\lvert S_{\phi \rightarrow ff}^{(1)} \big\rvert^2 &= \varphi^2 \mpl^2  \sum_k \lvert \mathcal{P}_k \rvert^2 \cdot \big\lvert \overline{\mathcal{M}}^{(1)}_{\phi \rightarrow ff}(k) \big\rvert^2 \Big[(2 \pi)^4 \delta(k \omega - p_1^0 - p_2^0) \delta^{(3)}(\vec{p}_1+\vec{p}_2) \Big]^2, \label{eq:s1}
\end{align} 
\item[(2)] the gravitational production through the graviton mediated process,
\begin{align}
\big\lvert S_{\phi \rightarrow ff}^{(2)}\big\rvert^2 &= \frac{\rho_\phi^2}{\mpl^4}  \sum_k \lvert \mathcal{P}_k^{2n} \rvert^2 \cdot \big\lvert \overline{\mathcal{M}}^{(2)}_{\phi \rightarrow ff} (k)\big\rvert^2  \Big[(2 \pi)^4 \delta(k \omega - p_1^0 - p_2^0) \delta^{(3)}(\vec{p}_1+\vec{p}_2) \Big]^2.\label{eq:s2}
\end{align}
\eit
Above $\lvert \overline{\mathcal{M}}(k)\rvert^2$ represents the spin-averaged amplitude square for a given process and it depends on the Fourier mode number $k$ of the background inflaton field. In our model, the first-order amplitudes result from the direct production of the SM Higgs boson and vector DM pairs, i.e.,
\begin{align}
\big|\overline{\mathcal{M}}^{(1)}_{\phi \rightarrow h_i h_i}\big|^2 &\equiv  g_{h \phi}^2,
&\big|\overline{\mathcal{M}}^{(1)}_{\phi \rightarrow XX} \big|^2 &\equiv |\mathcal{C}_X^\phi|^2 \frac{(k \omega)^4}{4\mpl^4} \bigg( 1 -\frac{4m_X^2}{(k \omega)^2}+\frac{12m_X^4}{(k \omega)^4}\bigg).
\end{align}
Whereas the second-order amplitudes in our work correspond to the gravitational production of the SM and DM particles through the graviton exchange, and for a final state massive scalar $S$ or vector $V$ particles are given by
\begin{align}
\big|\overline{\mathcal{M}}^{(2)}_{\phi \rightarrow SS}\big|^2 &\equiv  1+ \frac{4 m_S^2}{(k \omega)^2} +\frac{4m_S^4}{(k \omega)^4}, 
&\big|\overline{\mathcal{M}}^{(2)}_{\phi \rightarrow VV}\big|^2 &\equiv  1 -\frac{4m_V^2}{(k \omega)^2}+\frac{12m_V^4}{(k \omega)^4}.
\end{align}
where we have summed over polarizations of final states.

In order to simplify \eqref{eq:s1} and \eqref{eq:s2} one has to square two delta functions. To that end, we assume that the process of particle creation from the vacuum takes place in a box of volume $V$ with time duration $T$. At the very end of the computations, the regulators $V$ and $T$ are removed by taking the limit $V, \,T \rightarrow \infty$.  The square bracket in \eqref{eq:s1} and \eqref{eq:s2} can be written as
\begin{align}
\Big[(2 \pi)^4 \delta(k \omega - p_1^0 - p_2^0) \delta^{(3)}(\vec{p}_1+\vec{p}_2) \Big]^2 &= (2 \pi)^4 \delta(k \omega - p_1^0 - p_2^0) \delta^{(3)}(\vec{p}_1+\vec{p}_2) \times (2 \pi)^4 \delta^{(4)} (0),
\end{align}
where the regulators $V$ and $T$ are defined as
\begin{align}
(2 \pi)^4 \delta^{(4)} (0)& = \int {\rm d}^4 x e^{- i 0 \cdot x} = V T,
& (2 \pi)^3 \delta^{(3)} (0) &= \int {\rm d}^3 x e^{- i 0 \cdot \vec{x}} = V.
\end{align}
This result allows us to rewrite \eqref{eq:trans_prob} corresponding to \eqref{eq:s1} 
\begin{align}
P^{(1)}_{\phi\to ff}(p_1, p_2) &= \varphi^2 \mpl^2   \sum_k \lvert \mathcal{P}_k \rvert^2 \cdot  \frac{\big\lvert \overline{\mathcal{M}}^{(1)}_{\phi \rightarrow ff}(k) \big\rvert^2  }{2p_{1}^0 V \,2 p_{2}^0 V}  (2 \pi)^4 \delta (k \omega - p_{1}^0 - p_{2}^0) \delta^{(3)}(\vec{p}_1+\vec{p}_2) V T,
\end{align}
for the direct production through the inflaton interactions, and to \eqref{eq:s2} 
\begin{align}
P^{(2)}_{\phi\to ff}(p_1, p_2) &= \frac{\rho_\phi^2}{\mpl^4}  \sum_k \lvert \mathcal{P}_k^{2n}\rvert^2 \cdot  \frac{\big\lvert \overline{\mathcal{M}}^{(2)}_{\phi \rightarrow ff}(k) \big\rvert^2 }{2p_{1}^0 V \,2 p_{2}^0 V}  (2 \pi)^4 \delta (k \omega - p_{1}^0 - p_{2}^0) \delta^{(3)}(\vec{p}_1+\vec{p}_2) V T,
\end{align}
for the indirect, gravitational production of final state particles $f$ through the graviton exchange, respectively. 

To get the total probability one has to sum over each outgoing momenta. In the continuum limit, this reduces to multiplying $P (p_1, p_1)/T $ by a factor $ V  d^3 \vec{p}_1/(2 \pi)^3\,  V d^3 \vec{p}_2/(2 \pi)^3$. 
Note that the energy gain of created particles in volume $V$ and time ${\rm d} t$ can be calculated as
\begin{align}
{\rm d}E^{(i)} (p_1,p_2)= (p_1^0 + p_2^0)\frac{V {\rm d}^3 \vec{p}_1}{(2 \pi)^3} \frac{V {\rm d}^3 \vec{p}_2}{(2 \pi)^3} \frac{P^{(i)}_{\phi\to ff}(p_1, p_1)}{T} {\rm d}t,
\end{align}
where $i =1,2$ for the direct production and from the graviton mediation, respectively.
Thus, the total energy gain per volume and time for the $\phi \rightarrow ff$ process for the direct production through the inflaton field is given by
\begin{align}
&\frac{{\rm d}E^{(1)}}{V {\rm d} t}=  \varphi^2 \mpl^2 \sum_k \lvert \mathcal{P}_k \rvert^2 \cdot  \big\lvert \overline{\mathcal{M}}^{(1)}_{\phi \rightarrow ff}(k) \big\rvert^2  \int \frac{d^3 \vec{p}_1}{(2 \pi)^3 2 p_1^0} \int  \frac{d^3 \vec{p}_2}{(2 \pi)^3 2 p_2^0} (p_1^0 + p_2^0)  (2 \pi)^4 \delta (k \omega - p_{1}^0 - p_{2}^0) \delta^{(3)}(\vec{p}_1+\vec{p}_2).
\end{align}
Assuming the final particles have the same mass the resulting energy gain reads
\begin{align}
\frac{{\rm d}E^{(1)}}{V {\rm d} t}
&= \frac{\varphi^2 \mpl^2}{8 \pi}  \sum_{k=1}^\infty  k \omega \lvert \mathcal{P}_k \rvert^2 \cdot \big\lvert \overline{\mathcal{M}}^{(1)}_{\phi \rightarrow ff}(k) \big\rvert^2  \sqrt{1\!-\! \frac{4 m_f^2}{(k \omega)^2}}.
\end{align}
For the graviton mediated production we get the following result:
\begin{align}
\frac{{\rm d}E^{(2)}}{V {\rm d} t}
&\! =\!\frac{\rho_\phi^2}{8\pi \mpl^4} \! \sum_{k=1}^\infty  k \omega \lvert \mathcal{P}^{2n}_k \rvert^2 \cdot \big\lvert \overline{\mathcal{M}}^{(2)}_{\phi \rightarrow ff}(k) \big\rvert^2  \sqrt{1\!-\! \frac{4 m_f^2}{(k \omega)^2}}.
\end{align}
For the final state vector DM, the rate of production, or in other words, the collision term for DM number density due to the direct production and through the graviton mediation are
\begin{align}
\mathcal{D}^{(1)}_{\phi\to XX} &= \frac{ |\mathcal{C}_X^\phi|^2}{32 \pi}  \bigg(\frac{\rho_\phi}{\Lambda^4}\bigg)^{\!\!1/n}\, \sum_{k=1}^\infty   \lvert \mathcal{P}_k \rvert^2 (k \omega)^4\bigg( 1 -\frac{4m_X^2}{(k \omega)^2}+\frac{12m_X^4}{(k \omega)^4}\bigg) \sqrt{1\!-\! \frac{4 m_X^2}{(k \omega)^2}},	\label{eq:D1phiXX}		\\
\mathcal{D}^{(2)}_{\phi\to XX} &= \frac{\rho_\phi^2}{8\pi \mpl^4} \sum_{k=1}^\infty   \lvert \mathcal{P}^{2n}_k \rvert^2 \bigg( 1 -\frac{4m_X^2}{(k \omega)^2}+\frac{12m_X^4}{(k \omega)^4}\bigg)  \sqrt{1\!-\! \frac{4 m_X^2}{(k \omega)^2}}. \label{eq:energy_loss}
\end{align}

\subsection[BeqnPhi]{Inflaton decay rate}
The continuity equation,
\begin{align}
\dot{\rho}_\phi + 3 H(1+w) \rho_\phi = 0 ,
\end{align} 
indicates that the energy density of the inflaton field decreases due to the Universe's expansion. On the other hand, the inflaton loses its energy during reheating while the SM radiation and DM sector are gaining energy. Due to the energy conservation, the total energy density gained by the SM and DM particles must be equal to the energy loss of the inflaton field. Therefore the following time-averaged Boltzmann equation for the inflaton energy density could be written as
\begin{align}
\dot{\rho}_\phi + 3 H(1+\bar{w}) \rho_\phi = - \langle \Gamma_\phi \rangle \rho_\phi,
\end{align}
where $\langle \Gamma_\phi \rangle$ and $\bar{w}$ are the time average of the inflaton decay rate, $\Gamma_\phi$, and the equation-of-state parameter defined in (\ref{eq:width_def}) and (\ref{eq:w}), respectively.
For the considered interactions, the total decay rate for the inflaton field can be written as a sum of several contributions
\begin{align}
\Gamma_\phi &= \Gamma^{(1)}_{\phi \rightarrow hh} + \Gamma^{(1)}_{\phi \rightarrow XX} +  \Gamma^{(2)}_{\phi \rightarrow hh} +   \Gamma^{(2)}_{\phi \rightarrow \bar\psi\psi}+  \Gamma^{(2)}_{\phi \rightarrow VV}+ \Gamma^{(2)}_{\phi \rightarrow XX} ,
\end{align}
where the superscript $(1)$ and $(2)$ indicate the direct production through contact interaction with the inflaton field and the indirect gravitational production through the graviton exchange between the inflaton field and final state particles, respectively. Above $\psi$ and $V$ denote the SM fermions and gauge bosons, respectively. 
The inflaton partial decay rates through direct interactions in our model are
\begin{align}
\Gamma^{(1)}_{\phi \rightarrow hh} &=\frac{g_{h \phi}^2}{8\pi}  \frac{\mpl^4}{\rho_\phi}   \bigg(\frac{\rho_\phi}{\Lambda^4} \bigg)^{1/n} \sum_{i=0}^3  \sum_{k=0}^{\infty} k \omega  \lvert \mathcal{P}_k \rvert^2  \sqrt{1- \frac{4m_{h_i}^2}{(k \omega)^2}},			\\
\Gamma^{(1)}_{\phi \rightarrow XX} &=\frac{ |\mathcal{C}_X^\phi|^2}{32\pi} \frac{1}{\rho_\phi} \bigg(\frac{\rho_\phi}{\Lambda^4}\bigg)^{\!\!1/n}\, \sum_{k=1}^\infty   \lvert \mathcal{P}_k \rvert^2 (k\omega)^5\bigg( 1 -\frac{4m_X^2}{(k \omega)^2}+\frac{12m_X^4}{(k \omega)^4}\bigg) \sqrt{1\!-\! \frac{4 m_X^2}{(k \omega)^2}},
\end{align}
whereas the inflaton partial decay rates through the graviton mediation are
\begin{align}
\Gamma^{(2)}_{\phi \rightarrow hh}  &=\frac{1}{8 \pi} \frac{\rho_\phi}{\mpl^4}    \sum_{i=0}^3  \sum_{k=0}^{\infty} k \omega  \lvert \mathcal{P}_k^{2n} \rvert^2 \bigg( 1+ \frac{4 m_{h_i}^2}{(k \omega)^2} +\frac{4m_{h_i}^4}{(k \omega)^4}\bigg) \sqrt{1- \frac{4 m_{h_i}^2}{(k \omega)^2} },\\
\Gamma^{(2)}_{\phi \rightarrow \bar\psi\psi}&= \frac{1}{2 \pi}  \frac{\rho_\phi}{\mpl^4}   \sum_{\psi} \sum_{k=0}^{\infty} k \omega   \lvert \mathcal{P}_k^{2n} \rvert^2 \frac{m_\psi^2}{(k \omega)^2}  \bigg(1-\frac{4 m_\psi^2}{(k \omega)^2} \bigg)^{\!3/2},		\\
\Gamma^{(2)}_{\phi \rightarrow VV}  &= \frac{1}{8 \pi} \frac{\rho_\phi}{\mpl^4}   \sum_{V} \sum_{k=0}^{\infty}  k \omega  \lvert \mathcal{P}_k^{2n} \rvert^2 \bigg( 1 -\frac{4m_V^2}{(k \omega)^2}+\frac{12m_V^4}{(k \omega)^4}\bigg) \sqrt{1\!-\! \frac{4 m_V^2}{(k \omega)^2}},	\\
\Gamma^{(2)}_{\phi \rightarrow XX}  &= \frac{1}{8 \pi}  \frac{\rho_\phi}{\mpl^4}     \sum_{k=0}^{\infty}  k \omega  \lvert \mathcal{P}_k^{2n} \rvert^2  \bigg( 1 -\frac{4m_X^2}{(k \omega)^2}+\frac{12m_X^4}{(k \omega)^4}\bigg) \sqrt{1\!-\! \frac{4 m_X^2}{(k \omega)^2}}.
\end{align}
The gravitational inflaton decay width to massive fermions was also calculated in~\cite{Clery:2021bwz}. 

Note that for the direct inflaton decay rates, the $\phi \rightarrow XX$ rate is strongly suppressed relative to the SM Higgs boson decay channel. Thus, in the main text above neglecting the $\Gamma^{(1)}_{\phi \rightarrow XX}$ term in the Boltzmann equation for the inflaton energy density is justified. Moreover, the inflaton field cannot transfer its energy to massless vectors and fermions through gravitational interactions. Furthermore, for the gravitational inflaton decay rates to the SM particles we note that $\Gamma^{(2)}_{\phi \rightarrow \bar\psi\psi} \ll \Gamma^{(2)}_{\phi \rightarrow hh} \sim \Gamma^{(2)}_{\phi \rightarrow VV}$.
From the above results, it is straightforward to find the dominant source of reheating, i.e., whether it is due to the direct interactions of the Higgs field with the inflaton through $g_{h\phi}$ term or it is due to gravitational interactions between the inflaton field and the SM. To obtain the condition on the inflaton-Higgs direct coupling $g_{h\phi}$ such that the direct interactions dominate the reheating process to that of due to gravitational interactions we require $\Gamma^{(1)}_{\phi \rightarrow hh} \gtrsim\Gamma^{(2)}_{\phi \rightarrow hh}$. This implies that the reheating occurs dominantly due to $\phi \rightarrow hh$ channel if the inflaton-Higgs coupling, $g_{h \phi}$, satisfy
\begin{align}
g_{h \phi} \gtrsim \frac{\rho_\phi }{\mpl^4}   \left(\frac{\Lambda^4}{\rho_\phi} \right)^{\frac{1}{2 n}} \sim \frac{\rho_\phi }{\mpl^4}, \label{eq:ghphi_con}
\end{align}
where we neglect corrections proportional to phase space factors and the last approximation follows from the fact that $\rho_\phi\sim \Lambda^4$ at the onset of reheating phase and it gradually decreases during the reheating period. 
Note that the above inequality puts the strongest constraint on the value of the $g_{h \phi}$ coupling at the beginning of reheating when $\rho_\phi$ is the largest and scales as $ \rho_\phi \sim \Lambda^4$. The efficiency of the gravitational interactions quickly decreases with the Universe's expansion. Consequently, such interactions can be relevant for the dynamics of the radiation sector only at the highest possible energy scales, i.e., at the start of reheating phase. Neglecting terms of order one and adopting our benchmark value of $\Lambda$, we find the following limit $g_{h \phi} \gtrsim 10^{-11}$. Thus, we can conclude that for the inflaton-Higgs coupling $g_{h \phi} \in (10^{-5}, 10^{-10})$ as adopted in this work, the inflaton decay to pairs of SM Higgs bosons always dominates over graviton-mediated contributions.

\twocolumngrid

\bibliographystyle{aabib}
\bibliography{bib_vdm_reh}{} 

\providecommand{\href}[2]{#2}\begingroup\raggedright\begin{thebibliography}{10}

\bibitem{Guth:1980zm}
A.~H. Guth, ``{The Inflationary Universe: A Possible Solution to the Horizon
  and Flatness Problems},''
  \href{http://dx.doi.org/10.1103/PhysRevD.23.347}{{\em Phys. Rev. D}
  {\bfseries 23} (1981) 347--356}.

\bibitem{Linde:1981mu}
A.~D. Linde, ``{A New Inflationary Universe Scenario: A Possible Solution of
  the Horizon, Flatness, Homogeneity, Isotropy and Primordial Monopole
  Problems},'' \href{http://dx.doi.org/10.1016/0370-2693(82)91219-9}{{\em Phys.
  Lett. B} {\bfseries 108} (1982) 389--393}.

\bibitem{Baumann:2009ds}
D.~Baumann,
  \href{http://dx.doi.org/10.1142/9789814327183_0010}{``{Inflation},''}
\newblock 2011.
\newblock \href{http://arxiv.org/abs/0907.5424}{{\tt arXiv:0907.5424}}.

\bibitem{Planck:2018jri}
{\bfseries Planck} Collaboration, Y.~Akrami {\em et~al.}, ``{Planck 2018
  results. X. Constraints on inflation},''
  \href{http://dx.doi.org/10.1051/0004-6361/201833887}{{\em Astron. Astrophys.}
  {\bfseries 641} (2020) A10}, [\href{http://arxiv.org/abs/1807.06211}{{\tt
  arXiv:1807.06211}}].

\bibitem{Albrecht:1982mp}
A.~Albrecht, P.~J. Steinhardt, M.~S. Turner, and F.~Wilczek, ``{Reheating an
  Inflationary Universe},''
  \href{http://dx.doi.org/10.1103/PhysRevLett.48.1437}{{\em Phys. Rev. Lett.}
  {\bfseries 48} (1982) 1437}.

\bibitem{Dolgov:1982th}
A.~D. Dolgov and A.~D. Linde, ``{Baryon Asymmetry in Inflationary Universe},''
  \href{http://dx.doi.org/10.1016/0370-2693(82)90292-1}{{\em Phys. Lett. B}
  {\bfseries 116} (1982) 329}.

\bibitem{Abbott:1982hn}
L.~F. Abbott, E.~Farhi, and M.~B. Wise, ``{Particle Production in the New
  Inflationary Cosmology},''
  \href{http://dx.doi.org/10.1016/0370-2693(82)90867-X}{{\em Phys. Lett. B}
  {\bfseries 117} (1982) 29}.

\bibitem{Kofman:1994rk}
L.~Kofman, A.~D. Linde, and A.~A. Starobinsky, ``{Reheating after inflation},''
  \href{http://dx.doi.org/10.1103/PhysRevLett.73.3195}{{\em Phys. Rev. Lett.}
  {\bfseries 73} (1994) 3195--3198},
  [\href{http://arxiv.org/abs/hep-th/9405187}{{\tt hep-th/9405187}}].

\bibitem{Shtanov:1994ce}
Y.~Shtanov, J.~H. Traschen, and R.~H. Brandenberger, ``{Universe reheating
  after inflation},'' \href{http://dx.doi.org/10.1103/PhysRevD.51.5438}{{\em
  Phys. Rev. D} {\bfseries 51} (1995) 5438--5455},
  [\href{http://arxiv.org/abs/hep-ph/9407247}{{\tt hep-ph/9407247}}].

\bibitem{Kofman:1997yn}
L.~Kofman, A.~D. Linde, and A.~A. Starobinsky, ``{Towards the theory of
  reheating after inflation},''
  \href{http://dx.doi.org/10.1103/PhysRevD.56.3258}{{\em Phys. Rev. D}
  {\bfseries 56} (1997) 3258--3295},
  [\href{http://arxiv.org/abs/hep-ph/9704452}{{\tt hep-ph/9704452}}].

\bibitem{Lebedev:2021tas}
O.~Lebedev, F.~Smirnov, T.~Solomko, and J.-H. Yoon, ``{Dark matter production
  and reheating via direct inflaton couplings: collective effects},''
  \href{http://dx.doi.org/10.1088/1475-7516/2021/10/032}{{\em JCAP} {\bfseries
  10} (2021) 032}, [\href{http://arxiv.org/abs/2107.06292}{{\tt
  arXiv:2107.06292}}].

\bibitem{Ahmed:2021fvt}
A.~Ahmed, B.~Grzadkowski, and A.~Socha, ``{Implications of time-dependent
  inflaton decay on reheating and dark matter production},''
  \href{http://dx.doi.org/10.1016/j.physletb.2022.137201}{{\em Phys. Lett. B}
  {\bfseries 831} (2022) 137201}, [\href{http://arxiv.org/abs/2111.06065}{{\tt
  arXiv:2111.06065}}].

\bibitem{Garcia:2020wiy}
M.~A.~G. Garcia, K.~Kaneta, Y.~Mambrini, and K.~A. Olive, ``{Inflaton
  Oscillations and Post-Inflationary Reheating},''
  \href{http://dx.doi.org/10.1088/1475-7516/2021/04/012}{{\em JCAP} {\bfseries
  04} (2021) 012}, [\href{http://arxiv.org/abs/2012.10756}{{\tt
  arXiv:2012.10756}}].

\bibitem{Co:2020xaf}
R.~T. Co, E.~Gonzalez, and K.~Harigaya, ``{Increasing Temperature toward the
  Completion of Reheating},''
  \href{http://dx.doi.org/10.1088/1475-7516/2020/11/038}{{\em JCAP} {\bfseries
  11} (2020) 038}, [\href{http://arxiv.org/abs/2007.04328}{{\tt
  arXiv:2007.04328}}].

\bibitem{Barman:2022tzk}
B.~Barman, N.~Bernal, Y.~Xu, and O.~Zapata, ``{Ultraviolet freeze-in with a
  time-dependent inflaton decay},''
  \href{http://dx.doi.org/10.1088/1475-7516/2022/07/019}{{\em JCAP} {\bfseries
  07} no.~07, (2022) 019}, [\href{http://arxiv.org/abs/2202.12906}{{\tt
  arXiv:2202.12906}}].

\bibitem{Banerjee:2022fiw}
A.~Banerjee and D.~Chowdhury, ``{Fingerprints of freeze-in dark matter in an
  early matter-dominated era},'' \href{http://arxiv.org/abs/2204.03670}{{\tt
  arXiv:2204.03670}}.

\bibitem{Kallosh:2013hoa}
R.~Kallosh and A.~Linde, ``{Universality Class in Conformal Inflation},''
  \href{http://dx.doi.org/10.1088/1475-7516/2013/07/002}{{\em JCAP} {\bfseries
  07} (2013) 002}, [\href{http://arxiv.org/abs/1306.5220}{{\tt
  arXiv:1306.5220}}].

\bibitem{Kallosh:2013yoa}
R.~Kallosh, A.~Linde, and D.~Roest, ``{Superconformal Inflationary
  $\alpha$-Attractors},'' \href{http://dx.doi.org/10.1007/JHEP11(2013)198}{{\em
  JHEP} {\bfseries 11} (2013) 198}, [\href{http://arxiv.org/abs/1311.0472}{{\tt
  arXiv:1311.0472}}].

\bibitem{Ema:2015dka}
Y.~Ema, R.~Jinno, K.~Mukaida, and K.~Nakayama, ``{Gravitational Effects on
  Inflaton Decay},''
  \href{http://dx.doi.org/10.1088/1475-7516/2015/05/038}{{\em JCAP} {\bfseries
  05} (2015) 038}, [\href{http://arxiv.org/abs/1502.02475}{{\tt
  arXiv:1502.02475}}].

\bibitem{Ema:2016hlw}
Y.~Ema, R.~Jinno, K.~Mukaida, and K.~Nakayama, ``{Gravitational particle
  production in oscillating backgrounds and its cosmological implications},''
  \href{http://dx.doi.org/10.1103/PhysRevD.94.063517}{{\em Phys. Rev. D}
  {\bfseries 94} no.~6, (2016) 063517},
  [\href{http://arxiv.org/abs/1604.08898}{{\tt arXiv:1604.08898}}].

\bibitem{Ema:2018ucl}
Y.~Ema, K.~Nakayama, and Y.~Tang, ``{Production of Purely Gravitational Dark
  Matter},'' \href{http://dx.doi.org/10.1007/JHEP09(2018)135}{{\em JHEP}
  {\bfseries 09} (2018) 135}, [\href{http://arxiv.org/abs/1804.07471}{{\tt
  arXiv:1804.07471}}].

\bibitem{Ema:2019yrd}
Y.~Ema, K.~Nakayama, and Y.~Tang, ``{Production of purely gravitational dark
  matter: the case of fermion and vector boson},''
  \href{http://dx.doi.org/10.1007/JHEP07(2019)060}{{\em JHEP} {\bfseries 07}
  (2019) 060}, [\href{http://arxiv.org/abs/1903.10973}{{\tt
  arXiv:1903.10973}}].

\bibitem{Mambrini:2021zpp}
Y.~Mambrini and K.~A. Olive, ``{Gravitational Production of Dark Matter during
  Reheating},'' \href{http://dx.doi.org/10.1103/PhysRevD.103.115009}{{\em Phys.
  Rev. D} {\bfseries 103} no.~11, (2021) 115009},
  [\href{http://arxiv.org/abs/2102.06214}{{\tt arXiv:2102.06214}}].

\bibitem{Haque:2021mab}
M.~R. Haque and D.~Maity, ``{Gravitational dark matter: Free streaming and
  phase space distribution},''
  \href{http://dx.doi.org/10.1103/PhysRevD.106.023506}{{\em Phys. Rev. D}
  {\bfseries 106} no.~2, (2022) 023506},
  [\href{http://arxiv.org/abs/2112.14668}{{\tt arXiv:2112.14668}}].

\bibitem{Lebedev:2022ljz}
O.~Lebedev and J.-H. Yoon, ``{On gravitational preheating},''
  \href{http://dx.doi.org/10.1088/1475-7516/2022/07/001}{{\em JCAP} {\bfseries
  07} no.~07, (2022) 001}, [\href{http://arxiv.org/abs/2203.15808}{{\tt
  arXiv:2203.15808}}].

\bibitem{Clery:2021bwz}
S.~Clery, Y.~Mambrini, K.~A. Olive, and S.~Verner, ``{Gravitational portals in
  the early Universe},''
  \href{http://dx.doi.org/10.1103/PhysRevD.105.075005}{{\em Phys. Rev. D}
  {\bfseries 105} no.~7, (2022) 075005},
  [\href{http://arxiv.org/abs/2112.15214}{{\tt arXiv:2112.15214}}].

\bibitem{Haque:2022kez}
M.~R. Haque and D.~Maity, ``{Gravitational Reheating},''
  \href{http://arxiv.org/abs/2201.02348}{{\tt arXiv:2201.02348}}.

\bibitem{Clery:2022wib}
S.~Clery, Y.~Mambrini, K.~A. Olive, A.~Shkerin, and S.~Verner, ``{Gravitational
  Portals with Non-Minimal Couplings},''
  \href{http://arxiv.org/abs/2203.02004}{{\tt arXiv:2203.02004}}.

\bibitem{Barman:2021ugy}
B.~Barman and N.~Bernal, ``{Gravitational SIMPs},''
  \href{http://dx.doi.org/10.1088/1475-7516/2021/06/011}{{\em JCAP} {\bfseries
  06} (2021) 011}, [\href{http://arxiv.org/abs/2104.10699}{{\tt
  arXiv:2104.10699}}].

\bibitem{Garcia:2022vwm}
M.~A.~G. Garcia, M.~Pierre, and S.~Verner, ``{Scalar Dark Matter Production
  from Preheating and Structure Formation Constraints},''
  \href{http://arxiv.org/abs/2206.08940}{{\tt arXiv:2206.08940}}.

\bibitem{Aoki:2022dzd}
S.~Aoki, H.~M. Lee, A.~G. Menkara, and K.~Yamashita, ``{Reheating and dark
  matter freeze-in in the Higgs-R$^{2}$ inflation model},''
  \href{http://dx.doi.org/10.1007/JHEP05(2022)121}{{\em JHEP} {\bfseries 05}
  (2022) 121}, [\href{http://arxiv.org/abs/2202.13063}{{\tt
  arXiv:2202.13063}}].

\bibitem{Garny:2015sjg}
M.~Garny, M.~Sandora, and M.~S. Sloth, ``{Planckian Interacting Massive
  Particles as Dark Matter},''
  \href{http://dx.doi.org/10.1103/PhysRevLett.116.101302}{{\em Phys. Rev.
  Lett.} {\bfseries 116} no.~10, (2016) 101302},
  [\href{http://arxiv.org/abs/1511.03278}{{\tt arXiv:1511.03278}}].

\bibitem{Tang:2017hvq}
Y.~Tang and Y.-L. Wu, ``{On Thermal Gravitational Contribution to Particle
  Production and Dark Matter},''
  \href{http://dx.doi.org/10.1016/j.physletb.2017.10.034}{{\em Phys. Lett. B}
  {\bfseries 774} (2017) 676--681},
  [\href{http://arxiv.org/abs/1708.05138}{{\tt arXiv:1708.05138}}].

\bibitem{Garny:2017kha}
M.~Garny, A.~Palessandro, M.~Sandora, and M.~S. Sloth, ``{Theory and
  Phenomenology of Planckian Interacting Massive Particles as Dark Matter},''
  \href{http://dx.doi.org/10.1088/1475-7516/2018/02/027}{{\em JCAP} {\bfseries
  02} (2018) 027}, [\href{http://arxiv.org/abs/1709.09688}{{\tt
  arXiv:1709.09688}}].

\bibitem{Bernal:2018qlk}
N.~Bernal, M.~Dutra, Y.~Mambrini, K.~Olive, M.~Peloso, and M.~Pierre, ``{Spin-2
  Portal Dark Matter},''
  \href{http://dx.doi.org/10.1103/PhysRevD.97.115020}{{\em Phys. Rev. D}
  {\bfseries 97} no.~11, (2018) 115020},
  [\href{http://arxiv.org/abs/1803.01866}{{\tt arXiv:1803.01866}}].

\bibitem{Redi:2020ffc}
M.~Redi, A.~Tesi, and H.~Tillim, ``{Gravitational Production of a Conformal
  Dark Sector},'' \href{http://dx.doi.org/10.1007/JHEP05(2021)010}{{\em JHEP}
  {\bfseries 05} (2021) 010}, [\href{http://arxiv.org/abs/2011.10565}{{\tt
  arXiv:2011.10565}}].

\bibitem{Chianese:2020yjo}
M.~Chianese, B.~Fu, and S.~F. King, ``{Impact of Higgs portal on
  gravity-mediated production of superheavy dark matter},''
  \href{http://dx.doi.org/10.1088/1475-7516/2020/06/019}{{\em JCAP} {\bfseries
  06} (2020) 019}, [\href{http://arxiv.org/abs/2003.07366}{{\tt
  arXiv:2003.07366}}].

\bibitem{Kolb:2017jvz}
E.~W. Kolb and A.~J. Long, ``{Superheavy dark matter through Higgs portal
  operators},'' \href{http://dx.doi.org/10.1103/PhysRevD.96.103540}{{\em Phys.
  Rev. D} {\bfseries 96} no.~10, (2017) 103540},
  [\href{http://arxiv.org/abs/1708.04293}{{\tt arXiv:1708.04293}}].

\bibitem{Chianese:2020khl}
M.~Chianese, B.~Fu, and S.~F. King, ``{Interplay between neutrino and gravity
  portals for FIMP dark matter},''
  \href{http://dx.doi.org/10.1088/1475-7516/2021/01/034}{{\em JCAP} {\bfseries
  01} (2021) 034}, [\href{http://arxiv.org/abs/2009.01847}{{\tt
  arXiv:2009.01847}}].

\bibitem{Aghanim:2018eyx}
{\bfseries Planck} Collaboration, N.~Aghanim {\em et~al.}, ``{Planck 2018
  results. VI. Cosmological parameters},''
  \href{http://dx.doi.org/10.1051/0004-6361/201833910}{{\em Astron. Astrophys.}
  {\bfseries 641} (2020) A6}, [\href{http://arxiv.org/abs/1807.06209}{{\tt
  arXiv:1807.06209}}].

\bibitem{Espinosa:2015qea}
J.~R. Espinosa, G.~F. Giudice, E.~Morgante, A.~Riotto, L.~Senatore, A.~Strumia,
  and N.~Tetradis, ``{The cosmological Higgstory of the vacuum instability},''
  \href{http://dx.doi.org/10.1007/JHEP09(2015)174}{{\em JHEP} {\bfseries 09}
  (2015) 174}, [\href{http://arxiv.org/abs/1505.04825}{{\tt
  arXiv:1505.04825}}].

\bibitem{Ichikawa:2008iq}
K.~Ichikawa, T.~Suyama, T.~Takahashi, and M.~Yamaguchi, ``{Non-Gaussianity,
  Spectral Index and Tensor Modes in Mixed Inflaton and Curvaton Models},''
  \href{http://dx.doi.org/10.1103/PhysRevD.78.023513}{{\em Phys. Rev. D}
  {\bfseries 78} (2008) 023513}, [\href{http://arxiv.org/abs/0802.4138}{{\tt
  arXiv:0802.4138}}].

\bibitem{Garcia:2020eof}
M.~A.~G. Garcia, K.~Kaneta, Y.~Mambrini, and K.~A. Olive, ``{Reheating and
  Post-inflationary Production of Dark Matter},''
  \href{http://dx.doi.org/10.1103/PhysRevD.101.123507}{{\em Phys. Rev. D}
  {\bfseries 101} no.~12, (2020) 123507},
  [\href{http://arxiv.org/abs/2004.08404}{{\tt arXiv:2004.08404}}].

\bibitem{Co:2022bgh}
R.~T. Co, Y.~Mambrini, and K.~A. Olive, ``{Inflationary Gravitational
  Leptogenesis},'' \href{http://arxiv.org/abs/2205.01689}{{\tt
  arXiv:2205.01689}}.

\bibitem{Harigaya:2013vwa}
K.~Harigaya and K.~Mukaida, ``{Thermalization after/during Reheating},''
  \href{http://dx.doi.org/10.1007/JHEP05(2014)006}{{\em JHEP} {\bfseries 05}
  (2014) 006}, [\href{http://arxiv.org/abs/1312.3097}{{\tt arXiv:1312.3097}}].

\bibitem{Quiros:1999jp}
M.~Quiros, ``{Finite temperature field theory and phase transitions},''
  \href{http://arxiv.org/abs/hep-ph/9901312}{{\tt hep-ph/9901312}}.

\bibitem{Ichikawa:2008ne}
K.~Ichikawa, T.~Suyama, T.~Takahashi, and M.~Yamaguchi, ``{Primordial Curvature
  Fluctuation and Its Non-Gaussianity in Models with Modulated Reheating},''
  \href{http://dx.doi.org/10.1103/PhysRevD.78.063545}{{\em Phys. Rev. D}
  {\bfseries 78} (2008) 063545}, [\href{http://arxiv.org/abs/0807.3988}{{\tt
  arXiv:0807.3988}}].

\bibitem{Kaneta:2022gug}
K.~Kaneta, S.~M. Lee, and K.-y. Oda, ``{Boltzmann or Bogoliubov? Approaches
  Compared in Gravitational Particle Production},''
  \href{http://arxiv.org/abs/2206.10929}{{\tt arXiv:2206.10929}}.

\bibitem{Dufaux:2006ee}
J.~F. Dufaux, G.~N. Felder, L.~Kofman, M.~Peloso, and D.~Podolsky,
  ``{Preheating with trilinear interactions: Tachyonic resonance},''
  \href{http://dx.doi.org/10.1088/1475-7516/2006/07/006}{{\em JCAP} {\bfseries
  07} (2006) 006}, [\href{http://arxiv.org/abs/hep-ph/0602144}{{\tt
  hep-ph/0602144}}].

\bibitem{Abolhasani:2009nb}
A.~A. Abolhasani, H.~Firouzjahi, and M.~M. Sheikh-Jabbari, ``{Tachyonic
  Resonance Preheating in Expanding Universe},''
  \href{http://dx.doi.org/10.1103/PhysRevD.81.043524}{{\em Phys. Rev. D}
  {\bfseries 81} (2010) 043524}, [\href{http://arxiv.org/abs/0912.1021}{{\tt
  arXiv:0912.1021}}].

\bibitem{Gelmini:2006pw}
G.~B. Gelmini and P.~Gondolo, ``{Neutralino with the right cold dark matter
  abundance in (almost) any supersymmetric model},''
  \href{http://dx.doi.org/10.1103/PhysRevD.74.023510}{{\em Phys. Rev. D}
  {\bfseries 74} (2006) 023510},
  [\href{http://arxiv.org/abs/hep-ph/0602230}{{\tt hep-ph/0602230}}].

\bibitem{Ellis:2021kad}
J.~Ellis, M.~A.~G. Garcia, D.~V. Nanopoulos, K.~A. Olive, and S.~Verner,
  ``{BICEP/Keck constraints on attractor models of inflation and reheating},''
  \href{http://dx.doi.org/10.1103/PhysRevD.105.043504}{{\em Phys. Rev. D}
  {\bfseries 105} no.~4, (2022) 043504},
  [\href{http://arxiv.org/abs/2112.04466}{{\tt arXiv:2112.04466}}].

\bibitem{BICEP:2021xfz}
{\bfseries BICEP, Keck} Collaboration, P.~A.~R. Ade {\em et~al.}, ``{Improved
  Constraints on Primordial Gravitational Waves using Planck, WMAP, and
  BICEP/Keck Observations through the 2018 Observing Season},''
  \href{http://dx.doi.org/10.1103/PhysRevLett.127.151301}{{\em Phys. Rev.
  Lett.} {\bfseries 127} no.~15, (2021) 151301},
  [\href{http://arxiv.org/abs/2110.00483}{{\tt arXiv:2110.00483}}].

\bibitem{Tristram:2021tvh}
M.~Tristram {\em et~al.}, ``{Improved limits on the tensor-to-scalar ratio
  using BICEP and Planck data},''
  \href{http://dx.doi.org/10.1103/PhysRevD.105.083524}{{\em Phys. Rev. D}
  {\bfseries 105} no.~8, (2022) 083524},
  [\href{http://arxiv.org/abs/2112.07961}{{\tt arXiv:2112.07961}}].

\bibitem{deSalas:2015glj}
P.~F. de~Salas, M.~Lattanzi, G.~Mangano, G.~Miele, S.~Pastor, and O.~Pisanti,
  ``{Bounds on very low reheating scenarios after Planck},''
  \href{http://dx.doi.org/10.1103/PhysRevD.92.123534}{{\em Phys. Rev. D}
  {\bfseries 92} no.~12, (2015) 123534},
  [\href{http://arxiv.org/abs/1511.00672}{{\tt arXiv:1511.00672}}].

\end{thebibliography}\endgroup
\end{document}